%% file: template.tex
\documentclass[journal]{vgtc}                     

\onlineid{1062}

\vgtccategory{Data Transformations}

\title{EVOLVE: Efficient Learned Volume Compression with Variable-Rate Encoding on a Cross-Domain Database}

\author{%
  \authororcid{Kaiyuan Tang}{0009-0001-3512-0112},
  \authororcid{Maizhe Yang}{0009-0002-8120-7339}, and
  \authororcid{Chaoli Wang}{0000-0002-0859-3619}
}

\authorfooter{
  \item The authors are with the Department of Computer Science and Engineering, University of Notre Dame, Notre Dame, IN 46556, USA.
  \item E-mail: \{ktang2, myang9, chaoli.wang\}@nd.edu.
}

\abstract{%
Large-scale scientific simulations generate volumetric data at rates that far outpace advances in storage and network bandwidth, making effective lossy compression increasingly critical. However, conventional compressors often struggle to preserve fine structural details at high compression ratios (CRs), and implicit neural representations (INRs) require costly per-volume optimization and produce models with fixed CRs. To respond, we present EVOLVE, an autoencoder (AE)-based volume-compression framework \pin{that targets high CRs for offline compression}, with three key contributions. First, we construct a large-scale cross-domain database of 6,376 volumes from 21 scientific simulations, curated via perceptual hashing to ensure diversity, \hot{enabling the optimized model to extract features that generalize across volumes within the covered scientific simulation domains}. Second, we reexamine the design space of AE-based compressors and incorporate several macro- and micro-designs into a vanilla AE to develop EVOLVE, which substantially improves the expressive power and compression capability. Third, we develop a learnable gain mechanism with a three-stage training strategy to enable variable-rate encoding, allowing a single model to support continuous CR adjustment at inference time. Experiments on multiple unseen \hot{scientific simulation} datasets demonstrate that EVOLVE achieves substantially higher CRs than conventional compressors at comparable reconstruction quality, while delivering compression speeds that are orders of magnitude faster than INR-based methods, highlighting its promise as a strong alternative for compressing scientific data. \hot{The code, model weights, and results are available on our project page at \url{https://evolve-vis.github.io}.}
}


\teaser{
\centering
\begin{tabular}{c@{\hspace{0.02in}}c@{\hspace{0.02in}}c@{\hspace{0.02in}}c}
\includegraphics[height=0.1875\textwidth]{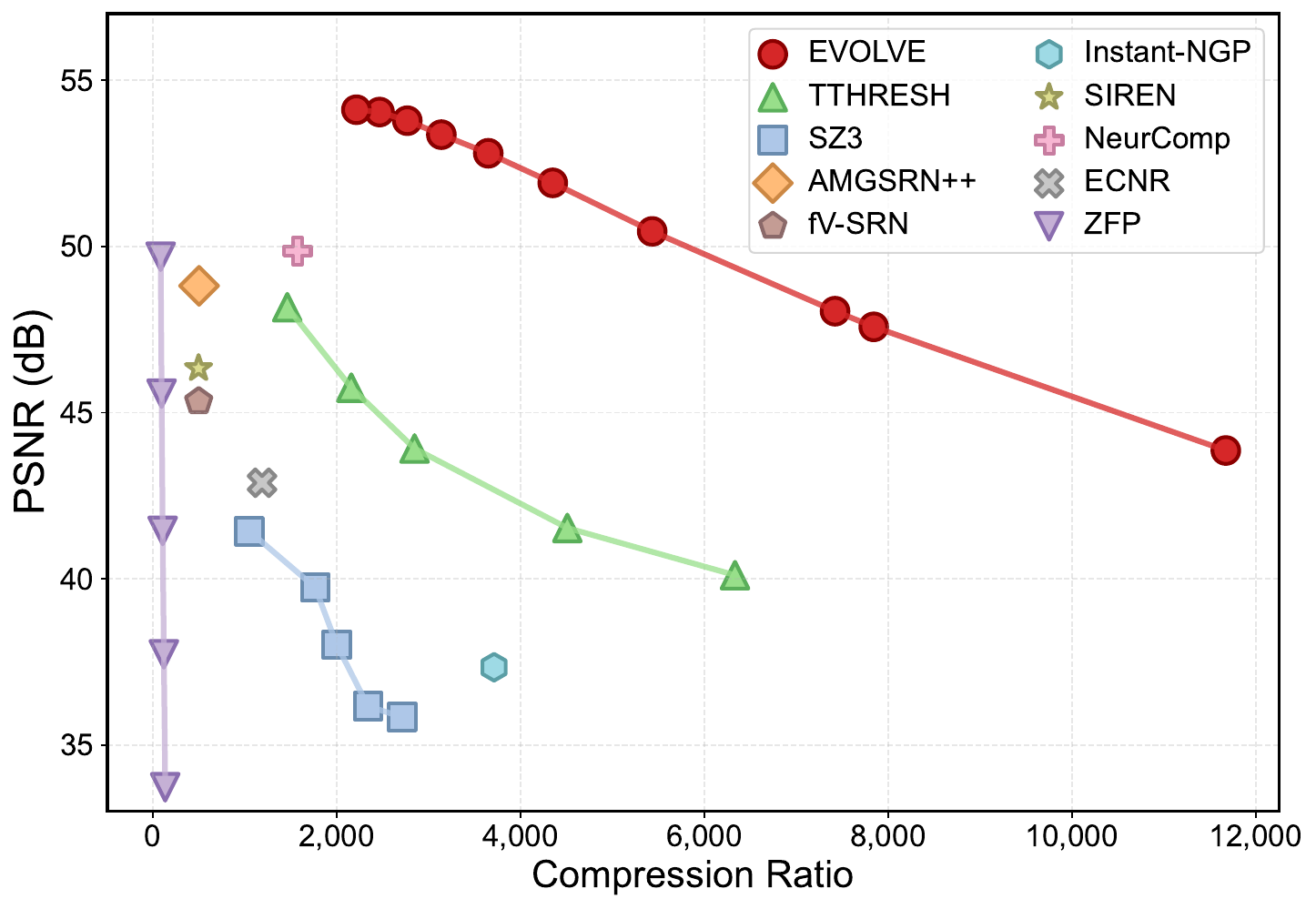} &
\includegraphics[height=0.1875\textwidth]{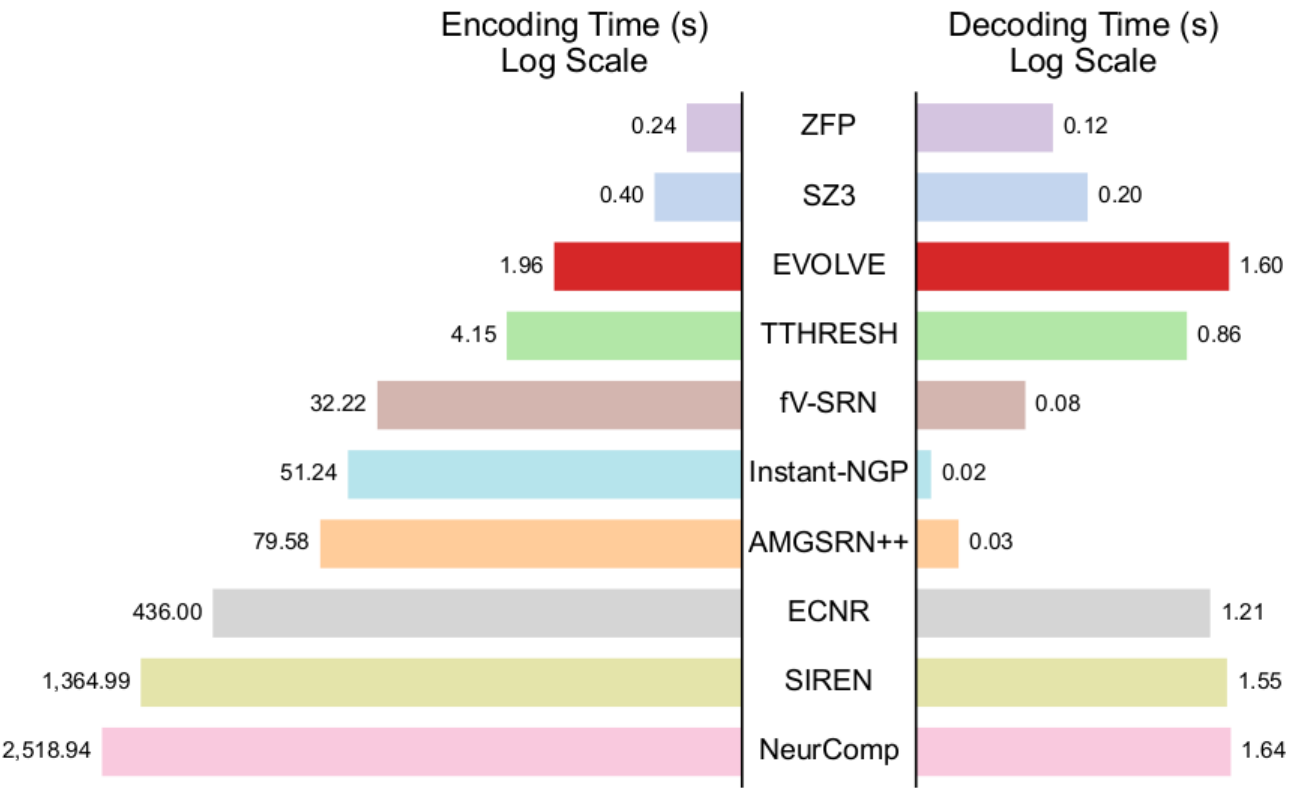} &
\includegraphics[height=0.1875\textwidth]{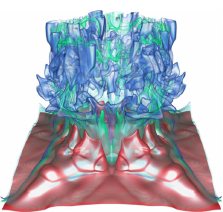} &
\includegraphics[height=0.1875\textwidth]{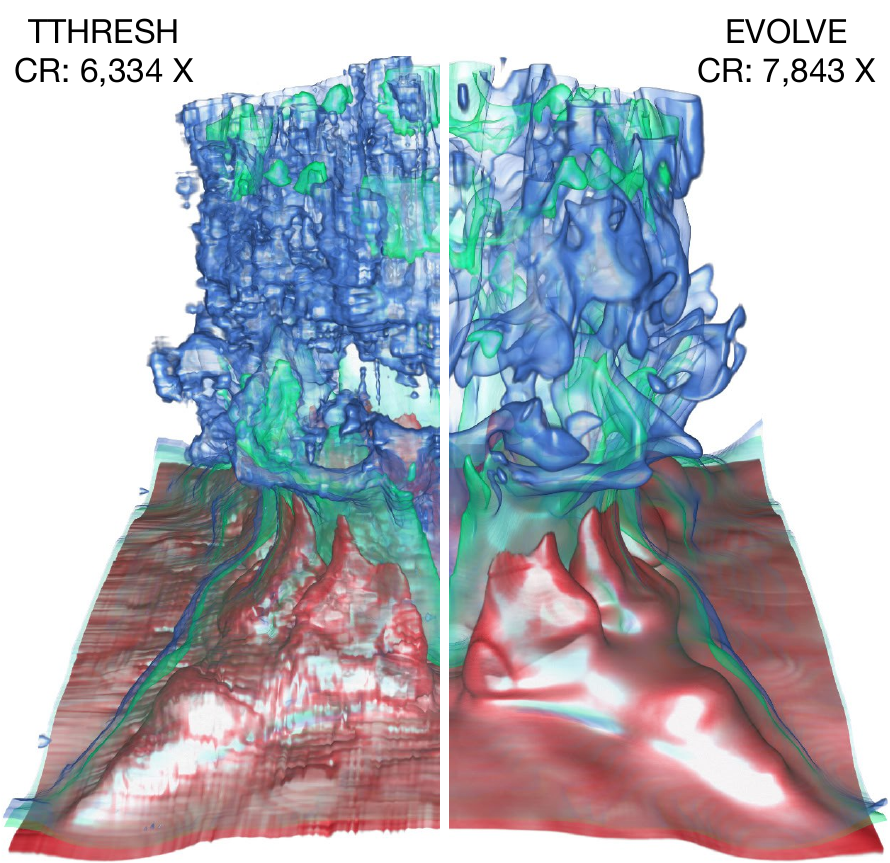} \\
\mbox{\small (a) rate-distortion curves} &
\mbox{\small (b) encoding/decoding time} &
\mbox{\small (c) GT} &
\mbox{\small (d) TTHRESH vs.\ EVOLVE} \\
\end{tabular}
\vspace{-0.125in}
\caption{
After training, EVOLVE, \pin{as an offline compression method, can achieve} a competitive compression ratio (CR) on the unseen \textsf{ionization (H+)} dataset (141 MB). Unlike prior learning-based methods that require storing one model per CR setting, a single trained EVOLVE model supports a wide range of variable-rate encoding, as shown in (a). The encoding/decoding time plot in (b) uses a log scale to improve the visibility of extremely small values. \hot{A more comprehensive comparison of encoding/decoding time is given in Tables~\ref{tab:traditional-results} and~\ref{tab:neural-baselines}.} All evaluations are conducted on an NVIDIA 4090 GPU.
}
\label{fig:teaser}
}

\graphicspath{{figs/}{figures/}{pictures/}{images/}{./}} 

\usepackage{newtxtext}     
\usepackage{bm}
\usepackage{graphicx}
\usepackage{algorithm}
\usepackage{algorithmic}
\usepackage{amsfonts}
\usepackage{booktabs}         
\usepackage{amsmath}
\usepackage{comment}
\usepackage{caption}
\usepackage{multirow}

\usepackage{color} 
\usepackage{anyfontsize}
\usepackage{soul}
\usepackage[normalem]{ulem} 

\newcommand{\hot}[1]{{\color{black} #1}}
\newcommand{\pin}[1]{{\color{black} #1}}

\newenvironment{myitemize}{
\begin{itemize}
 \setlength{\itemsep}{1pt}
 \setlength{\parskip}{0pt}
 \setlength{\parsep}{0pt}}{\end{itemize}
}

\begin{document}


\firstsection{Introduction}
\maketitle
\input{intro.tex}

\vspace{-0.075in}
\input{related.tex}

\vspace{-0.075in}
\input{method.tex}

\vspace{-0.075in}
\input{result.tex}

\vspace{-0.075in}
\input{conclusion.tex}


\appendix 
\crefalias{section}{appendix} 

\input{appendix}

\clearpage 
\acknowledgments{This research was supported in part by the U.S.\ National Science Foundation through grants IIS-2101696, OAC-2104158, IIS-2401144, and CCF-2550610, and the U.S.\ Department of Energy through grant DE-SC0023145. We thank the anonymous reviewers for their insightful comments.}
\vspace{-0.075in}
\bibliographystyle{abbrv-doi-hyperref-narrow}

\bibliography{template-abbv}

\end{document}

%% file: intro.tex
Modern scientific simulations and instruments produce volumetric data at rates that far outpace the growth of storage and network capacity.
Large-scale astrophysics, climate science, and turbulent combustion simulations routinely generate extreme-scale data, while the cost of storing and moving such data continues to grow.
For this reason, data archival and transmission have become major bottlenecks in scientific workflows, limiting both the scale of simulations that can be preserved and the speed at which results can be shared and analyzed.

As a viable solution, lossy compression addresses this challenge by significantly reducing data size while preserving essential features for visualization and analysis.
Conventional error-bounded lossy compressors, such as ZFP~\cite{Lindstrom-TVCG14}, TTHRESH~\cite{Ballester-TVCG20}, and SZ3~\cite{Liang-TBD23}, rely on predefined mathematical transforms or predictors.
While effective at moderate compression ratios (CRs), say a few hundred, these methods suffer severe distortion at high CRs (over 1,000).
\pin{Besides these general-purpose compressors, non-neural learning-based solutions, such as hierarchical vector quantization~\cite{Schneider-VIS03} and sparse dictionary learning~\cite{Gobbetti-CGF12, Marton-CGF19, Diaz-CG20}, learn codebooks or dictionaries directly from the data, which have mostly been integrated into real-time compression-domain rendering pipelines.}

Implicit neural representations (INRs)~\cite{Sitzmann-NeurIPS20} treat compression as function approximation and support random access.
They yield high CRs while preserving good data quality. 
However, INR-based approaches have significant drawbacks: fully connected INRs~\cite{Sitzmann-NeurIPS20, Lu-CGF21, Han-TVCG24} may require several hours of training time per volume; grid-based INRs~\cite{Weiss-CGF22, Wu-TVCG24, Wurster-TVCG24} accelerate training but achieve lower CRs; and a single trained model corresponds to only one specific CR.

Autoencoder (AE) methods, such as AE-SZ~\cite{Liu-CLUSTER21} and IDLat~\cite{Shen-TVCG23}, map volume blocks to compact latent representations, yet they exhibit notable limitations as compression tools.
First, their CRs do not demonstrate significant advantages over state-of-the-art conventional compressors~\cite{Ballester-TVCG20, Liang-TBD23}.
Second, these methods are typically trained on small-scale datasets (e.g., a few volume blocks sampled from a single time-varying volumetric dataset). 
While they can generalize to other timesteps within the same dataset, cross-dataset generalization remains an open challenge.
Third, as with INR methods, the CRs of AEs are determined by fixed latent-space dimensions, thereby requiring different model architectures for different target rates.
\pin{Due to these limitations, prior research has predominantly leveraged AEs as visual analysis tools~\cite{Han-TVCG20, Shen-TVCG23} rather than as competitive compression tools.}

To address these challenges, we rethink the application of AEs to volumetric data compression and propose an AE-based framework, named EVOLVE (\underline{E}fficient Learned \underline{VOL}ume Compression with \underline{V}ariable-Rate \underline{E}ncoding).
\pin{Unlike methods that incorporate compression-domain rendering, EVOLVE targets high CRs for offline purposes, where the full volume is reconstructed before visualization and analysis.
Under this setting, EVOLVE aims to encode and decode various volumes with a single shared model (see Section~\ref{sec:deployment}).}
We introduce three key innovations to overcome the existing limitations.
First, to achieve higher CRs than conventional compressors at equivalent reconstruction quality, we incorporate advanced {\em context-aware entropy modeling} that combines \hot{spatial-wise~\cite{Minnen-NeurIPS18, Cheng-CVPR20, He-CVPR21} and channel-wise context~\cite{Minnen-ICIP20}}, fusing multi-source information to estimate latent-variable distributions accurately.
Second, to enable cross-dataset generalization, we construct a large-scale training database comprising thousands of volumes spanning diverse scientific simulation domains, \hot{allowing the model to learn {\em transferable feature representations} and, after being optimized on sufficiently representative volumes from a given domain, it can support compress unseen data without per-volume optimization}.
Third, to support flexible CRs with a single model, we adopt a {\em learnable gain mechanism} that enables continuous rate variation, coupled with a {\em three-stage training strategy} to ensure consistent performance across the entire rate-distortion spectrum.

We evaluate EVOLVE on multiple scientific simulation datasets against conventional compressors, INR methods, and prior AE approaches.
At equivalent reconstruction quality, EVOLVE attains substantially higher CRs than both conventional and AE-based methods (see Figure~\ref{fig:teaser}), while compressing orders of magnitude faster than INR methods on encoding speed.
Moreover, a single EVOLVE model spans a continuous rate-distortion range, eliminating the need to train a separate model for each target CR, and generalizes well to volumes unseen during training.
\hot{That said, EVOLVE is validated primarily on curated scientific simulation data; extending it to other domains, such as medical or microscopy volumes, would require retraining on representative data.}

In summary, this paper makes the following contributions.
\begin{myitemize}
\item
We propose EVOLVE, an AE-based volumetric data compression framework that achieves high compression efficiency through advanced context-aware entropy modeling and significantly outperforms existing methods at equivalent reconstruction quality, \hot{while supporting continuous single-model variable-rate compression without requiring multiple models per setting}.
\item
We construct the first large-scale volume database spanning diverse scientific domains, \hot{enabling a single trained compression model to generalize to unseen volumes from the covered scientific simulation domains without per-volume optimization}.
\item
Extensive experiments show that EVOLVE achieves state-of-the-art compression performance across multiple datasets while demonstrating strong generalization and practical efficiency \hot{for unseen simulation data}.
\end{myitemize}

%% file: related.tex
\vspace{-0.05in}
\section{Related Work}
\label{sec:related}

\subsection{Lossy Volumetric Data Compression}

Lossy compression has been critical for managing the massive storage requirements and addressing I/O challenges posed by large-scale scientific simulations.
\hot{Many schemes embed compressed representations into the rendering pipeline, decoding only the portion needed per frame on demand.
Treib et al.~\cite{Treib-TVCG12} developed a GPU-decodable wavelet codec for terascale turbulence, Nystad et al.~\cite{Nystad-HPG12} introduced the fixed-rate block-local texture format ASTC with constant-time random access, and Schneider and Westermann~\cite{Schneider-VIS03} proposed a hierarchical vector-quantization scheme.
Gobbetti et al.~\cite{Gobbetti-CGF12} developed COVRA, which models octree data blocks at multiple resolutions as sparse combinations of learned dictionary atoms, and Marton et al.~\cite{Marton-CGF19} and D\'iaz et al.~\cite{Diaz-CG20} later scaled this direction to massive time-varying data with out-of-core GPU streaming of variable-rate sparse codes.
\pin{Among them, works including~\cite{Schneider-VIS03, Gobbetti-CGF12, Marton-CGF19, Diaz-CG20} are considered learning-based solutions that optimize dictionaries and sparse codes from the data.}
Such methods prioritize random access and decoding throughput at render time.}
\pin{For more details about compression-domain volume rendering, we refer readers to the surveys~\cite{Balsa-CGF14, Beyer-CGF15}.}
\hot{A complementary line of work targets offline storage and transmission, decompressing the full volume before any visualization or analysis.
Lindstrom~\cite{Lindstrom-TVCG14} developed ZFP, which applies customized block-level transforms with fast I/O access; Liang et al.~\cite{Liang-TBD23} developed SZ3, an error-bounded framework built on an improved Lorenzo predictor; and Ballester-Ripoll et al.~\cite{Ballester-TVCG20} proposed TTHRESH, using tensor decomposition via higher-order SVD.
Yan et al.~\cite{Yan-TVCG24} further proposed TopoSZ to preserve topological features under user-specified error bounds.
Like these methods, EVOLVE is designed for offline compression, reconstructing the full volume before use rather than for transient render-time decoding (for more discussion, refer to Section~\ref{sec:deployment}).}

\vspace{-0.05in}
\subsection{Deep Learning for Volume Visualization}

Deep learning solutions~\cite{Han-TVCG20, Han-TVCG22-STNet, Tang-TVCG25-StyleRFVolVis,Tang-TVCG25-iVRGS,Tang-TVCG26,Ai-TVCG26, Jeon-VIS26} have been increasingly applied to volume generation, scene representation and data compression~\cite{Wang-TVCG23}.
Since SIREN~\cite{Sitzmann-NeurIPS20} introduced periodic activation functions to capture high-frequency details, INRs have emerged as a promising paradigm for representing data as continuous functions parameterized by neural networks.
NeurComp~\cite{Lu-CGF21} was the first to apply SIREN-based architectures with residual connections to volumetric scalar field compression, utilizing network weight quantization to achieve high CRs.
\hot{To accelerate training and inference, grid-based approaches replace a large part of the fully connected layers with trainable feature grids (e.g., the multiresolution hash encoding of Instant-NGP~\cite{Muller-TOG22}) that are queried by a lightweight multilayer perceptron (MLP), trading larger model sizes for orders-of-magnitude faster optimization and rendering.}
fV-SRN~\cite{Weiss-CGF22} uses a coarse grid of learnable latent features, together with a small MLP, to enable interactive volume rendering via custom CUDA TensorCore kernels.
APMGSRN~\cite{Wurster-TVCG24} improves reconstruction quality through multiple spatially adaptive feature grids that dynamically allocate network parameters to high-error regions.
For time-varying data, KD-INR~\cite{Han-TVCG24} employs knowledge distillation to sequentially compress individual timesteps into a single coherent model, while ECNR~\cite{Tang-PVis24} utilizes a Laplacian pyramid for multiscale spatiotemporal decomposition with parallel MLPs. 
Other examples include STSR-INR~\cite{Tang-CG24}, Meta-INR~\cite{Yang-PVISVN25}, MC-INR~\cite{Son-VISSP25}, and Lossless-INR~\cite{Tang-VISSP26}. 
F-Hash~\cite{Sun-TVCG26} further extends multiresolution hash encoding to the spatiotemporal domain through a feature-based tesseract encoding, accelerating convergence on time-varying volumes.
However, INR methods require per-volume training, limiting their practical applicability for general-purpose compression.

AE-based methods offer an alternative by learning generalizable latent representations.
Deep Fluids~\cite{Kim-CGF19} employs a generative AE to learn compact latent representations of fluid flows from parameterized simulations.
FlowNet~\cite{Han-TVCG20} uses a sparse stacked AE to learn implicit feature descriptors from streamlines or stream surfaces for clustering and selection.
AE-SZ~\cite{Liu-CLUSTER21} integrates a convolutional AE as a predictor within the SZ framework to improve prediction accuracy.
IDLat~\cite{Shen-TVCG23} generates latent representations guided by spatial importance maps, ensuring higher reconstruction quality in user-specified regions.
While these methods demonstrate the potential of AEs for scientific data, they are typically trained on small-scale datasets from single simulations and achieve CRs around a few hundred at comparable quality, which does not significantly outperform state-of-the-art conventional compressors.
Our EVOLVE addresses these limitations by training on a database with thousands of volumes to achieve cross-dataset generalization, while substantially surpassing both prior AE methods and conventional compressors in CR at equivalent reconstruction quality.

\vspace{-0.05in}
\subsection{\pin{Learned Neural Data Compression}}

End-to-end learned compression has achieved remarkable success in image coding, surpassing traditional codecs such as JPEG~\cite{Wallace-CACM91} in rate-distortion performance.
Unlike conventional approaches that rely on handcrafted transforms and entropy coders, learned methods jointly optimize the encoder, decoder, and entropy model through gradient descent, enabling the network to discover data-adaptive representations.
Ball{\'{e}} et al.~\cite{Balle-ICLR17, Balle-ICLR18} established the foundational variational AE framework with nonlinear transforms and introduced the scale hyperprior to capture spatial dependencies in latent representations.
Subsequent work focused on improving entropy modeling through context models.
Minnen et al.~\cite{Minnen-NeurIPS18} combined the hyperprior with autoregressive context modeling to outperform traditional codecs, and subsequent research~\cite{Minnen-ICIP20, He-CVPR21, He-CVPR22} further improved the accuracy of entropy estimation and decoding efficiency.
For variable-rate compression, Cui et al.~\cite{Cui-CVPR21} proposed asymmetric gain units that rescale latent magnitudes, enabling continuous bitrate adjustment within a single model. 
These techniques have been extended to video compression~\cite{Lu-CVPR19, Sheng-TMM23} and Gaussian splatting~\cite{Wang-NeurIPS24, Chen-ECCV24, Tang-VIS26, Pan-CVPR26}, demonstrating the broad applicability of learned entropy modeling.
In this work, we extend 2D learned compression techniques to 3D volumetric data by adapting context models for spatial volumes and training on a large-scale volume database to achieve \hot{generalization across the diverse scientific simulation domains it covers}.


%% file: method.tex
\section{EVOLVE}

This section presents the three core contributions of EVOLVE.
First, we describe how we collect and curate our high-quality database of thousands of volumes for training in Section~\ref{sec:database-curation}.
Second, in Section~\ref{sec:evolve-roadmap}, we provide an ablation-based roadmap that progressively upgrades an initially underperforming vanilla AE into EVOLVE, ultimately outperforming existing state-of-the-art conventional compressors.
Third, we describe how we achieve variable-rate encoding in EVOLVE, enabling multiple CRs within a single model for more flexible deployment, as detailed in Section~\ref{sec:variable-rate}.

\vspace{-0.05in}
\subsection{Database Curation}
\label{sec:database-curation}


\pin{Training an effective learned volume compression model requires a large-scale, diverse, and high-quality database: models trained on narrow data distributions tend to overfit dataset-specific spatial statistics rather than learn transferable representations.}
Therefore, we curate our database by collecting time-varying scientific simulation data from multiple established repositories, \hot{including the Open SciVis Datasets~\cite{Klacansky-OSV}, IEEE SciVis Contest archives~\cite{SciVis-Contest}, ETH Z{\"u}rich visualization datasets~\cite{ETH-CGL-Data}, and Well~\cite{Ohana-NeurIPS24}, as well as our in-house simulation data}.
The resulting database comprises 21 datasets totaling 9,921 volumes across various scientific domains, before further filtering.
This breadth ensures that the training data covers a wide range of physical quantities and spatial structures.
\hot{Its full composition is provided in Appendix~\ref{sec:database-detail}.}

Time-varying simulations often exhibit significant redundancy across consecutive frames, which can be suboptimal for training.
For example, \textsf{five-jet} contains 2,000 timesteps, yet the dynamics evolve slowly, making adjacent frames nearly indistinguishable.
Directly using all 9,921 volumes can lead the model to memorize over-represented local patterns, reducing training efficiency and harming out-of-distribution performance, as observed in prior large-scale database curation studies~\cite{Lee-ACL22, Abbas-NeurIPS23}. 
To improve the quality and diversity of our database, we design a curation process to identify and remove redundant volumes before training.
Specifically, we adopt pHash~\cite{Zauner-SPIE11}, a perceptual hash-based strategy, to quantify structural similarity and identify near-duplicate samples.
Specifically, we apply a 3D discrete cosine transform (DCT) to each volume, binarize the low-frequency DCT coefficients into a perceptual hash, and measure pairwise similarity using the Hamming distance between hashes.
Inspired by nearest neighbor search~\cite{Webster-arXiv23, Oquab-TMLR24}, we define the {\em nearest-neighbor similarity} (NNS) for each volume as the similarity to its closest neighbor and use the 95th-percentile of NNS within each dataset as a diversity metric.
In our corpus, datasets such as \textsf{five-jet} and \textsf{supercurrent} initially exhibit very high redundancy, with the 95th-percentile NNS exceeding 0.95, indicating that many frames are near-duplicates.
We perform deduplication for datasets whose 95th-percentile NNS exceeds 0.85 by clustering highly similar volumes with a union–find algorithm and retaining one representative per cluster. 
We repeat this filtering until the dataset-level NNS falls below 0.85.
\hot{After each filtering iteration, we further validate the results with a visual inspection step that operates only on the clusters identified by deduplication: we render each retained representative side-by-side with its flagged near-duplicates using random transfer functions.
Since the goal is to detect structural near-identity between candidate duplicates rather than to produce perceptually meaningful visualizations, randomly sampled transfer functions are sufficient.}
After filtering, the final training set contains 6,376 volumes with substantially improved diversity in data distribution.

\hot{{\bf Train/validation/test split.}
All 6,376 curated volumes are used exclusively for training, while design decisions are made on a small validation subset of the training volumes, with its own 95\%/5\% train/validation split (see Section~\ref{sec:evolve-roadmap}).
For evaluation, we deliberately avoid splitting along the temporal axis, as holding out intermediate timesteps of highly correlated sequences would inflate test performance. All test volumes (see Table~\ref{tab:test_dataset}) are drawn from unseen simulations or variables and are disjoint from the 9,921 collected volumes.}

\vspace{-0.05in}
\subsection{Ablation-Based Roadmap}
\label{sec:evolve-roadmap}

In this section, we describe the design process of EVOLVE through a systematic architectural ablation roadmap.
We conduct this ablation study using a small subset of the 6,376 training database to identify effective design choices.
Specifically, for each dataset listed in Table~\ref{tab:dataset}, we randomly sample 5\% of the selected volumes, resulting in a subset of 319 volumes. 
We then use 95\% of them (303 volumes) for training and the remaining 5\% (16 volumes) for validation.
Experimenting first on such a small subset of volumes allows us to compare different architectural variants while avoiding the optimization noise introduced by long training cycles on the large volume database.

\begin{figure}[!ht]
\centering
\includegraphics[width=\linewidth]{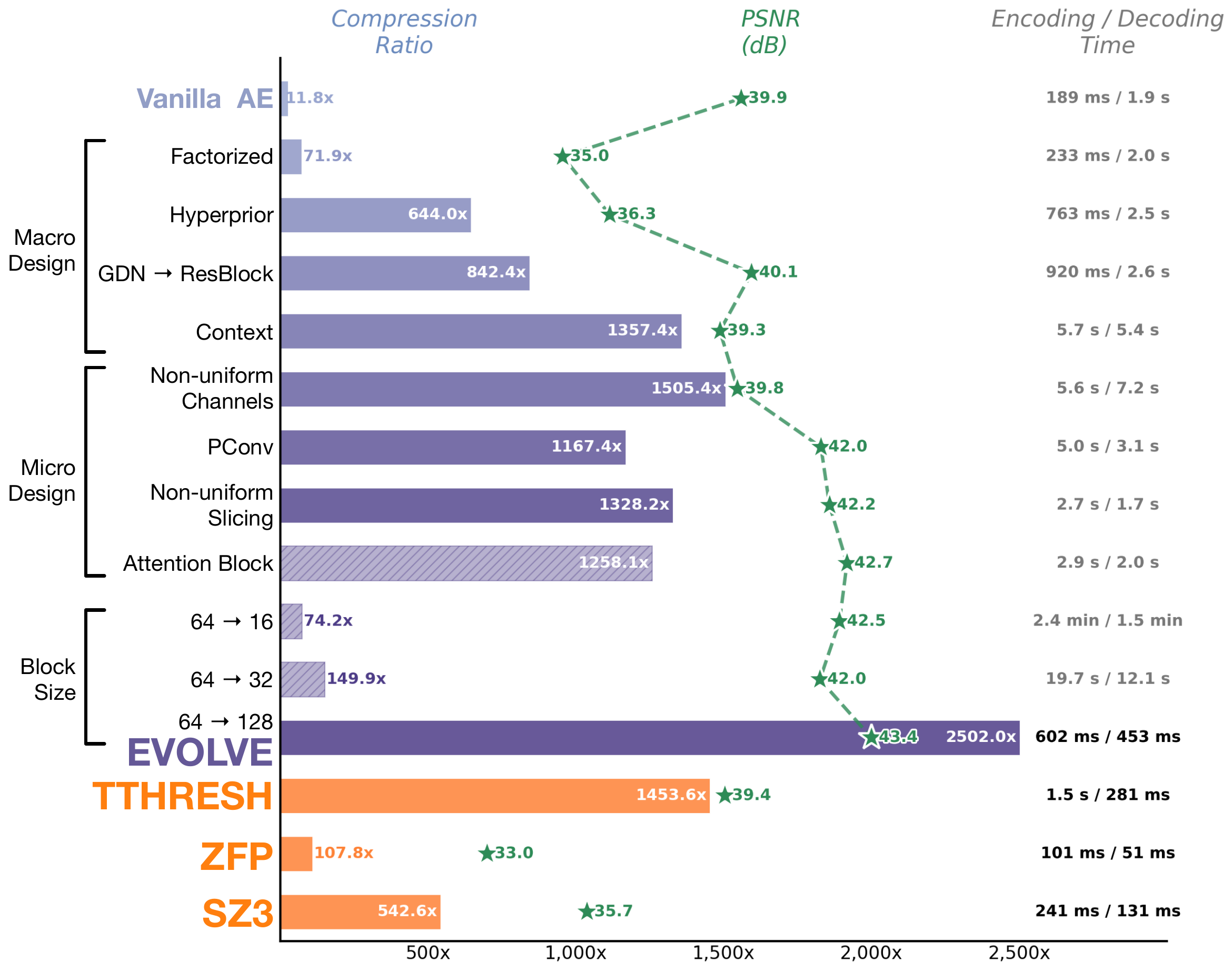}
\vspace{-.25in}
\caption{We modernize a vanilla AE towards EVOLVE through progressive improvements. All evaluations are conducted on an NVIDIA RTX 4090 workstation. A hatched bar indicates that the corresponding modification is not adopted. Note that although only a 303-volume subset is used for training in this roadmap study, the final EVOLVE model already achieves performance that surpasses conventional compression methods.}
\label{fig:roadmap}
\end{figure}

We start with a vanilla convolutional AE that has strided convolution layers for downsampling and deconvolution layers for upsampling, with \emph{generalized divisive normalization} (GDN)~\cite{Balle-GDN} as the normalizing function, following common design choices in prior AE-based compressors~\cite{Liu-CLUSTER21, Shen-TVCG23}.
To avoid excessive memory consumption, we adopt the same block-based processing strategy~\cite{Han-TVCG-SSR, Han-TVCG22-STNet} in which the network operates on fixed-size blocks cropped from the target volume. 
We set the block resolution to 64$\times$64$\times$64 as our starting point.
During inference, the decoded blocks are merged using a weighted blending scheme, which leverages spatial overlap to avoid boundary discontinuities.
The vanilla AE is optimized using a mean squared error (MSE) loss computed between the ground truth (GT) and the decoded volumes.
We then progressively modernize each component of this vanilla AE toward our final EVOLVE model, as summarized in Figure~\ref{fig:roadmap}.


\begin{figure*}[htb]
\centering
$\begin{array}{c@{\hspace{0.1in}}c@{\hspace{0.1in}}c@{\hspace{0.1in}}c}
 \includegraphics[height=1.45in]{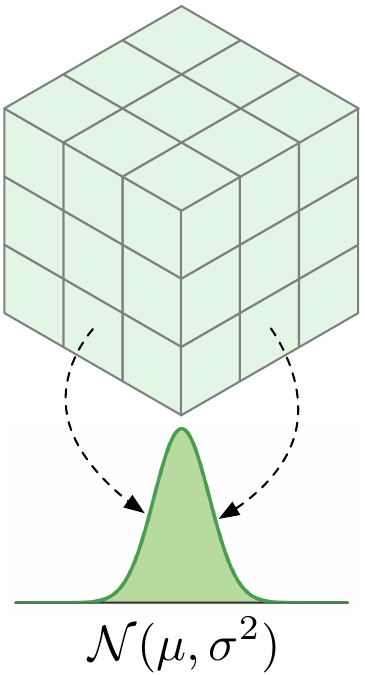}&
  \includegraphics[height=1.45in]{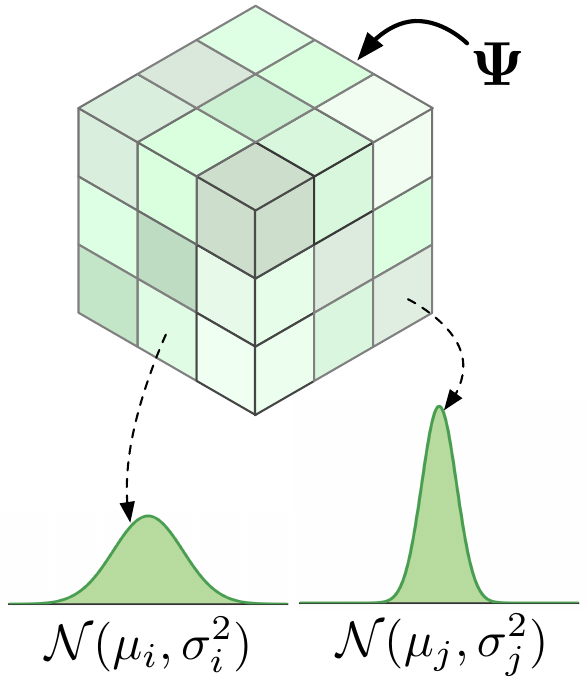}&
   \includegraphics[height=1.45in]{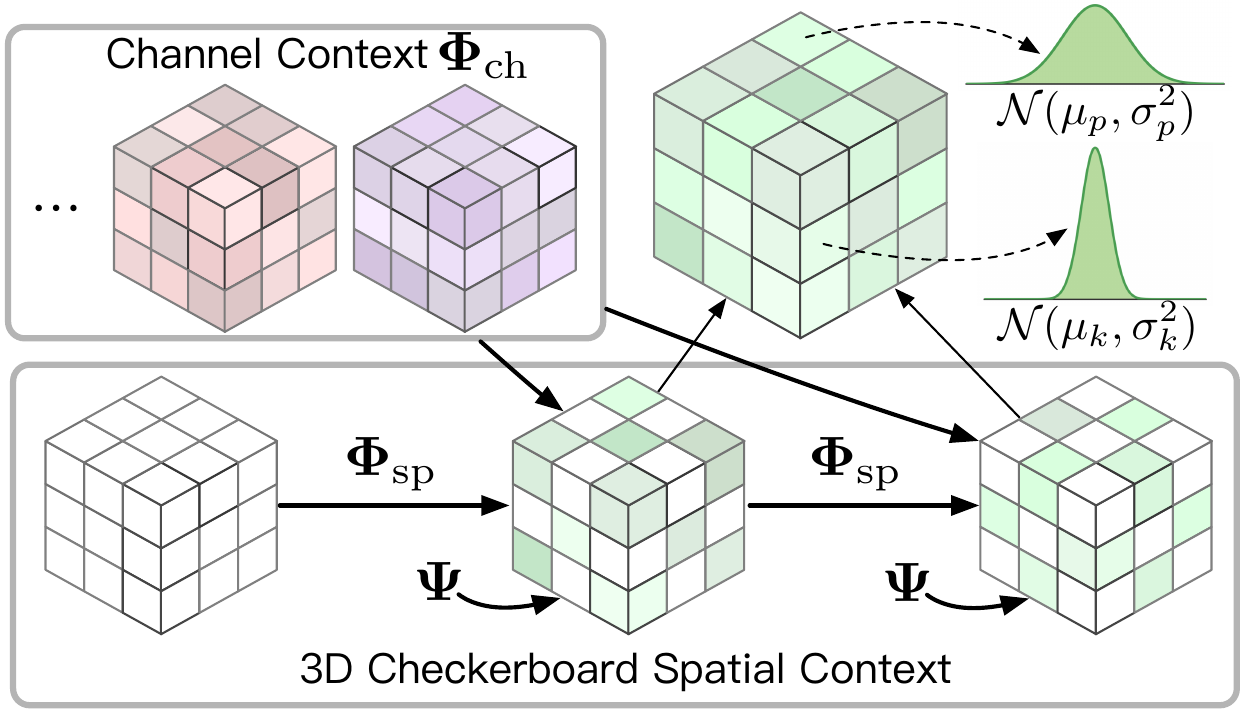}&
 \includegraphics[height=1.45in]{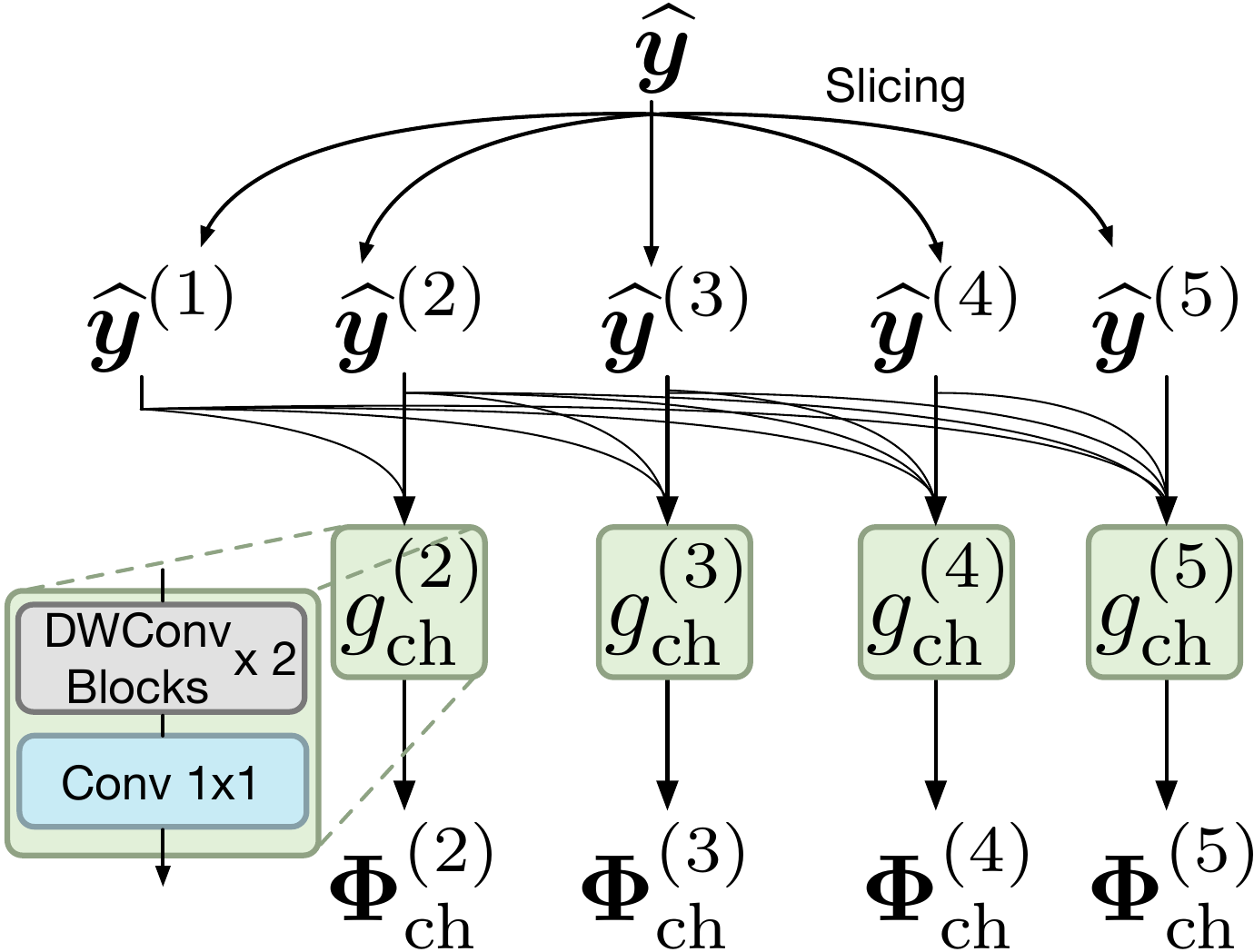}\\
 \mbox{\small (a) factorized model} &\mbox{\small (b) hyperprior model} &\mbox{\small (c) our context model} &\mbox{\small (d) channel-context architecture}\\
\end{array}$
\vspace{-.125in}
\caption{Comparison of different entropy models and our proposed context model.
(a) Factorized model with independent Gaussian assumptions.
(b) Hyperprior model that conditions latent distributions on hyperprior $\bm{\Psi}$.
(c) Our context model that further combines the 3D checkerboard spatial and channel context to achieve a more accurate probability estimation.
(d) Channel-context architecture that decodes latent channels progressively, conditioning each slice on previously decoded ones.} 
\label{fig:diff-entropy-encoding}
\end{figure*}

\begin{figure*}[!ht]
\centering
\includegraphics[width=\linewidth]{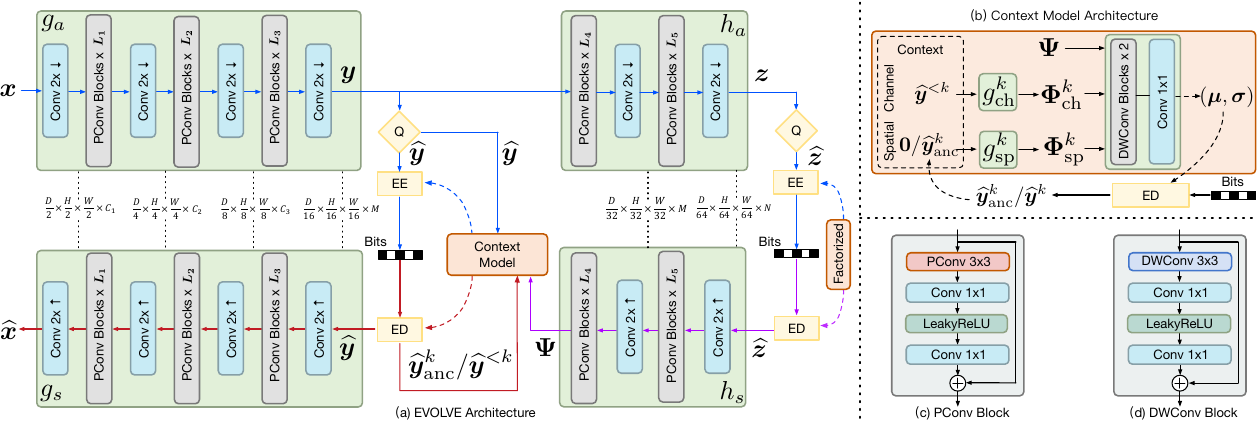}
\vspace{-.25in}
\caption{
(a) EVOLVE architecture. 
Q, EE, and ED denote quantization, entropy encoding, and entropy decoding. 
Blue and red lines indicate the encoding and decoding data paths, while the purple ones are shared by both paths. 
Dashed arrows denote probability estimation.
(b) Context model fuses hyperprior $\bm{\Psi}$, channel-wise context from previously decoded slices $\hat{\mathbf{y}}^{<k}$, and spatial context from decoded anchor positions $\hat{\mathbf{y}}_\mathrm{anc}^k$ through an aggregation network to predict distribution parameters $(\bm{\mu}, \bm{\sigma})$ for probability modeling.
(c) and (d) Detailed structures of PConv and DWConv blocks.
}
\label{fig:workflow}
\end{figure*}

\vspace{-0.05in}
\subsubsection{Training Techniques}

All model variants in the roadmap are trained for 500 epochs on a small subset of volumes, using a learning rate of 0.0001 and a batch size of 16.
\pin{Because voxel statistics vary significantly across scientific
domains, direct optimization on a large database suffers from instability
and slow convergence, so we adopt several modern training techniques.}
In particular, we employ a linear warmup~\cite{Goyal-arXiv17} during the first 20 epochs to stabilize early-stage optimization under heterogeneous data distributions, and use the AdamW optimizer~\cite{Loshchilov-ICLR19} for its robustness to scale variation and improved generalization through decoupled weight decay. 
In addition, common data augmentation techniques described in~\cite{Krizhevsky-NeurIPS12}, such as random cropping, flipping, and rotation, are applied to enhance robustness to spatial variability and improve generalization across diverse volumetric structures.

\vspace{-0.05in}
\subsubsection{Macro Design}

We denote the input volume as $\bm{x}\in\mathbb{R}^{H\times W\times D}$.
The encoder of the vanilla AE performs a non-linear analysis transform $g_a(\cdot)$ to produce a latent representation
$\bm{y}$.
The decoder then applies a non-linear synthesis transform $g_s(\cdot)$ to generate the reconstructed volume
$\hat{\bm{x}}$.
Although the latent representation $\bm{y}$ is typically more compact than the original volume $\bm{x}$, it still contains substantial redundancy that can be further exploited for compression.
Motivated by this, we introduce a series of macro-level design choices to improve compression performance.

{\bf Factorized entropy model.}
The most straightforward way to compress the latent representation $\bm{y}$ is quantization, which can be expressed as $\hat{\bm{y}} = Q(\bm{y})$, where $Q(\cdot)$ denotes the quantization operator.
If a probability model $p_{\hat{\bm{y}}}(\hat{\bm{y}})$ is given, entropy coding techniques, such as arithmetic coding~\cite{arithmetic-coder}, can then be used to encode the quantized codes into a compact bitstream losslessly. 
In practice, under a factorized entropy model, $p_{\hat{\bm{y}}}(\hat{\bm{y}})$ is represented by learnable channel-wise Gaussian parameters $(\bm{\mu}, \bm{\sigma})$ as illustrated in Figure~\ref{fig:diff-entropy-encoding}(a) and optimized during training.
Since arithmetic coding is near-optimal, the entropy of $\hat{\bm{y}}$ provides an accurate estimate of the coding rate. 
Consequently, we can formulate a rate-distortion loss
\begin{equation}
\mathcal{L}_{RD} = \mathcal{R}(\hat{\bm{y}}) + \lambda \, D(\bm{x}, \hat{\bm{x}}),
\label{eqn:factorized}
\end{equation}
where $\mathcal{R}(\hat{\bm{y}})=\mathbb{E}\!\left[-\log_2 p_{\hat{\bm{y}}}(\hat{\bm{y}})\right]$ is the rate term that indicates the average number of bits used when saving $\hat{\bm{y}}$ as bitstreams.
$D(\cdot)$ is the distortion term that is computed using the MSE loss between the GT and reconstructed data.
$\lambda$ is a Lagrange multiplier, a hyperparameter that controls the tradeoff between reconstruction quality and CR, and is set to 300 in our roadmap experiment.
Note that the quantization operation is not differentiable.
Therefore, we approximate quantized latent $\hat{\bm{y}}$ by adding a uniform noise $\mathcal{U}\!\left(-\tfrac{1}{2}, \tfrac{1}{2}\right)
$ to the extracted $\bm{y}$ during training.
As shown in Figure~\ref{fig:roadmap}, the factorized entropy model improves the CR, but leads to a degradation in reconstruction quality due to the introduction of the additional rate term $\mathcal{R}(\hat{\bm{y}})$ in Equation~\ref{eqn:factorized}.

{\bf Hyperprior entropy model.}
In a factorized entropy model, latent elements within the same channel are assumed to follow a shared distribution across spatial positions, ignoring spatially varying statistics.
In contrast, the hyperprior entropy model~\cite{Balle-ICLR17} introduces additional side information to more accurately estimate the probability of latent elements at different spatial locations, as shown in Figure~\ref{fig:diff-entropy-encoding}(b).
Prior work, such as IDLat~\cite{Shen-TVCG23}, falls into this category by adopting a hyperprior entropy model to improve CR. 
In particular, as illustrated in Figure~\ref{fig:workflow}(a), the hyperprior entropy model introduces an additional hyper-latent $\bm{z}$ extracted from $\bm{y}$ via a hyper-analysis transform $h_a$.
The quantized hyper-latent $\hat{\bm{z}}$ is entropy coded and transmitted as side information.
By performing a hyper-synthesis transform $h_s$, the decoded hyperprior $\bm{\Psi}$ is used to condition the probability model of $\hat{\bm{y}}$, resulting in spatially varying probability modeling.
Similar to $\hat{\bm{y}}$, the quantized hyper-latent $\hat{\bm{z}}$ is also entropy coded and saved as bitstreams after encoding.  
Accordingly, an additional rate term $\mathcal{R}(\hat{\bm{z}})$ is introduced into Equation~\ref{eqn:factorized} to optimize the bit cost associated with this part.
The hyperprior entropy model improves the CR from 71.94$\times$ of the factorized entropy model to 644.0$\times$, with a slight improvement in reconstruction quality.

{\bf GDN $\rightarrow$ residual block.} 
We replace the GDN~\cite{Balle-GDN} functions used in many AE-based compressors, such as AE-SZ~\cite{Liu-CLUSTER21} and IDLat~\cite{Shen-TVCG23}, with residual blocks~\cite{He-CVPR16}. 
GDN was initially designed to provide local normalization together with a point-wise nonlinearity.
However, we empirically observe that with a large number of model parameters, incorporating multiple GDNs slows convergence, whereas residual blocks with skip connections facilitate stronger gradient flow and greater nonlinear expressiveness during training.
Therefore, we replace GDN layers with residual blocks in our architecture.
This substitution leads to a marked improvement in reconstruction quality and CR, though computational cost increases due to the added residual layers.

{\bf Context model.}
IDLat~\cite{Shen-TVCG23} adopted the hyperprior entropy model to achieve better compression performance than the factorized entropy model by enabling spatially varying probability modeling.
However, the hyperprior model still assumes independence among latent elements given the side information and thus fails to fully exploit local dependencies between neighboring latent variables.
In volumetric data, voxel values at nearby locations are often very close and highly correlated, indicating substantial local redundancy that can be removed to achieve more effective compression.

To respond, we propose to design a {\em context model}~\cite{Minnen-ICIP20, He-CVPR21, Sheng-TMM23, Wang-NeurIPS24, Chen-ECCV24, Tang-VIS26} that explicitly captures such local dependencies in the latent space.
As illustrated in Figure~\ref{fig:diff-entropy-encoding}(c), we design our context model with two components: \emph{3D checkerboard spatial context} and \emph{channel context}.
The 3D checkerboard spatial context partitions the latent grid into anchor and non-anchor positions and predicts the probability distributions of non-anchor elements conditioned on already decoded anchor elements.
This design enables local spatial dependency modeling while maintaining efficient parallel decoding.
In addition to spatial dependencies, the context model also captures inter-channel correlations through a {\em channel-context architecture} that slices the latent into different channel groups and models them progressively, conditioning each group on previously decoded ones, as illustrated in Figure~\ref{fig:diff-entropy-encoding}(d).
By jointly leveraging masked spatial and channel context, the proposed architecture enables more accurate probability estimation by conditioning on neighboring decoded information, thereby achieving better compression performance.

The architecture of the context model is illustrated in Figure~\ref{fig:workflow}(b).
When decoding the $k$-th channel group, the context module receives three types of conditioning information: the hyperprior $\bm{\Psi}$, previously decoded channel groups $\hat{\bm{y}}^{<k}$, and spatial anchor elements from the current channel group.
The spatial anchor elements are either a zero vector for decoding anchor positions or the decoded anchors $\hat{\bm{y}}_{\text{anc}}^{k}$ for decoding non-anchor positions.
The spatial context feature $\bm{\Phi}_{\text{sp}}^{k}$ is extracted by $g_{\text{sp}}^{k}$, which is a single-layer masked convolution.
The channel context feature $\bm{\Phi}_{\text{ch}}^{k}$ is extracted by $g_{\text{ch}}^{k}$, which consists of two depthwise convolution (DWConv) blocks followed by a 1$\times$1 convolution.
Since the dimensionality of $\hat{\bm{y}}^{<k}$ varies with $k$, separate context transforms $g_{\text{sp}}^{k}$ and $g_{\text{ch}}^{k}$ are employed for different channel groups.
We adopt DWConv blocks to reduce computational cost, while the final 1$\times$1 convolution enables effective channel-level feature fusion.
Finally, the hyperprior, spatial context, and channel context features are then concatenated and fed into a lightweight aggregation network to estimate the Gaussian distribution parameters $(\bm{\mu}, \bm{\sigma})$ for entropy coding.

Our context model plays a critical role in improving compression performance and distinguishes EVOLVE from prior AE-based volume compression methods~\cite{Liu-CLUSTER21, Shen-TVCG23}.
While incorporating context modeling inevitably increases encoding and decoding latency due to its autoregressive nature, this design enables substantial improvements in CR while preserving reconstruction quality (see Figure~\ref{fig:roadmap}), making it a key component of our framework.

\vspace{-0.05in}
\subsubsection{Micro Design}

\textbf{Non-uniform channel allocation.}
Prior work~\cite{Shen-TVCG23} typically assigns the same number of channels to all network stages (i.e., $C_1=C_2=C_3$ in Figure~\ref{fig:workflow}(a)).
In contrast, we observe that allocating more channels to deeper layers—where spatial resolution is progressively reduced—significantly improves reconstruction quality.
Increasing channel width in deeper layers enhances the network’s capacity to model complex, high-level features in a more compact representation space, leading to more faithful reconstructions.
To control computational cost, we reduce the channel width in shallow layers, where spatial resolution is higher, yielding larger savings in computation overhead.
As shown in Figure~\ref{fig:roadmap}, this design improves both CR and reconstruction quality.

\textbf{PConv in residual blocks.}
We replace standard convolutions with advanced partial convolutions (PConv)~\cite{Chen-CVPR23} in the original residual blocks.
PConv applies a parametric filter only to a few input channels, leaving the rest untouched.
This substitution significantly reduces computational complexity without sacrificing much of the representation capacity, leading to faster encoding and decoding.

\textbf{Non-uniform slicing in latent channels.}
In the channel-context architecture, a common practice~\cite{Minnen-ICIP20} is to uniformly split latent channels into multiple groups. 
However, transform-based compression algorithms, such as TTHRESH~\cite{Ballester-TVCG20}, demonstrate that coefficients derived from high-dimensional data are not equally important and achieve competitive compression by truncating the long tail of minor coefficients.
%
Inspired by this, we split $\hat{\bm{y}}$ into uneven groups along the channel dimension.
Since a network with comparable capacity models each group, this allocation implies that smaller groups benefit from denser parametric representation. 
This treatment effectively encourages the framework to learn fine-grained dependencies among critical components while efficiently handling redundant components with coarse-grained modeling.
Consequently, it also allows us to minimize the total number of channel groups required without compromising reconstruction quality, thereby improving both compression efficiency and processing speed.

\textbf{Attention block.}
Attention mechanisms~\cite{Vaswani-NeurIPS17} have demonstrated strong performance in large-scale models, and we also explore incorporating attention blocks into our architecture.
While attention provides a modest improvement in reconstruction quality, it introduces a disproportionate increase in computational cost.
As the marginal quality gain does not justify the added complexity, this modification is not adopted in the final model, as indicated by the hatched bar in Figure~\ref{fig:roadmap}.

\vspace{-0.05in}
\subsubsection{Block Size}

All preceding experiments use a 64$\times$64$\times$64 block size for training. We conclude the roadmap by examining the effect of block size on compression performance.
Smaller blocks require more patches to cover the full volume, leading to redundant boundary regions and limited receptive fields, which degrade both CR and reconstruction quality.
Increasing the block size to 128$\times$128$\times$128 allows the model to capture longer-range spatial dependencies within a single forward pass.
This final adjustment yields the complete EVOLVE model, which achieves the best performance across all metrics.
However, increasing the block size substantially elevates both training and inference memory consumption due to the cubic growth of feature map resolution and intermediate activations.
Given our current computational resources, we restrict the maximum block size to 128$\times$128$\times$128, providing a favorable tradeoff between compression performance and training cost.

\vspace{-0.05in}
\subsection{Variable-Rate Encoding}
\label{sec:variable-rate}

A key limitation of existing \hot{deep-learning-based} volume compressors~\cite{Lu-CGF21, Shen-TVCG23, Wurster-TVCG24} is that each trained model is tied to a single fixed CR.
Supporting different rate-distortion tradeoffs, therefore, requires training, storing, and switching among multiple models, which is cumbersome in practical scientific workflows.
To address this limitation, EVOLVE incorporates a \emph{learnable gain mechanism} that enables a single model to span a continuous range of CRs during inference, without retraining or maintaining multiple models.

{\bf Gain-based quantization regulation.}
The core idea is to modulate the quantization bin size via a learnable gain vector $\boldsymbol{\gamma} = [\gamma_0, \gamma_1, \ldots, \gamma_{A-1}]$, where each $\gamma_i$ is initialized with a substantially different magnitude to cover a wide range of CRs before optimization, and $A$ denotes the number of discrete quality levels.
During compression at quality level $i$, the latent representation $\bm{y}$ is scaled by the corresponding gain $\gamma_i$ prior to quantization and rescaled by its inverse afterward
\begin{equation}
    \hat{\bm{y}} = Q(\bm{y} \cdot \gamma_i) / \gamma_i,
    \label{eq:gain}
\end{equation}
where $Q(\cdot)$ denotes the quantization operator.
A large $\gamma_i$ amplifies the latent values, equivalent to a smaller quantization bin size, resulting in a lower CR and better preservation of fine details.
Conversely, a small $\gamma_i$ increases the effective bin size, leading to coarser quantization and higher CR.
Following~\cite{Cui-CVPR21}, we define a set of rate-distortion tradeoff parameters
$\boldsymbol{\lambda} = [\lambda_0, \lambda_1, \ldots, \lambda_{A-1}]$ in one-to-one correspondence with the gain vector $\boldsymbol{\gamma}$ during optimization.
Before training starts, the gain values in $\boldsymbol{\gamma}$ are initialized as
$\sqrt{\boldsymbol{\lambda}/\lambda_0}$ following~\cite{Tong-ICIP23}.
After the model is sufficiently trained across the {\em discrete} quality levels, {\em continuous} variable-rate encoding can be achieved at inference time by interpolating between adjacent $\boldsymbol{\gamma}$ values, enabling a single EVOLVE model to support a continuous range of CRs.

{\bf Three-stage training strategy.}
Training a variable-rate model end-to-end is challenging because the gain values interact with all other network weights, potentially destabilizing the optimization.
To ensure stable convergence across the full range of quality levels, we adopt a three-stage training strategy.
(1) In the first stage, the model is trained at a single, fixed highest quality level while the gain parameters are kept frozen.
This stage allows the encoder, decoder, and entropy model to converge to a strong baseline without being affected by rate variation.
(2) In the second stage, the gain parameters are unfrozen. The model is trained jointly across all $A$ quality levels: at each iteration, a quality level $i$ is selected by deterministically cycling through the levels, and the corresponding $\lambda_i$ replaces $\lambda$ in Equation~\ref{eqn:factorized}, with the network weights and gain parameters optimized simultaneously.
(3) In the third stage, the model is further finetuned by replacing the uniform-noise quantization approximation with the {\em straight-through estimator} (STE)~\cite{Yin-ICLR19}, ensuring that the learned gain parameters and network weights are well adapted to the discrete rounding operation used during inference.

\begin{figure*}[htb]
\centering
$\begin{array}{c@{\hspace{0.05in}}c@{\hspace{0.05in}}c@{\hspace{0.05in}}c@{\hspace{0.05in}}c}
\mbox{\small CR / PSNR} &
 \mbox{\small 6,047$\times$ / 49.31 dB} &
 \mbox{\small 2,948$\times$ / 41.10 dB} &
 \mbox{\small 2,714$\times$ / 42.98 dB} &
 \mbox{\small \hot{129$\times$ / 41.92 dB}} \\
 \includegraphics[width=0.19125\linewidth]{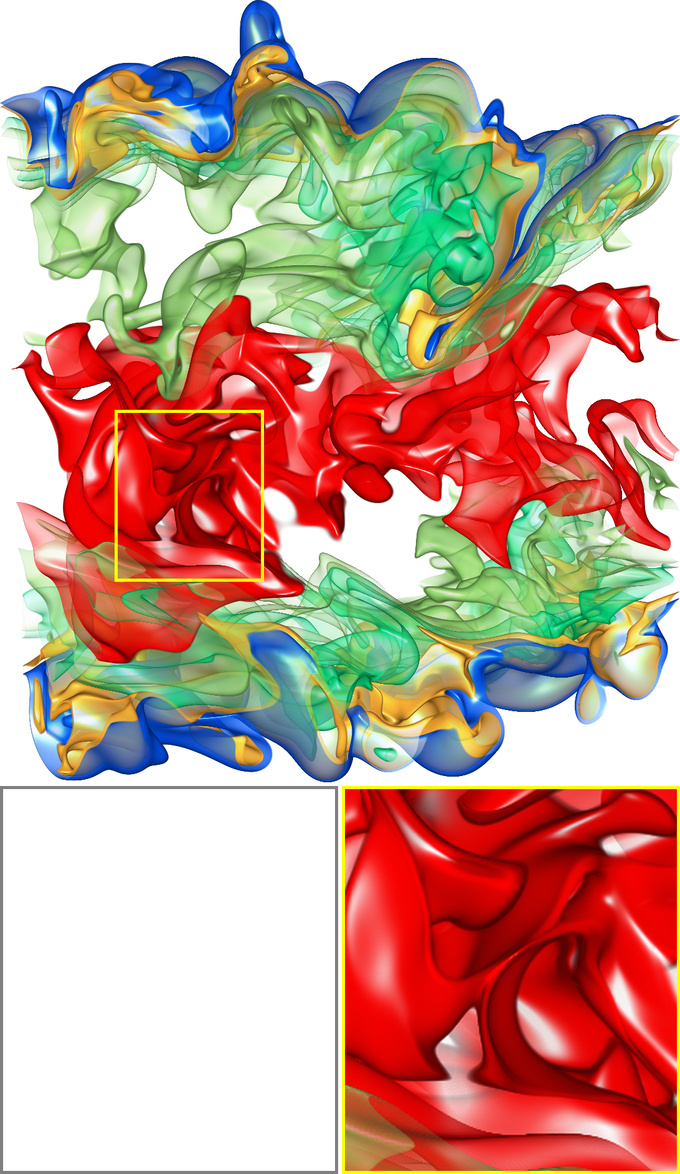}&
 \includegraphics[width=0.19125\linewidth]{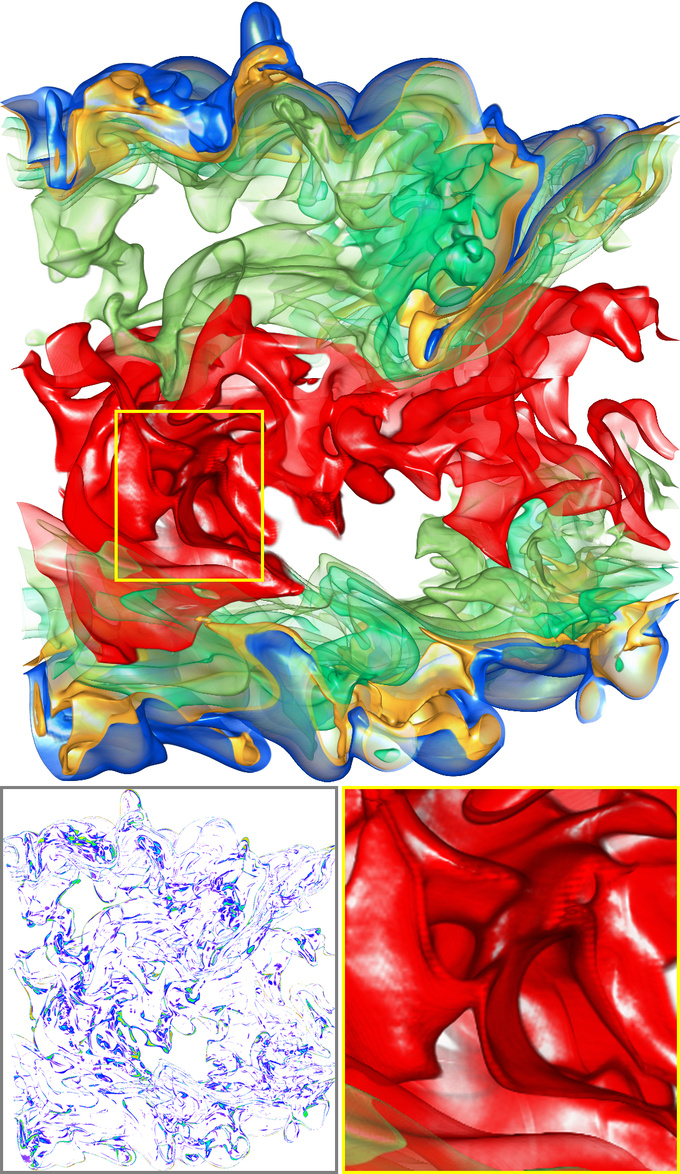}&
 \includegraphics[width=0.19125\linewidth]{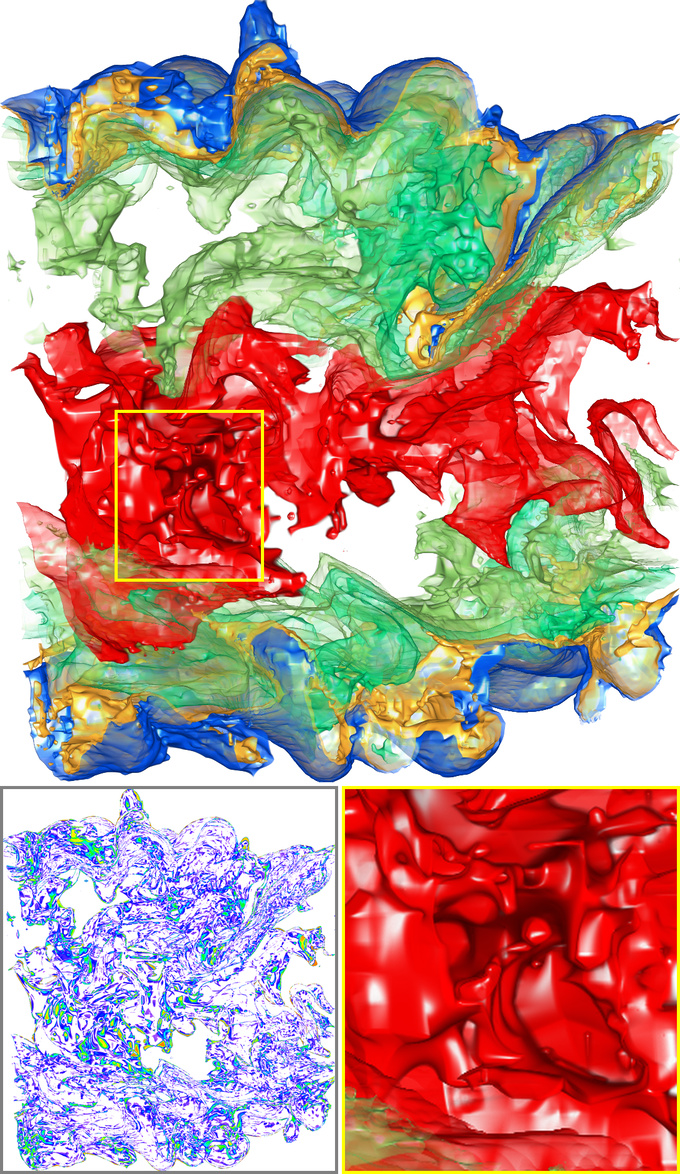}&
 \includegraphics[width=0.19125\linewidth]{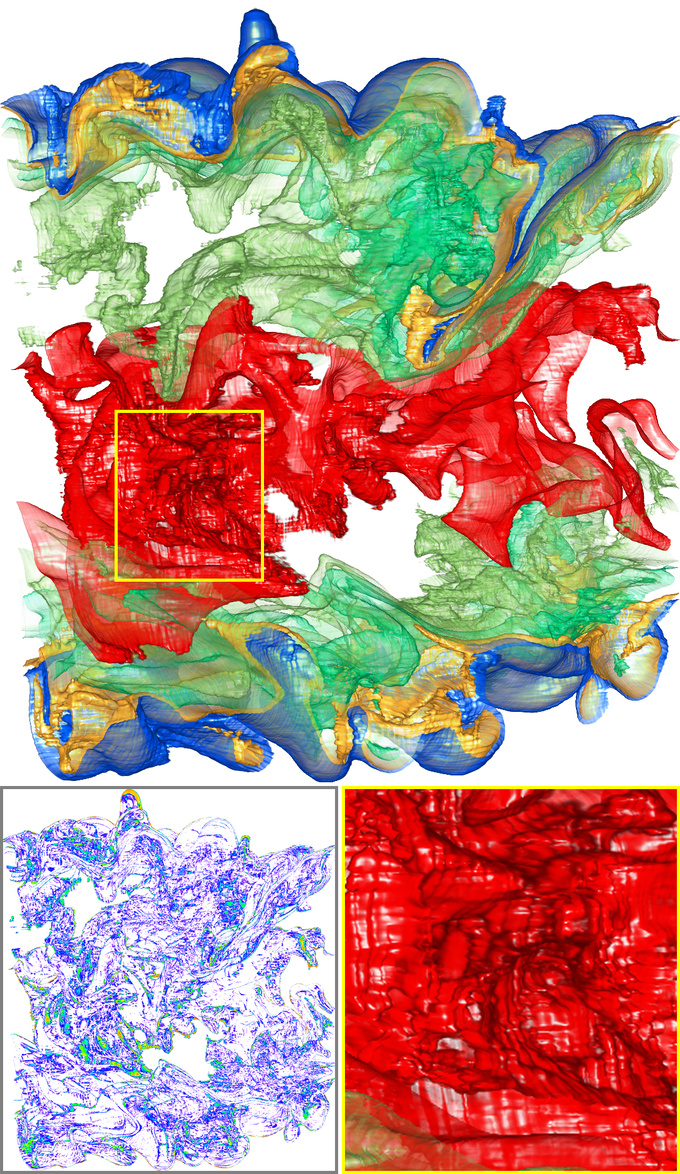}&
 \includegraphics[width=0.19125\linewidth]{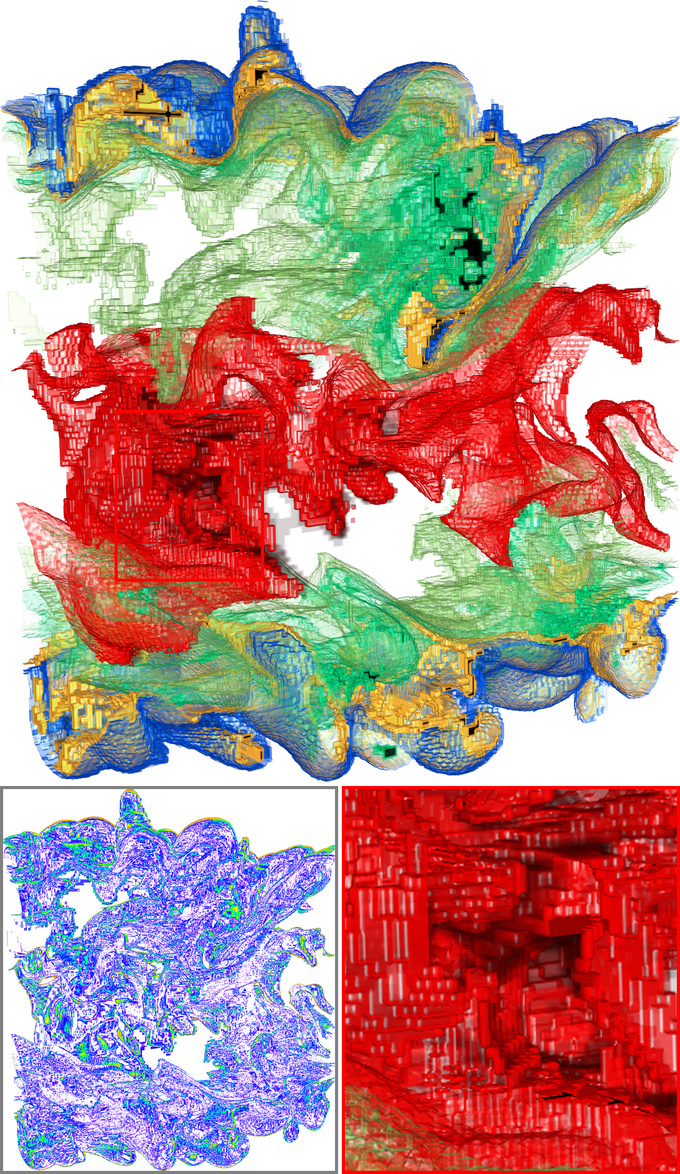}\\
 \mbox{\small CR / PSNR} &
 \mbox{\small 7,843$\times$ / 47.58 dB} &
 \mbox{\small 1,051$\times$ / 41.43 dB} &
 \mbox{\small 6,334$\times$ / 40.11 dB} &
 \mbox{\small \hot{107$\times$ / 41.44 dB}} \\
 \includegraphics[width=0.19125\linewidth]{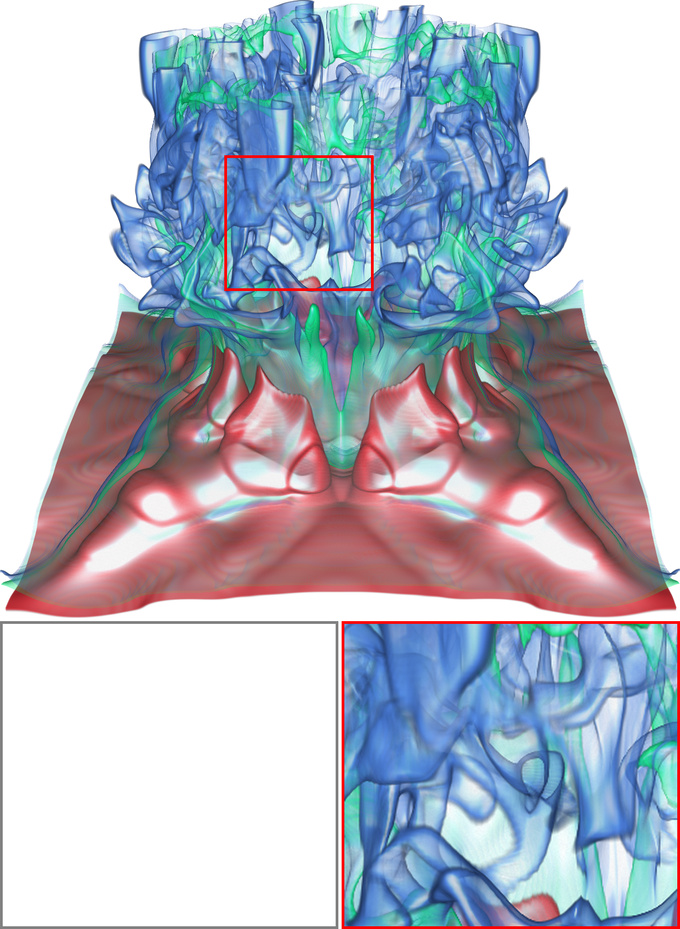}&
 \includegraphics[width=0.19125\linewidth]{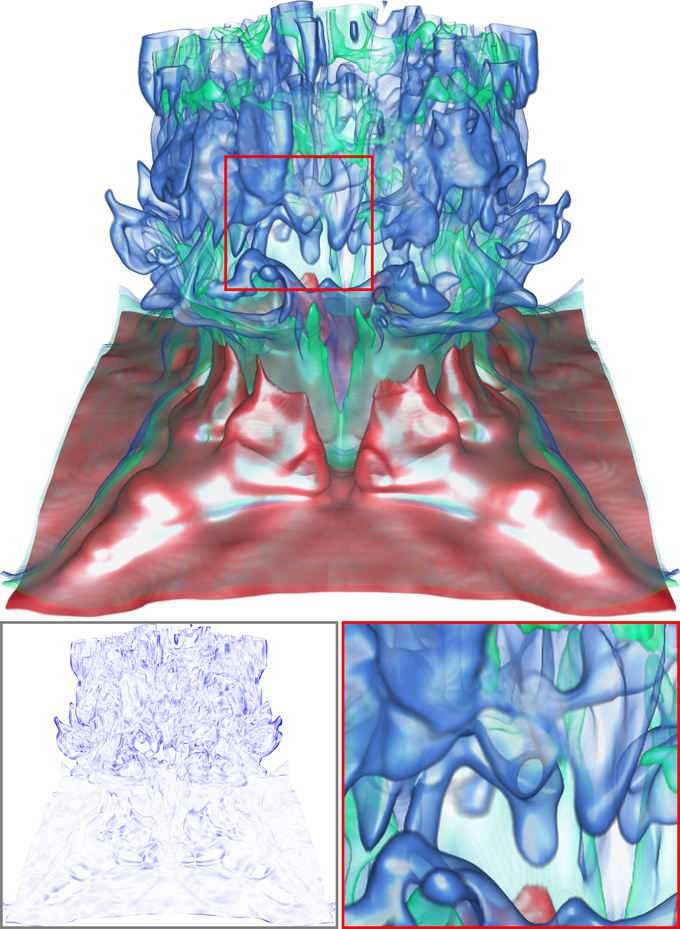}&
 \includegraphics[width=0.19125\linewidth]{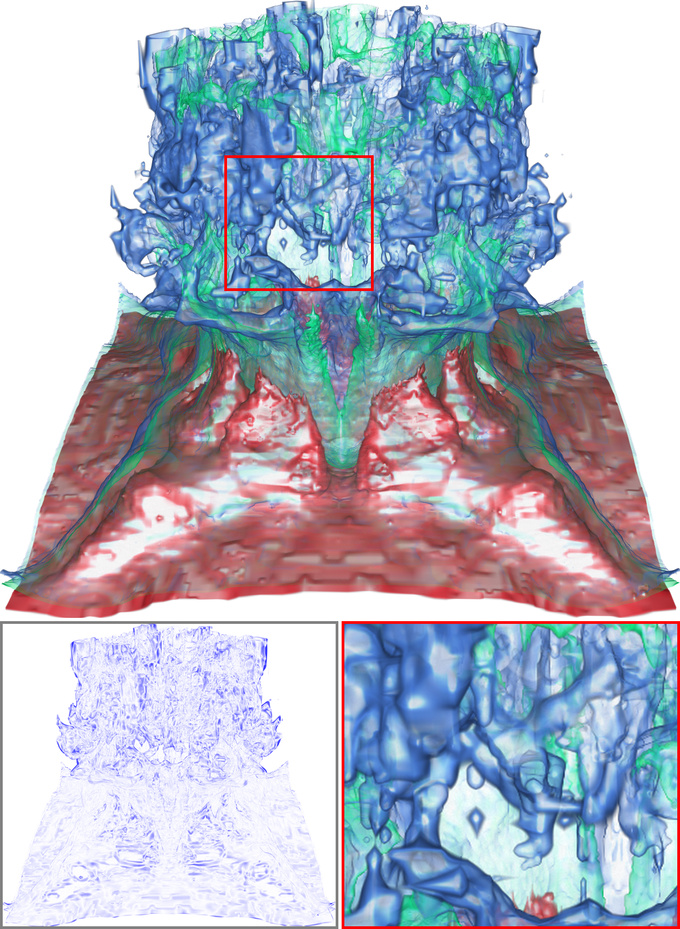}&
 \includegraphics[width=0.19125\linewidth]{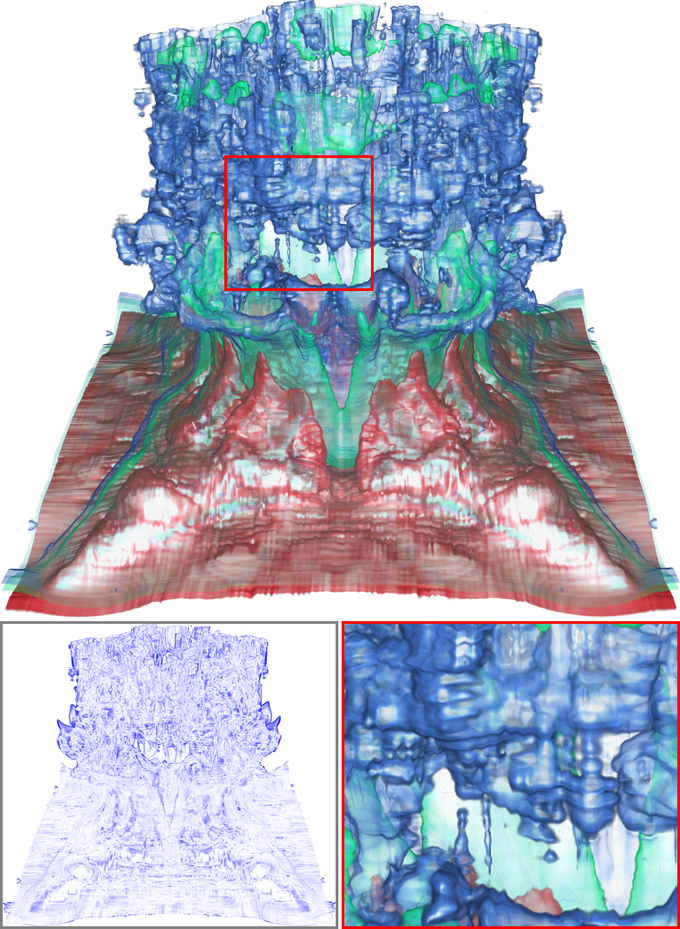}&
 \includegraphics[width=0.19125\linewidth]{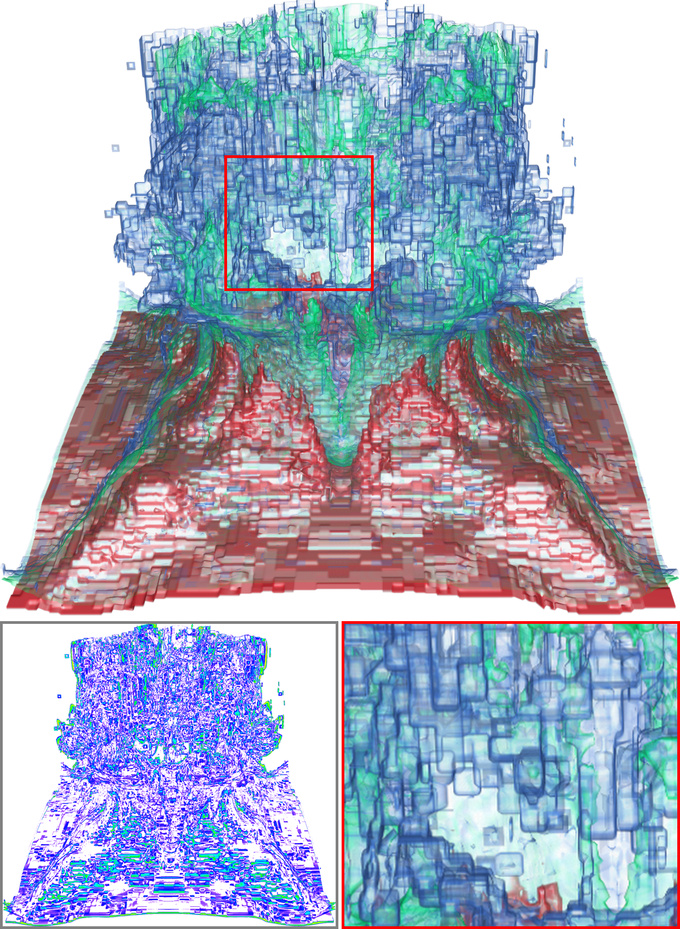}\\
 \mbox{\small GT} &
 \mbox{\small EVOLVE} &
 \mbox{\small SZ3} &
 \mbox{\small TTHRESH} &
 \mbox{\small ZFP}\\
\end{array}$
\vspace{-.125in}
\caption{Comparison of volume rendering results between EVOLVE and conventional lossy compressors. Top and Bottom: \textsf{combustion (MF)} and \textsf{ionization (H+)}. The difference image shows noticeable pixel differences relative to the GT in the CIELUV color space.}
\label{fig:traditional-baselines-vol}
\end{figure*}

\begin{table}[!ht]
\centering
\caption{Testing datasets used for compression performance comparison.}
\label{tab:test_dataset}
\vspace{-0.05in}
\resizebox{0.95\columnwidth}{!}{%
\begin{tabular}{cccc}
 & volume resolution  & \# timesteps &  \\ 
dataset & ($x \times y \times z$) & or ensembles & size \\ \hline
\textsf{asteroids} & 1,000$\times$1,000$\times$1,000 & 1 & 3.7 GB \\
\textsf{asteroids-T} & 500$\times$500$\times$500 & 220 & 102.5 GB \\
\textsf{combustion (MF)} & 480$\times$720$\times$120 & 1 & 158 MB \\
\textsf{gas} & 512$\times$512$\times$512 & 1 & 512 MB \\
\textsf{half-cylinder (VLM, 6,400)} & 640$\times$240$\times$80 & 1 & 47 MB \\
\textsf{ionization (H+)} & 600$\times$248$\times$248 & 1 & 141 MB \\
\textsf{ionization-T (H+)} & 600$\times$248$\times$248 & 200 & 27.5 GB \\
\textsf{isotropic} & 512$\times$512$\times$512 & 1 & 512 MB \\
\textsf{magnetic} & 512$\times$512$\times$512 & 1 & 512 MB \\
\textsf{Nyx-E} & 512$\times$512$\times$512 & 206 & 103.0 GB \\
\end{tabular}
}
\end{table}

\begin{figure*}[!t]
\centering
$\begin{array}{c@{\hspace{0.05in}}c@{\hspace{0.05in}}c@{\hspace{0.05in}}c@{\hspace{0.05in}}c}

\mbox{\small CR / PSNR} &
 \mbox{\small 9,803$\times$ / 47.16 dB} &
 \mbox{\small 2,482$\times$ / 40.85 dB} &
 \mbox{\small 2,401$\times$ / 43.44 dB} &
 \mbox{\small \hot{1,164$\times$ / 40.00 dB}} \\

 \includegraphics[width=0.19125\linewidth]{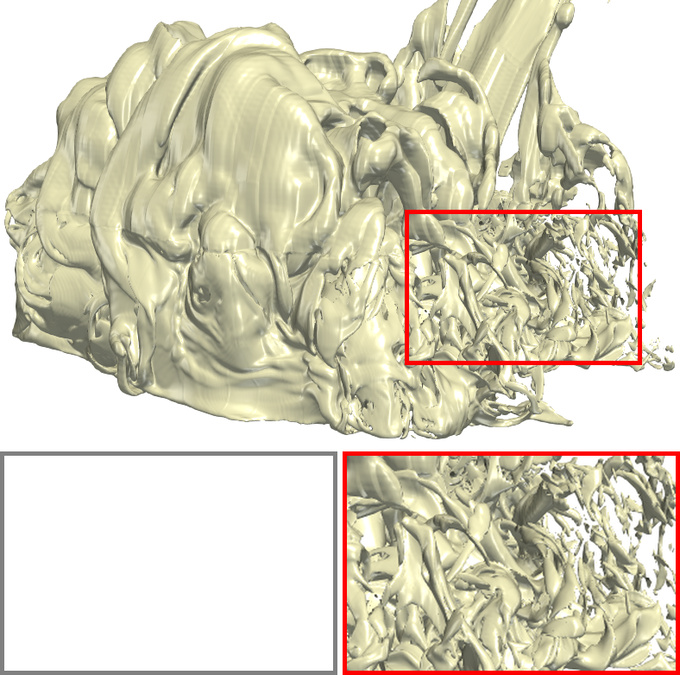}&
 \includegraphics[width=0.19125\linewidth]{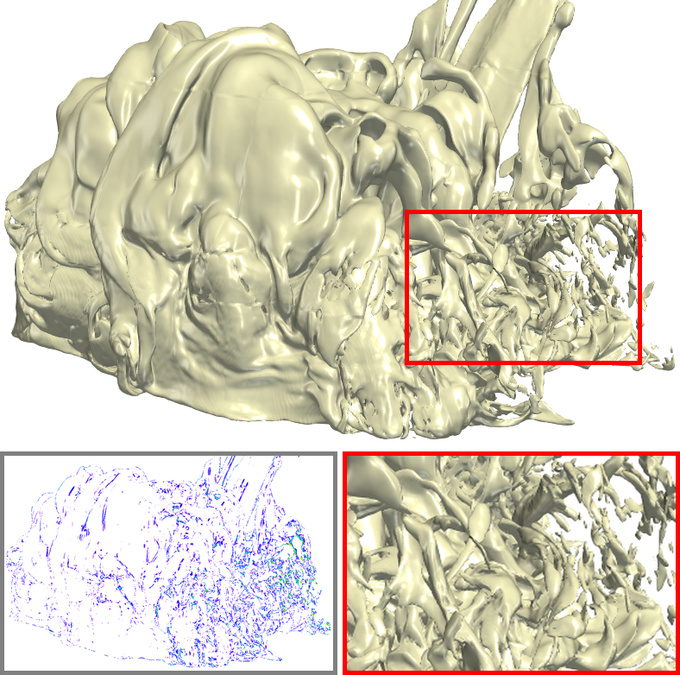}&
 \includegraphics[width=0.19125\linewidth]{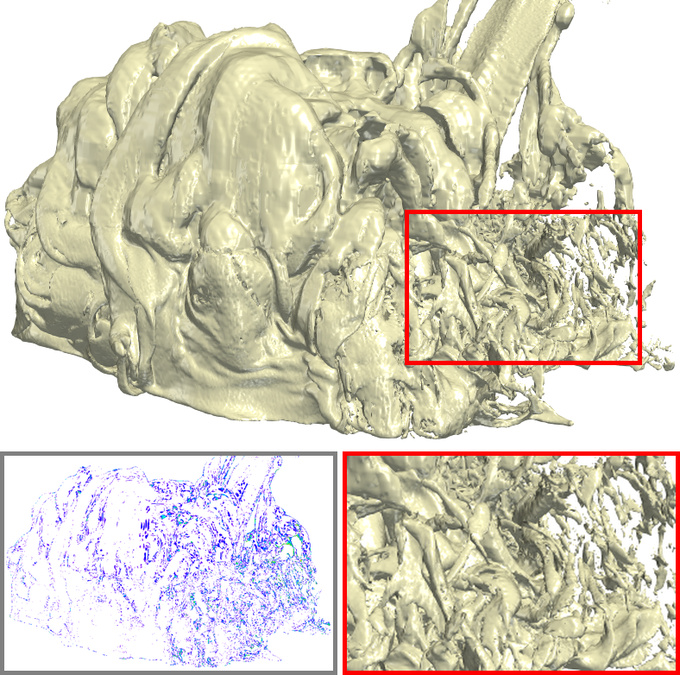}&
 \includegraphics[width=0.19125\linewidth]{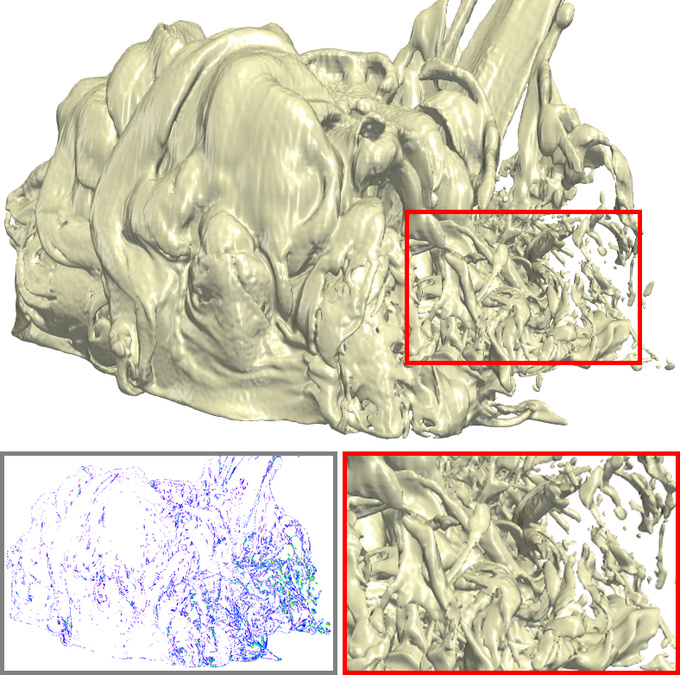}&
 \includegraphics[width=0.19125\linewidth]{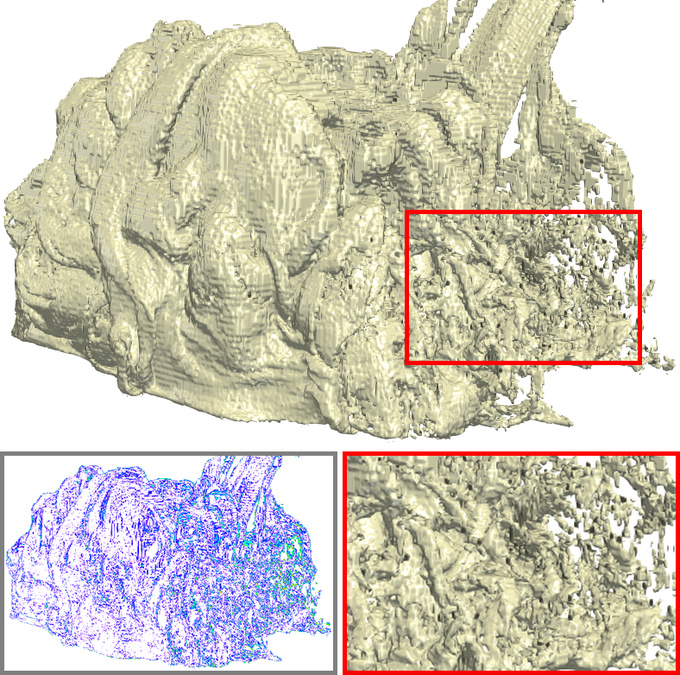}\\

\mbox{\small CR / PSNR} &
 \mbox{\small 2,846$\times$ / 45.18 dB} &
 \mbox{\small 985$\times$ / 40.03 dB} &
 \mbox{\small 870$\times$ / 43.51 dB} &
 \mbox{\small \hot{64$\times$ / 43.79 dB}} \\

 \includegraphics[width=0.19125\linewidth]{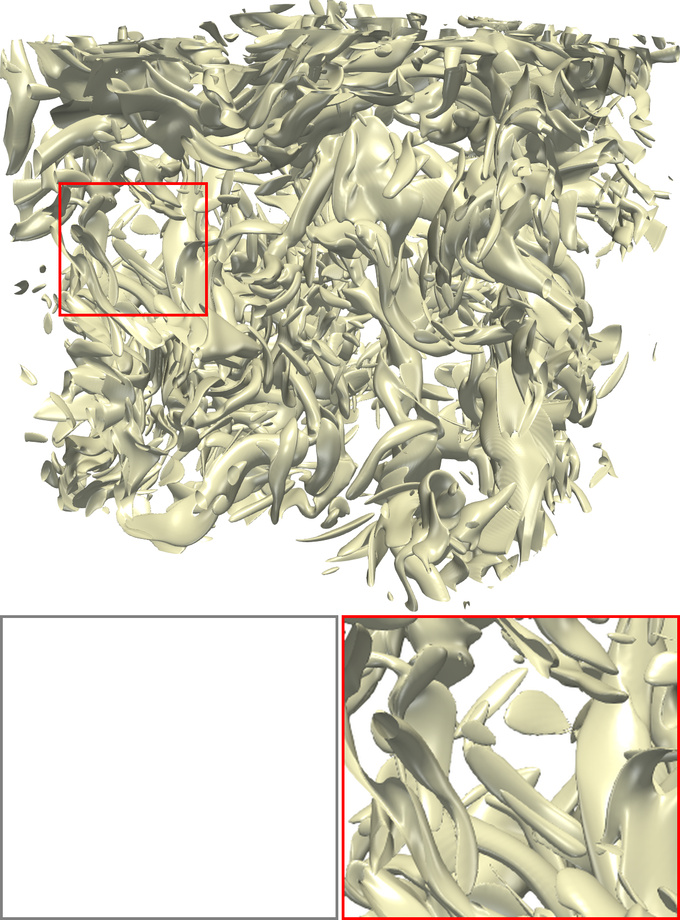}&
 \includegraphics[width=0.19125\linewidth]{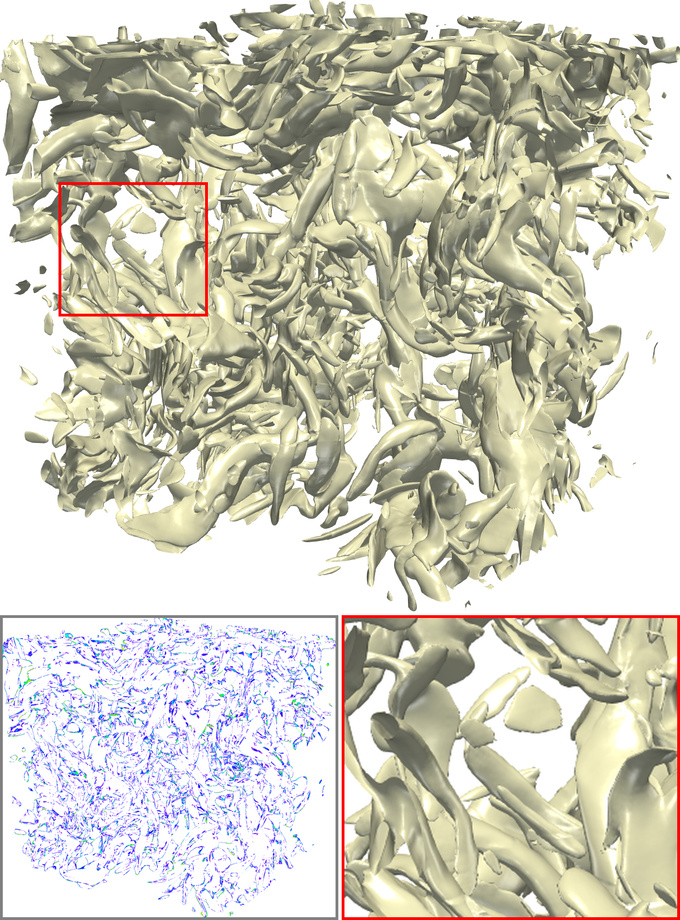}&
 \includegraphics[width=0.19125\linewidth]{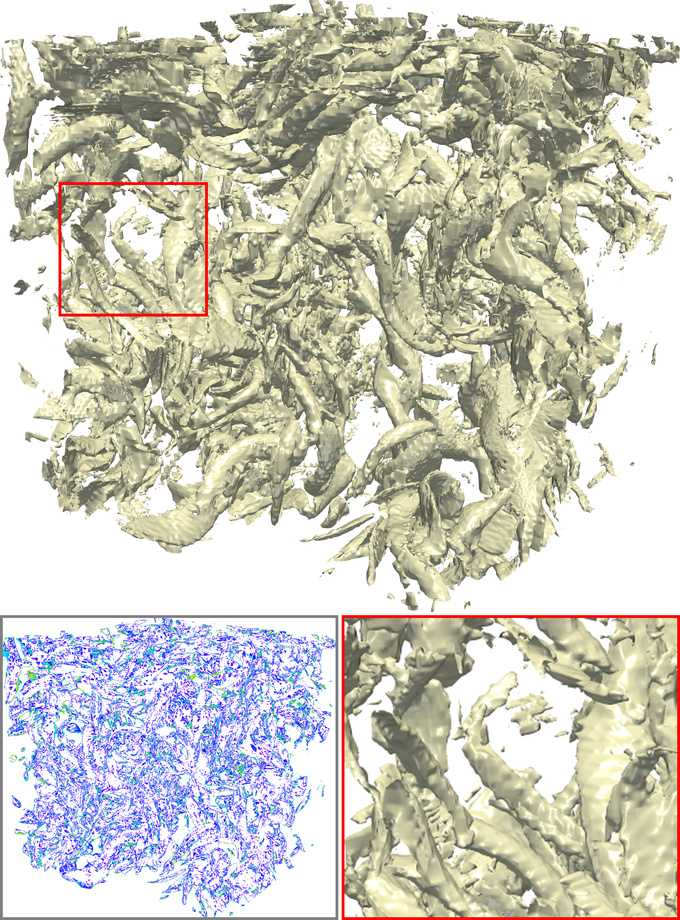}&
 \includegraphics[width=0.19125\linewidth]{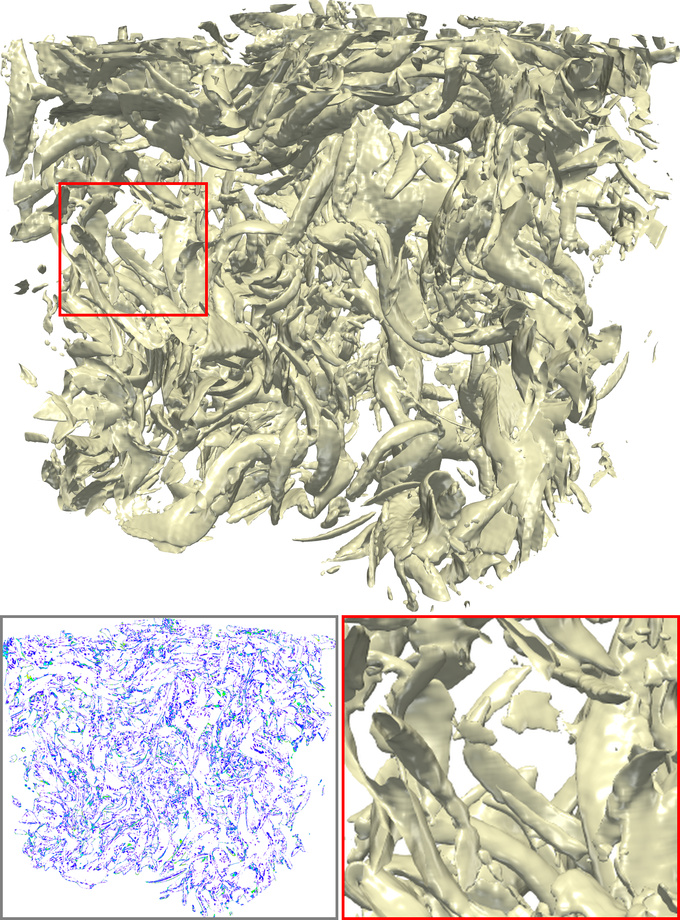}&
 \includegraphics[width=0.19125\linewidth]{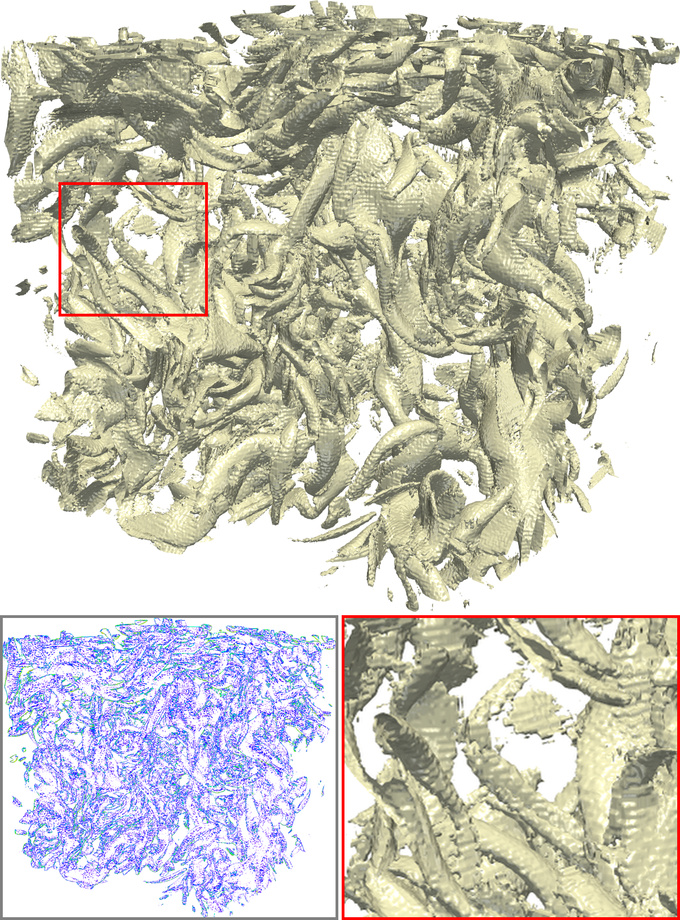}\\

 \mbox{\small GT} &\mbox{\small EVOLVE} &\mbox{\small SZ3} &\mbox{\small TTHRESH} &\mbox{\small ZFP}\\
\end{array}$
\vspace{-.125in}
\caption{Comparison of isosurface rendering results between EVOLVE and conventional lossy compressors. Top and bottom: \textsf{asteroids} and \textsf{isotropic}. The chosen isovalues are 0.2 and 0.7, respectively.}
\label{fig:traditional-baselines-iso}
\end{figure*}

%% file: result.tex
\section{Results and Discussion}

\subsection{Implementation Details}

{\bf Training details.}
We trained EVOLVE on the 6,376 selected volumes listed in Table~\ref{tab:dataset}. 
For model architecture, we set $[C_1, C_2, C_3]$ to $[96, 192, 384]$ and $[L_1, L_2, L_3, L_4, L_5]$ to $[2, 2, 4, 1, 3]$. 
We set the number of latent channels $M$ to 320 and the number of hyperlatent channels $N$ to 256.
For channel-context slicing, we split the latent channels unevenly into five groups with channel numbers $[16, 16, 32, 64, 192]$.
\hot{The resulting EVOLVE model contains 85.6M parameters ($\approx$327 MB in FP32).
Unlike INRs, where the network itself is the compressed representation of a single volume, the same shared EVOLVE model can compress both seen and unseen in-domain volumes without per-volume optimization, and the reported CRs account for the per-volume bitstream.}
We used the AdamW optimizer~\cite{Loshchilov-ICLR19} with an initial learning rate of $10^{-4}$ and a weight decay of 0.01 for the encoder and decoder parameters.
A separate optimizer with a learning rate of $10^{-3}$ was used for other parameters, including the entropy model, as they exhibit different gradient scales and require a higher learning rate for stable convergence.
We optimized our model for 1,600 epochs in total, with 800, 400, and 400 epochs for the three stages, respectively, and applied a linear warmup during the first 20 epochs at the beginning of each stage.
In the second stage, we set the number of discrete quality levels $A$ to 8 and $\bm{\lambda}$ to [100, 200, 400, 800, 1,600, 3,200, 6,400, 12,800].
The training was conducted on an NVIDIA H200 GPU using a batch size of 24. 
For data augmentation, we randomly sampled 128$\times$128$\times$128 blocks from each dataset. If any dimension was smaller than 128, it was resized to allow block extraction.
We additionally applied random axis-wise flipping and $90^\circ$ rotations, each with 50\% probability.

{\bf Baselines.}
For conventional lossy compressors, we chose ZFP~\cite{Lindstrom-TVCG14}, TTHRESH~\cite{Ballester-TVCG20}, and  SZ3~\cite{Liang-TBD23}, which represent state-of-the-art prediction-based and transform-based approaches.
\pin{Since compressors are sensitive to their operating modes, we document the exact settings.}
TTHRESH uses target-PSNR mode (\texttt{-p}), and SZ3 uses an absolute error bound (\texttt{-M ABS}) derived from the target PSNR as $\epsilon = r \cdot 10^{-\text{PSNR}/20}$, with $r$ being the data range.
ZFP is run in fixed-accuracy mode (\texttt{-a}), which adapts the per-block bit budget to a global error tolerance and generally attains higher CRs at a given quality than the fixed-rate mode (\texttt{-r}); even though it forgoes the fixed-size blocks and random access that fixed-rate mode provides.
All operating points are selected so that each compressor meets the high-fidelity threshold (PSNR$\geq$40 dB) for each dataset.
For \hot{deep-learning-based} compressors, we considered
fully-connected INRs (SIREN~\cite{Sitzmann-NeurIPS20}, NeurComp~\cite{Lu-CGF21}, and ECNR~\cite{Tang-PVis24}) that use MLPs to map coordinates to scalar values, 
grid-based INRs (Instant-NGP~\cite{Muller-TOG22}, fV-SRN~\cite{Weiss-CGF22}, and AMGSRN++~\cite{Wurster-PVIS25}) that combine learnable feature grids with lightweight decoders for faster training and inference, and 
an AE-based method (IDLat~\cite{Shen-TVCG23}) that learns lightweight latent representations through convolutional encoding. Note that we did not compare with time-varying INR compression methods (e.g., KD-INR~\cite{Han-TVCG24} and MoE-INR~\cite{Han-VIS25}), as EVOLVE focuses on static volume compression. 
\hot{For IDLat, we followed the original implementation and trained it on the same curated database as EVOLVE (see Table~\ref{tab:dataset}), using identical latent and hyperlatent dimensions for a fair comparison.
For INR-based methods, we control the CR by adjusting their network hyperparameters, such as the number of hidden neurons for fully connected INRs and the feature grid resolution for grid-based INRs.}

{\bf Testing datasets and metrics.}
All compression tests were conducted on a workstation with a single NVIDIA RTX 4090 GPU to ensure a fair comparison.
We used the datasets listed in Table~\ref{tab:test_dataset} for testing the compression performance of different methods. 
To prevent data leakage, the volumes in Table~\ref{tab:test_dataset} have no overlap with the 9,921 volumes in Table~\ref{tab:dataset}, which are from unseen variables (\textsf{combustion}, \textsf{half-cylinder}, \textsf{ionization}) or new datasets (\textsf{asteroids}, \textsf{gas}, \textsf{isotropic}, \textsf{magnetic}).
We assessed reconstruction quality using three complementary metrics: data-level peak signal-to-noise ratio (PSNR), image-level learned perceptual image patch similarity (LPIPS)~\cite{Zhang2018}, and surface-level Chamfer distance (CD)~\cite{Barrow1977}. 
LPIPS quantifies perceptual differences by computing distances between deep features extracted from rendered images using AlexNet. 
CD evaluates geometric fidelity by measuring the average nearest-neighbor distance between isosurfaces extracted from decompressed and GT volumes using the L2 norm.
\hot{The isovalues used for CD computation and isosurface rendering were manually selected per dataset to capture salient structures.}
Additional results with time-varying (\textsf{asteroids-T}, \textsf{ionization-T}) and ensemble (\textsf{Nyx-E}) datasets are presented in Appendices~\ref{sec:lse}~and~\ref{sec:tved}.

    \begin{table}[!ht]
    \centering
    \caption{PSNR (dB), LPIPS, and CD values, as well as encoding time (ET, in sec),
    decoding time (DT, in sec), and CR of EVOLVE and other conventional lossy compressors.
    $v$ is the chosen isovalue used to compute CD.}
    \label{tab:traditional-results}
    \vspace{-0.05in}
    \resizebox{1.0\columnwidth}{!}{%
    \begin{tabular}{c c c c c c c c}
    dataset & method & PSNR$\uparrow$ & LPIPS$\downarrow$ & CD$\downarrow$ & ET$\downarrow$ & DT$\downarrow$
  &
    CR$\uparrow$ \\
    \hline

    \multirow{4}{*}{\begin{tabular}{c}\textsf{asteroids}\\ $(v=0.2)$\end{tabular}}
    & \hot{ZFP} & \hot{40.00} & \hot{0.330} & \hot{0.68} & \hot{\textbf{2.08}} & \hot{\textbf{3.18}} & \hot{1,164} \\
    & TTHRESH & 43.44 & 0.093 & 0.46 & 159.59 & 31.91 & 2,401 \\
    & SZ3     & 40.85 & 0.096 & 0.36 & 10.09 & 4.24 & 2,482 \\
    & EVOLVE  & \textbf{47.16} & \textbf{0.073} & \textbf{0.33} & 47.71 & 48.33 & \textbf{9,803} \\

    \hline

    \multirow{4}{*}{\begin{tabular}{c}\textsf{combustion}\\ \textsf{(MF)}\\ $(v=0.5)$\end{tabular}}
    & \hot{ZFP} & \hot{41.92} & \hot{0.436} & \hot{0.68} & \hot{\textbf{0.25}} & \hot{\textbf{0.20}} & \hot{129} \\
    & TTHRESH & 42.98 & 0.352 & 0.73 & 5.50 & 1.07 & 2,714 \\
    & SZ3     & 41.10 & 0.338 & 0.92 & 0.57 & 0.26 & 2,948 \\
    & EVOLVE  & \textbf{49.31} & \textbf{0.146} & \textbf{0.44} & 2.25 & 2.30 & \textbf{6,047} \\

    \hline

    \multirow{4}{*}{\begin{tabular}{c}\textsf{ionization}\\ \textsf{(H+)}\\ $(v=0.1)$\end{tabular}}
    & \hot{ZFP} & \hot{41.44} & \hot{0.372} & \hot{0.58} & \hot{\textbf{0.24}} & \hot{\textbf{0.12}} & \hot{107} \\
    & TTHRESH & 40.11 & 0.379 & 0.92 & 4.15 & 0.86 & 6,334 \\
    & SZ3     & 41.43 & 0.300 & 0.37 & 0.40 & 0.20 & 1,051 \\
    & EVOLVE  & \textbf{47.58} & \textbf{0.203} & \textbf{0.36} & 1.96 & 1.60 & \textbf{7,843} \\

    \hline

    \multirow{4}{*}{\begin{tabular}{c}\textsf{isotropic}\\ $(v=0.7)$\end{tabular}}
    & \hot{ZFP} & \hot{43.79} & \hot{0.332} & \hot{0.49} & \hot{\textbf{1.27}} & \hot{\textbf{0.55}} & \hot{64} \\
    & TTHRESH & 43.51 & 0.220 & 0.50 & 17.63 & 3.89 & 870 \\
    & SZ3     & 40.03 & 0.301 & 0.79 & 2.43 & 1.55 & 985 \\
    & EVOLVE  & \textbf{45.18} & \textbf{0.151} & \textbf{0.42} & 6.63 & 6.14 & \textbf{2,846} \\
    \end{tabular}
    }
  \end{table}

\vspace{-0.05in}
\subsection{Comparison with Conventional Lossy Compressors}
\label{subsec:clc}

We compare EVOLVE with three representative conventional lossy compressors: ZFP, TTHRESH, and SZ3.
All methods are evaluated in a high-fidelity reconstruction setting (i.e., PSNR$\geq$40 dB) to ensure a fair comparison of compression efficiency without sacrificing reconstruction quality.
Table~\ref{tab:traditional-results} summarizes the averaged quantitative results across all datasets.
In terms of CR, EVOLVE consistently achieves substantially higher CR than conventional compressors on unseen volumes, while maintaining superior reconstruction quality as measured by PSNR, LPIPS, and CD.
In particular, on the \textsf{asteroids} dataset, EVOLVE reaches a CR of 9,803$\times$, which is \hot{3.95}$\times$ higher than the best conventional baseline, while simultaneously improving PSNR, LPIPS, and CD.
Regarding runtime performance, EVOLVE achieves encoding speeds comparable to, and in some cases faster than, those of conventional compressors.
The decoding stage is slower than the second-slowest baseline (TTHRESH) primarily due to additional operations, such as block merging.
See Section~\ref{sec:various-CR} for comparisons at varying CRs.
Figure~\ref{fig:traditional-baselines-vol} presents volume rendering comparisons on the \textsf{combustion} and \textsf{ionization} datasets. 
Although all compressors preserve the overall structure, the reconstructions from conventional methods exhibit noticeable artifacts and noise, particularly in the \textsf{combustion} zoom-in region. 
In contrast, EVOLVE yields high-fidelity renderings that better preserve fine-grained spatial details.
\pin{Similarly, for isosurface renderings shown in
Figure~\ref{fig:traditional-baselines-iso}, ZFP and SZ3 exhibit noticeable roughness and jagged artifacts on the \textsf{asteroids} dataset, and TTHRESH introduces subtle distortions, whereas EVOLVE produces smoother, more geometrically consistent isosurfaces.}

\begin{table}[t]
\centering
\caption{PSNR (dB), LPIPS, and CD values, as well as ET (sec), DT (sec), and CR of EVOLVE and other \hot{deep-learning-based} compressors. 
}
\label{tab:neural-baselines}
\vspace{-0.05in}
\resizebox{1.0\columnwidth}{!}{%
\small
\begin{tabular}{c c c c c c c c}
dataset & method & PSNR$\uparrow$ & LPIPS$\downarrow$ & CD$\downarrow$ & ET$\downarrow$ & DT$\downarrow$ & CR$\uparrow$ \\ \hline

\multirow{8}{*}{\begin{tabular}{c}\textsf{gas}\\ $(v=0.25)$\end{tabular}}
& SIREN       & 41.37 & 0.134 & 0.75 & 3,488 & 3.70 & 1,979 \\
& NeurComp    & 41.56 & 0.133 & 0.74 & 3,239 & 5.07 & 5,954 \\
& ECNR        & 43.35 & 0.148 & 0.77 & 2,543 & 5.34 & 1,714 \\
& Instant-NGP & 42.22 & 0.176 & 0.98 & 50.32 & \textbf{0.08} & 1,839 \\
& fV-SRN      & 40.24 & 0.167 & 0.89 & 31.45 & 0.26 & 936 \\
& AMGSRN++    & 40.78 & 0.180 & 1.23 & 81.31 & 0.12 & 1,575 \\
& IDLat       & 41.28 & 0.115 & 0.73 & 10.41 & 20.60 & 1,808 \\
& EVOLVE      & \textbf{45.62} & \textbf{0.113} & \textbf{0.71} & \textbf{6.05} & 6.05 & \textbf{10,517} \\ \hline

\multirow{8}{*}{\begin{tabular}{c}\textsf{half-cylinder}\\\textsf{(VLM, 6,400)}\\ $(v=0.6)$\end{tabular}}
& SIREN       & 40.44 & 0.150 & 1.26 & 234.2 & 0.27 & 633 \\
& NeurComp    & 42.29 & 0.118 & 0.73 & 235.5 & 0.52 & 1,044 \\
& ECNR        & 41.50 & 0.116 & 1.35 & 151.6 & 0.42 & 446 \\
& Instant-NGP & 41.35 & 0.199 & 0.88 & 48.04 & \textbf{0.01} & 168 \\
& fV-SRN      & 42.23 & 0.165 & 0.73 & 29.94 & 0.03 & 166 \\
& AMGSRN++    & 40.90 & 0.191 & 0.91 & 76.33 & \textbf{0.01} & 371 \\
& IDLat       & 42.17 & 0.119 & 0.65 & 1.73 & 3.17 & 1,114 \\
& EVOLVE      & \textbf{44.91} & \textbf{0.106} & \textbf{0.59} & \textbf{1.61} & 1.24 & \textbf{2,596} \\ \hline

\multirow{8}{*}{\begin{tabular}{c}\textsf{magnetic}\\ $(v=0.05)$\end{tabular}}
& SIREN       & 42.09 & 0.367 & 2.09 & 3,457 & 3.65 & 5,256 \\
& NeurComp    & 43.48 & 0.339 & 1.50 & 3,273 & 3.97 & 6,649 \\
& ECNR        & 43.29 & 0.342 & 1.31 & 1,566 & 5.73 & 3,047 \\
& Instant-NGP & 43.91 & 0.326 & 1.37 & 47.04 & \textbf{0.08} & 1,839 \\
& fV-SRN      & 42.16 & 0.351 & 1.92 & 29.72 & 0.24 & 1,813 \\
& AMGSRN++    & 44.23 & 0.312 & 1.23 & 75.69 & 0.11 & 1,575 \\
& IDLat       & 38.84 & 0.332 & 1.48 & 10.35 & 20.76 & 3,748 \\
& EVOLVE      & \textbf{45.20} & \textbf{0.306} & \textbf{1.04} & \textbf{5.95} & 5.88 & \textbf{10,774} \\ 

\end{tabular}
}
\end{table}

\begin{figure*}[htb]
\centering
$\begin{array}{c@{\hspace{0.05in}}c@{\hspace{0.05in}}c@{\hspace{0.05in}}c@{\hspace{0.05in}}c}
\mbox{\small CR / PSNR} &
 \mbox{\small 2,596$\times$ / 44.91 dB} &
 \mbox{\small 446$\times$ / 41.50 dB} &
 \mbox{\small 1,044$\times$ / 42.29 dB} &
 \mbox{\small 633$\times$ / 40.44 dB} \\
 \includegraphics[width=0.19\linewidth]{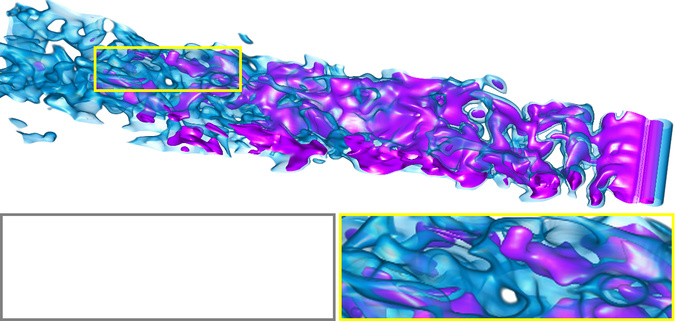}&
 \includegraphics[width=0.19\linewidth]{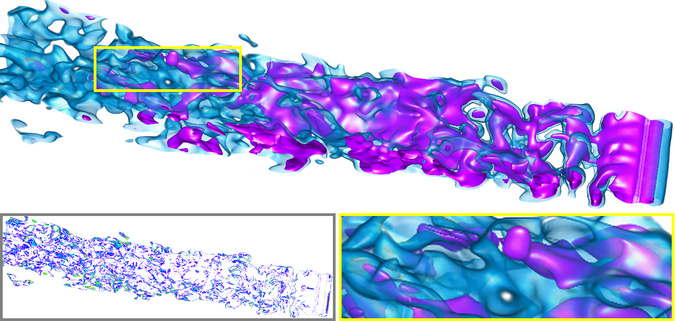}&
 \includegraphics[width=0.19\linewidth]{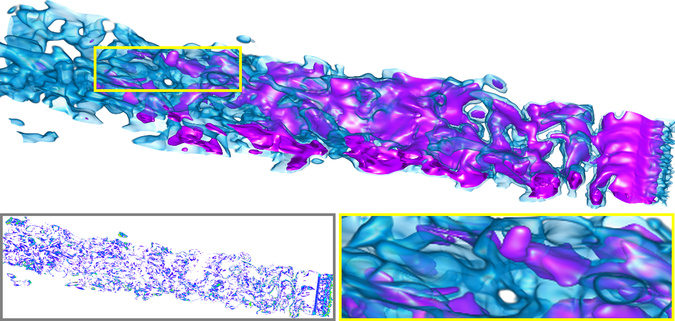}&
 \includegraphics[width=0.19\linewidth]{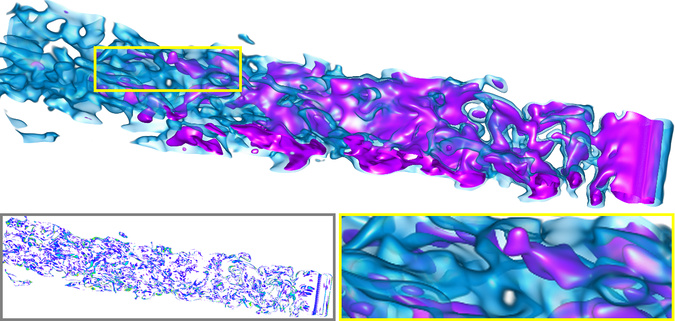}&
 \includegraphics[width=0.19\linewidth]{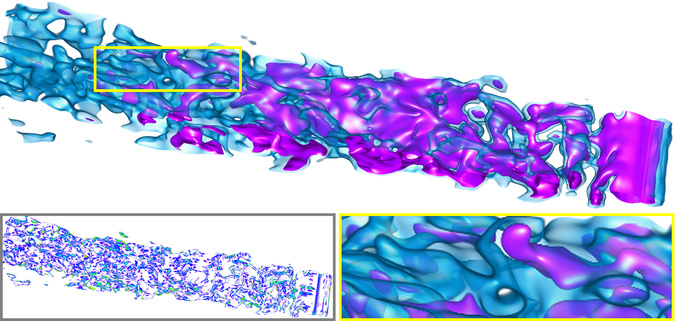}\\

 \mbox{\small GT} &
 \mbox{\small EVOLVE} &
 \mbox{\small ECNR} &
 \mbox{\small NeurComp} &
 \mbox{\small SIREN}\\
\end{array}$
\vspace{-.125in}
\caption{Comparison of volume rendering results between EVOLVE and fully-connected INRs using the \textsf{half-cylinder (VLM, 6,400)} dataset. 
}
\label{fig:fullyINR-vol}
\vspace{-0.05in}
\end{figure*}

\begin{figure*}[!t]
\centering
$\begin{array}{c@{\hspace{0.05in}}c@{\hspace{0.05in}}c@{\hspace{0.05in}}c@{\hspace{0.05in}}c}
\mbox{\small CR / PSNR} &
 \mbox{\small 10,774$\times$ / 45.20 dB} &
 \mbox{\small 1,575$\times$ / 44.23 dB} &
 \mbox{\small 1,813$\times$ / 42.16 dB} &
 \mbox{\small 1,839$\times$ / 43.91 dB} \\
 \includegraphics[width=0.19\linewidth]{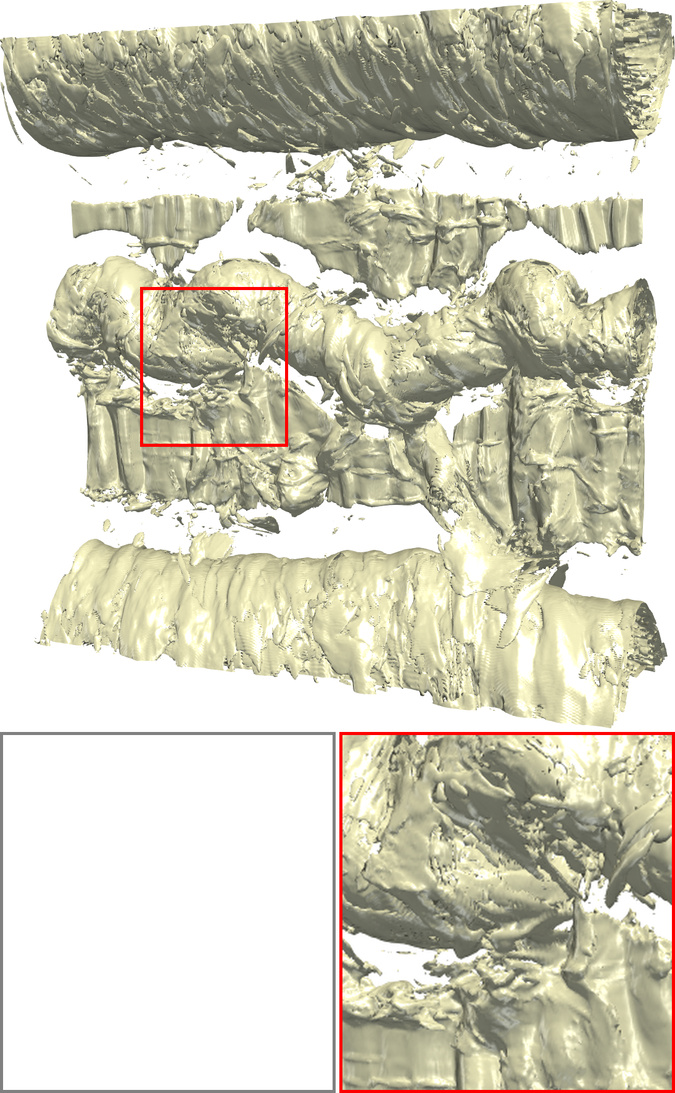}&
 \includegraphics[width=0.19\linewidth]{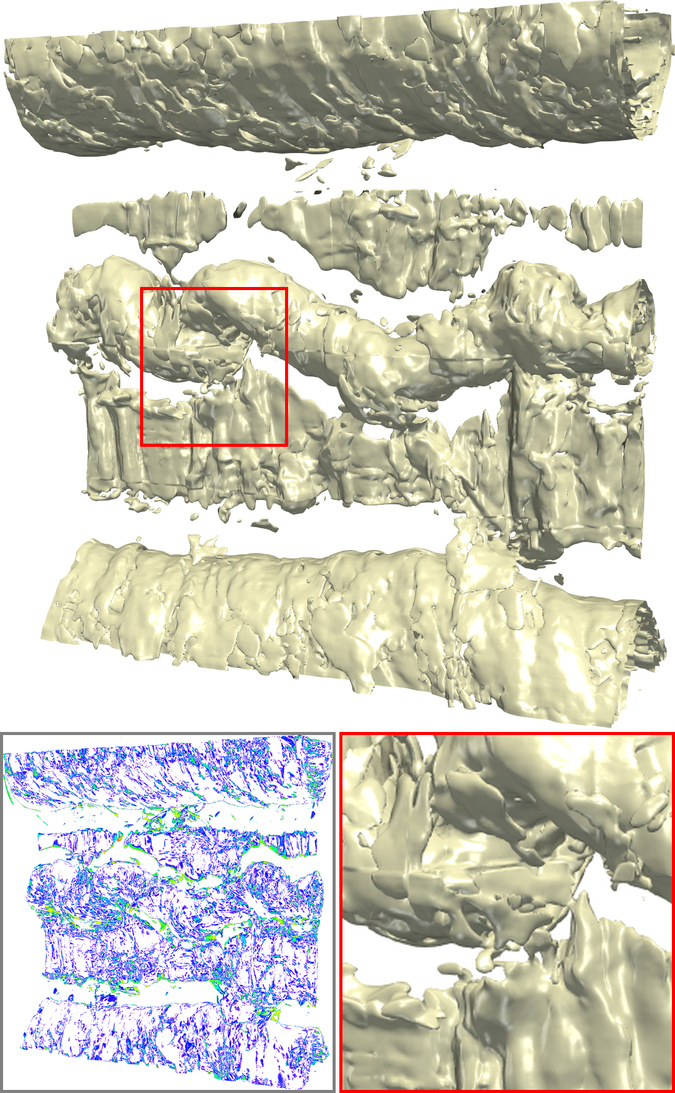}&
 \includegraphics[width=0.19\linewidth]{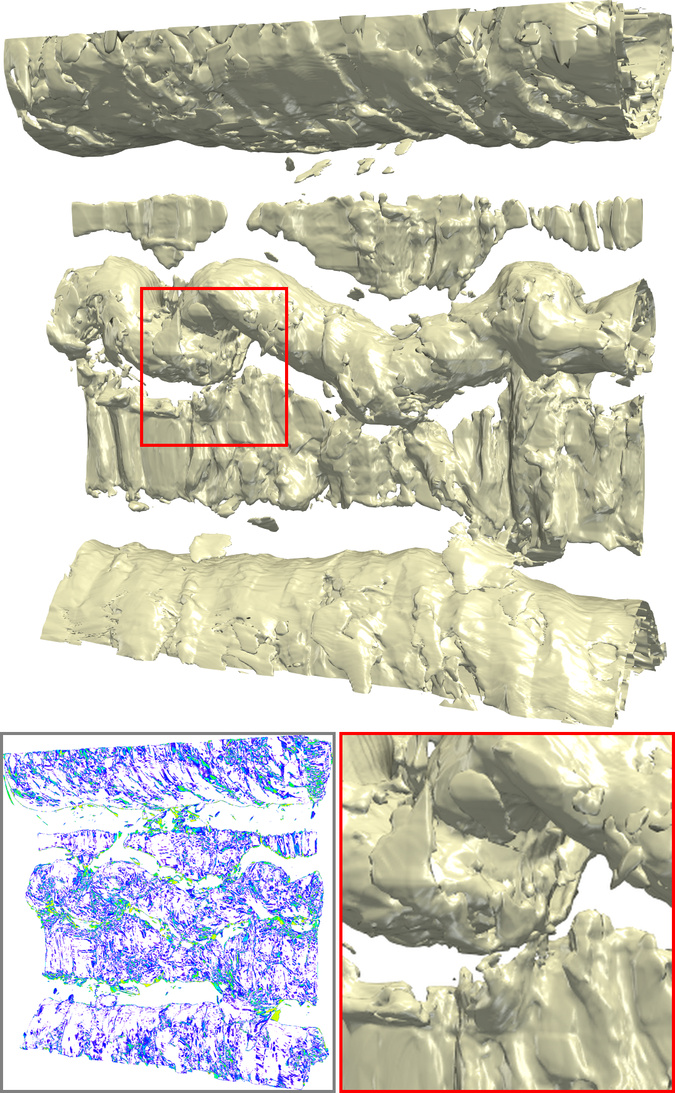}&
 \includegraphics[width=0.19\linewidth]{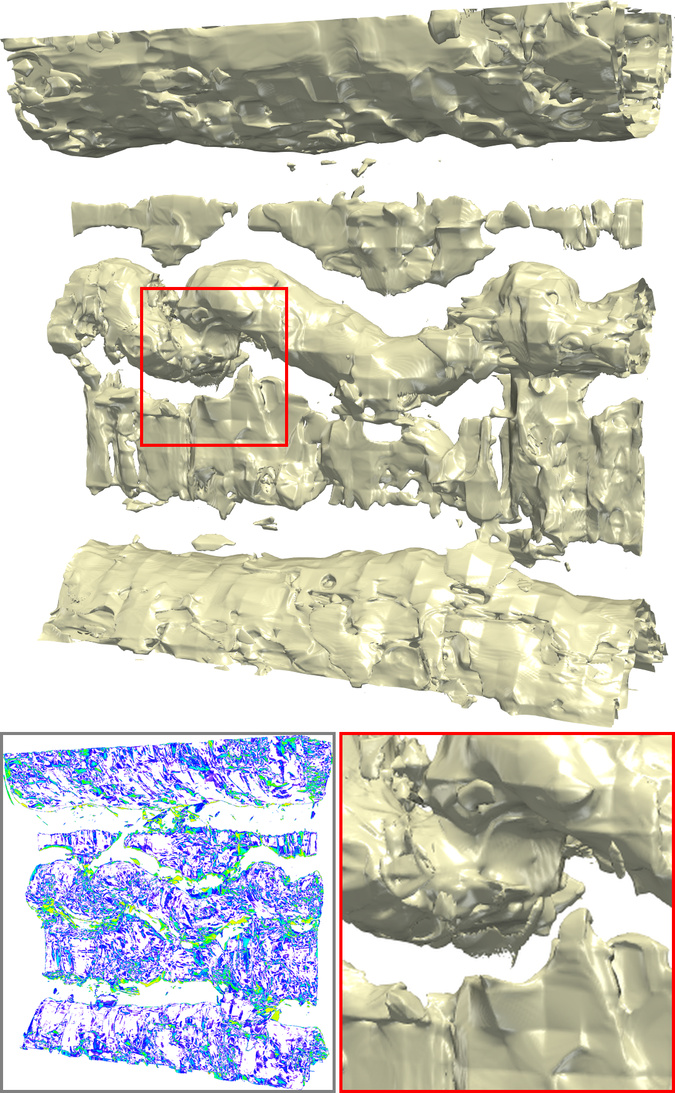}&
 \includegraphics[width=0.19\linewidth]{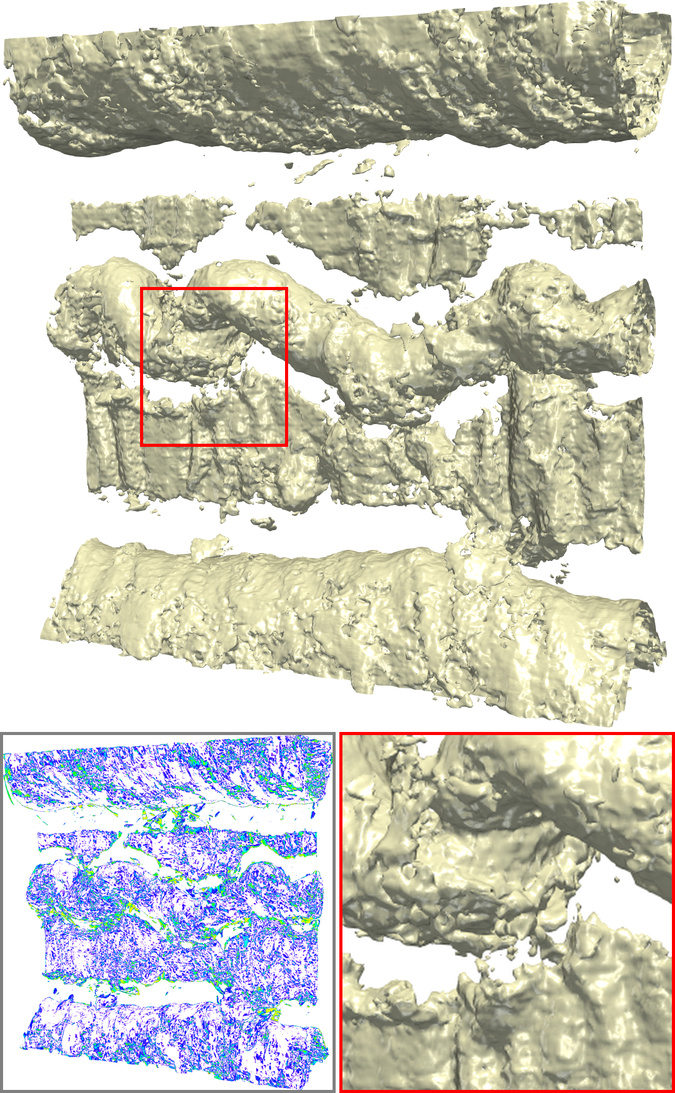}\\

 \mbox{\small GT} &
 \mbox{\small EVOLVE} &
 \mbox{\small AMGSRN++} &
 \mbox{\small fV-SRN} &
 \mbox{\small Instant-NGP}\\
\end{array}$
\vspace{-.125in}
\caption{Comparison of isosurface rendering results between EVOLVE and grid-based INRs using the \textsf{magnetic} dataset. The chosen isovalue is 0.05.}
\label{fig:gridINR-iso}
\vspace{-0.05in}
\end{figure*}

\begin{figure}[htb]
\centering
$\begin{array}{c@{\hspace{0.05in}}c@{\hspace{0.05in}}c}
\mbox{\small CR / PSNR} &
 \mbox{\small 10,517$\times$ / 45.62 dB} &
 \mbox{\small 1,808$\times$ / 41.28 dB} \\
 \includegraphics[width=0.27\linewidth]{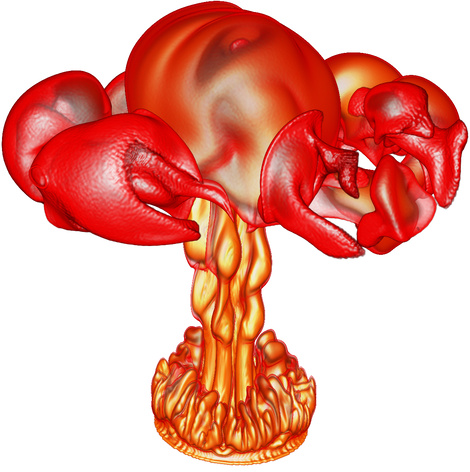}&
 \includegraphics[width=0.27\linewidth]{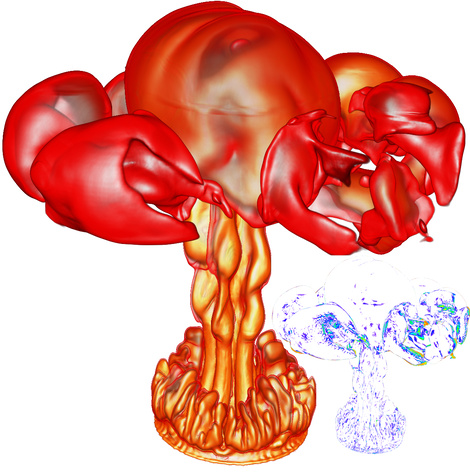}&
 \includegraphics[width=0.27\linewidth]{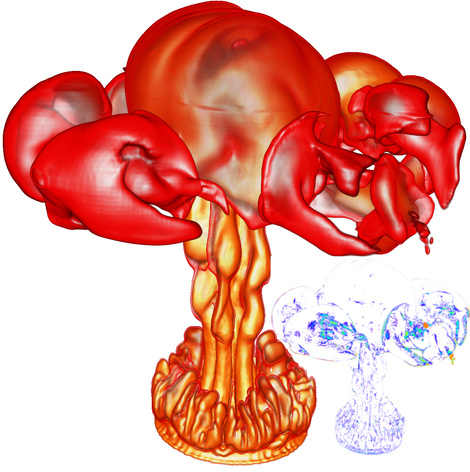}\\
 \mbox{\small GT} &
 \mbox{\small EVOLVE} &
 \mbox{\small IDLat}\\
 \end{array}$
 $\begin{array}{c@{\hspace{0.05in}}c}
  \includegraphics[width=0.27\linewidth]{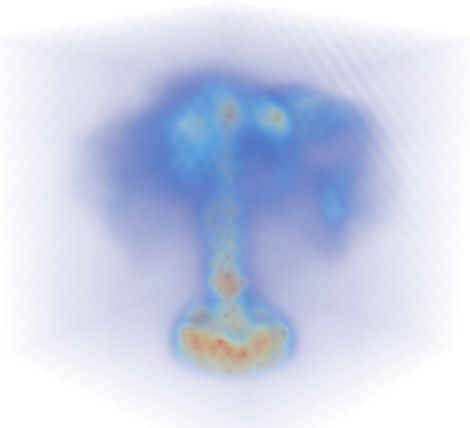}&
 \includegraphics[width=0.27\linewidth]{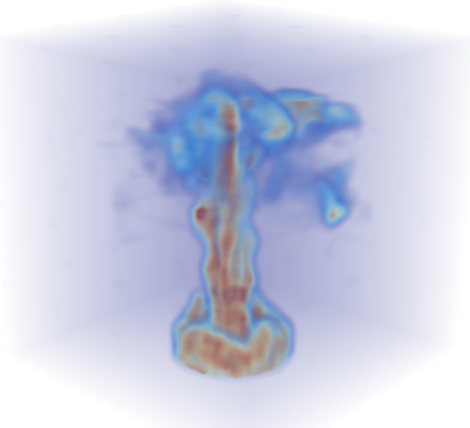}\\
 
 \mbox{\small EVOLVE scale} &
 \mbox{\small IDLat scale} \\
\end{array}$
\vspace{-.125in}
\caption{Top: Comparison of volume rendering results between EVOLVE and IDLat using the \textsf{gas} dataset. 
Bottom: Visualization of the predicted Gaussian scale parameter from the entropy model. Blue-to-red indicates increasing magnitude, corresponding to lower-to-higher bitrate allocation.}
\label{fig:IDLat-vol}
\end{figure}

\vspace{-0.05in}
\subsection{Comparison with \hot{Deep-Learning-Based} Compressors}
\label{subsec:lbc}

We compare EVOLVE with three categories of existing \hot{deep-learning-based} compressors: fully connected INRs (SIREN, NeurComp, ECNR), grid-based INRs (Instant-NGP, fV-SRN, AMGSRN++), and the AE-based compressor IDLat.
Following the same high-fidelity setting (PSNR$\geq$40 dB), we tune the network capacity for INR-based methods to meet this quality threshold.
For IDLat, since its CR is fixed once the model is trained on the same database as EVOLVE, we report its results directly without quality-based adjustment.
Table~\ref{tab:neural-baselines} summarizes the quantitative comparisons across three test datasets. EVOLVE consistently achieves the highest CR while maintaining superior PSNR, LPIPS, and CD on all datasets.
Moreover, thanks to its generalizability, EVOLVE performs compression via a feed-forward inference process without per-instance retraining, resulting in orders-of-magnitude faster encoding speeds than INR-based methods.
Compared with IDLat, the improved efficiency and compression quality of EVOLVE stem primarily from the proposed context model and computationally efficient network designs, which enable faster, higher-quality compression.
Specifically, on the \textsf{magnetic} dataset, EVOLVE achieves a CR of 10,774$\times$ at 45.20~dB, surpassing the best INR method (AMGSRN++, 1,575$\times$ at 44.23~dB) by nearly 7$\times$ in CR while also improving quality.

In terms of visual quality, Figure~\ref{fig:fullyINR-vol} compares the volume rendering results between EVOLVE and fully connected INRs. Under comparable visual fidelity, EVOLVE achieves nearly 2$\times$ higher CR than NeurComp while preserving sharper structural details.
For more challenging datasets such as \textsf{magnetic} (Figure~\ref{fig:gridINR-iso}), both grid-based INRs and EVOLVE inevitably lose some high-frequency details under high CR. However, EVOLVE achieves a CR that is 5.9$\times$ higher than that of fV-SRN, while avoiding the excessive over-smoothing artifacts observed in fV-SRN.
Against IDLat, EVOLVE achieves much higher CR while maintaining comparable visual quality (Figure~\ref{fig:IDLat-vol}).
The scale maps further reveal the entropy model's spatially adaptive behavior. Compared with IDLat, EVOLVE allocates fewer bits to the relatively homogeneous upper regions of the \textsf{gas} volume, while assigning a relatively high bitrate to structurally complex areas near the bottom. This indicates that the proposed context modeling can exploit spatial redundancy, enabling more efficient bit allocation and leading to high CR.

\vspace{-0.05in}
\subsection{Performance Evaluation under Various CRs}
\label{sec:various-CR}

The comparisons in Sections~\ref{subsec:clc}~and~\ref{subsec:lbc} evaluate all methods at a single operating point under the high-fidelity setting (PSNR$\geq$40~dB).
Since scientific workflows often require flexible tradeoffs between quality and compression, we further compare how different methods perform across a range of CRs on the \textsf{combustion (MF)} and \textsf{asteroids} datasets.
For EVOLVE, the variable-rate encoding mechanism (Section~\ref{sec:variable-rate}) enables a single trained model to produce a densely sampled rate-distortion curve by interpolating between learned gain values, without retraining or model switching.
For conventional compressors (SZ3, TTHRESH), different CRs are obtained by varying their quality or error-bound parameters.
For INR-based methods, each operating point requires a separately trained model with adjusted network capacity.
ZFP is excluded from this comparison as its achievable CRs (refer to Table~\ref{tab:traditional-results}) fall well below the evaluated range, and IDLat is excluded because its fixed latent space dimensionality determines its CR.

\begin{figure}[!t]                         
\centering 
$\begin{array}{c@{\hspace{0.005in}}c}
\includegraphics[width=0.42\linewidth]{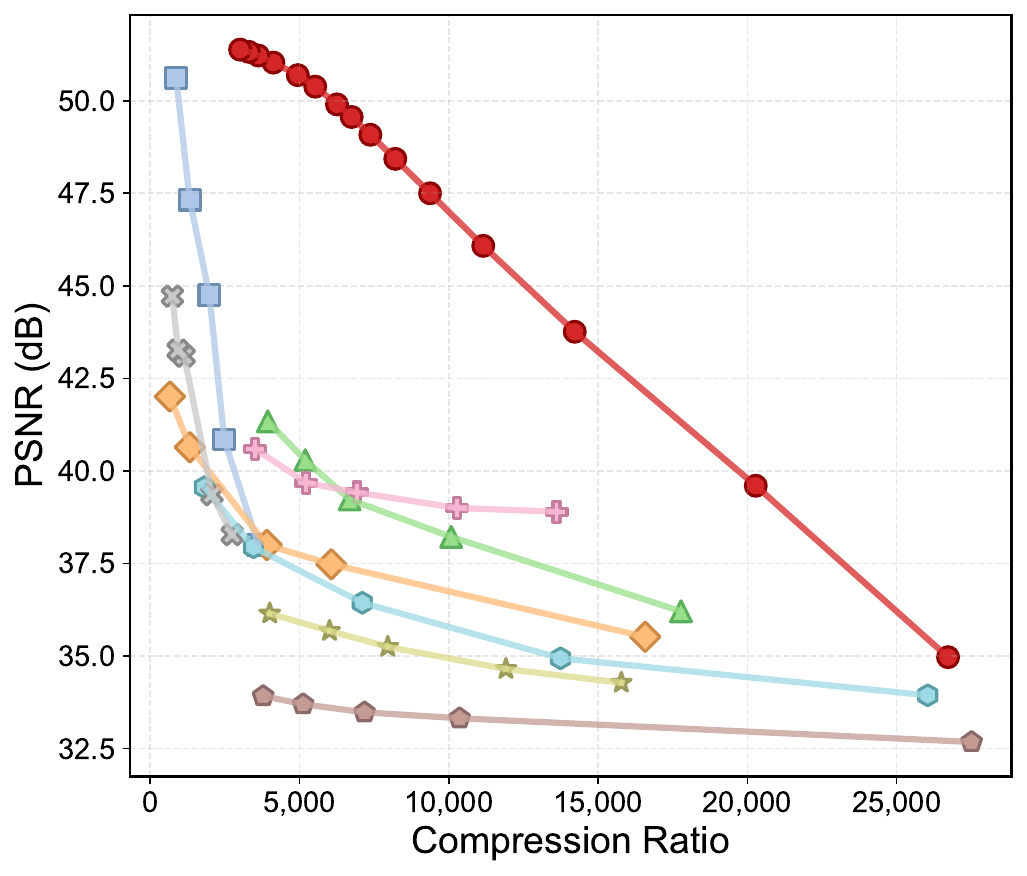}&
\includegraphics[width=0.42\linewidth]{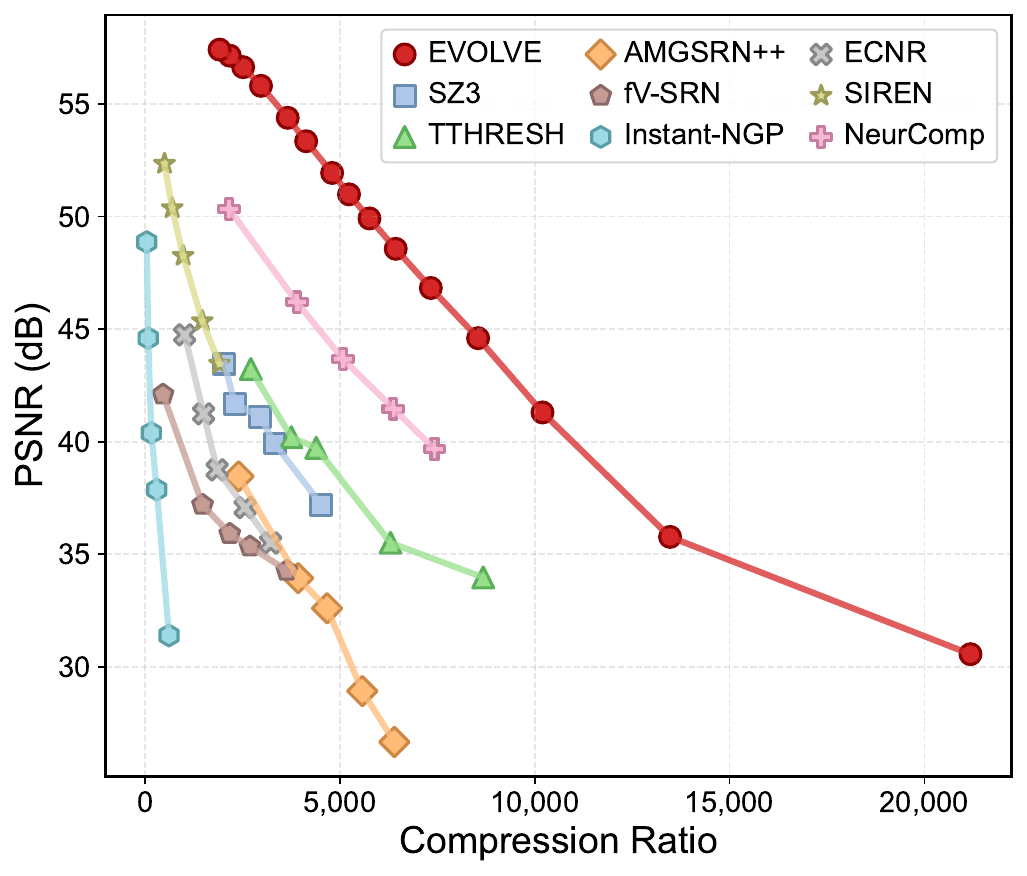}\\
\includegraphics[width=0.42\linewidth]{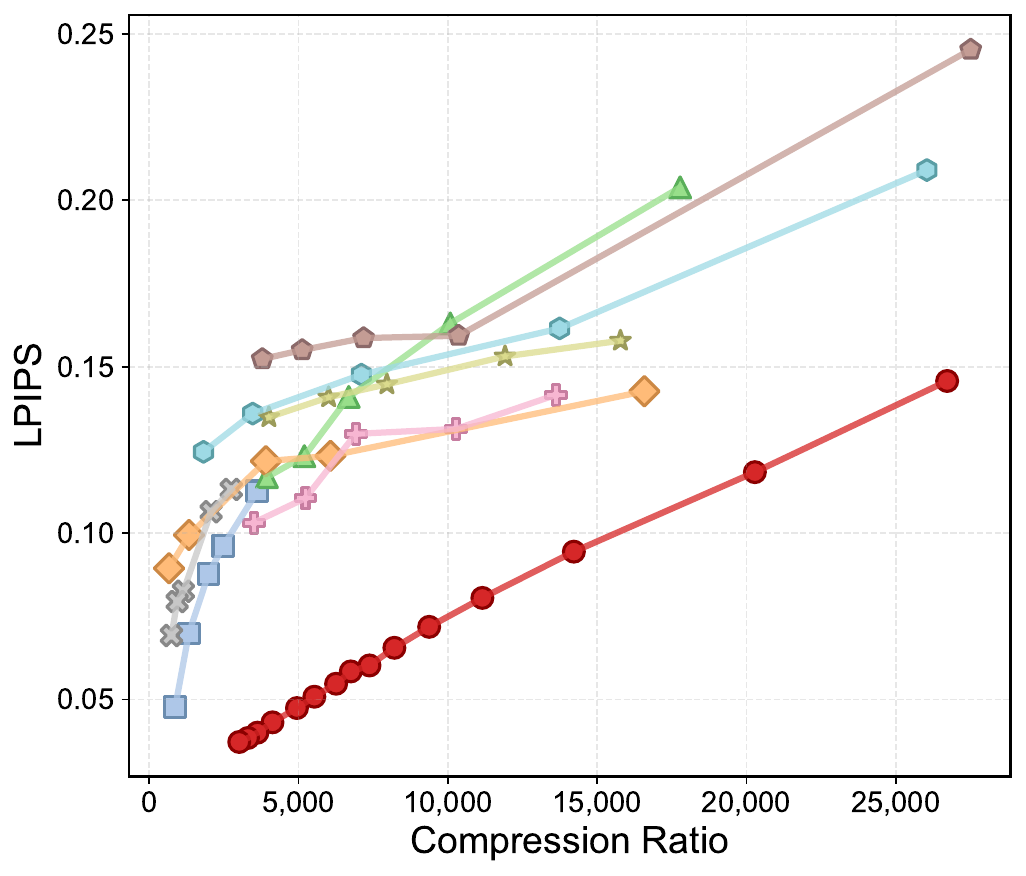}&
\includegraphics[width=0.42\linewidth]{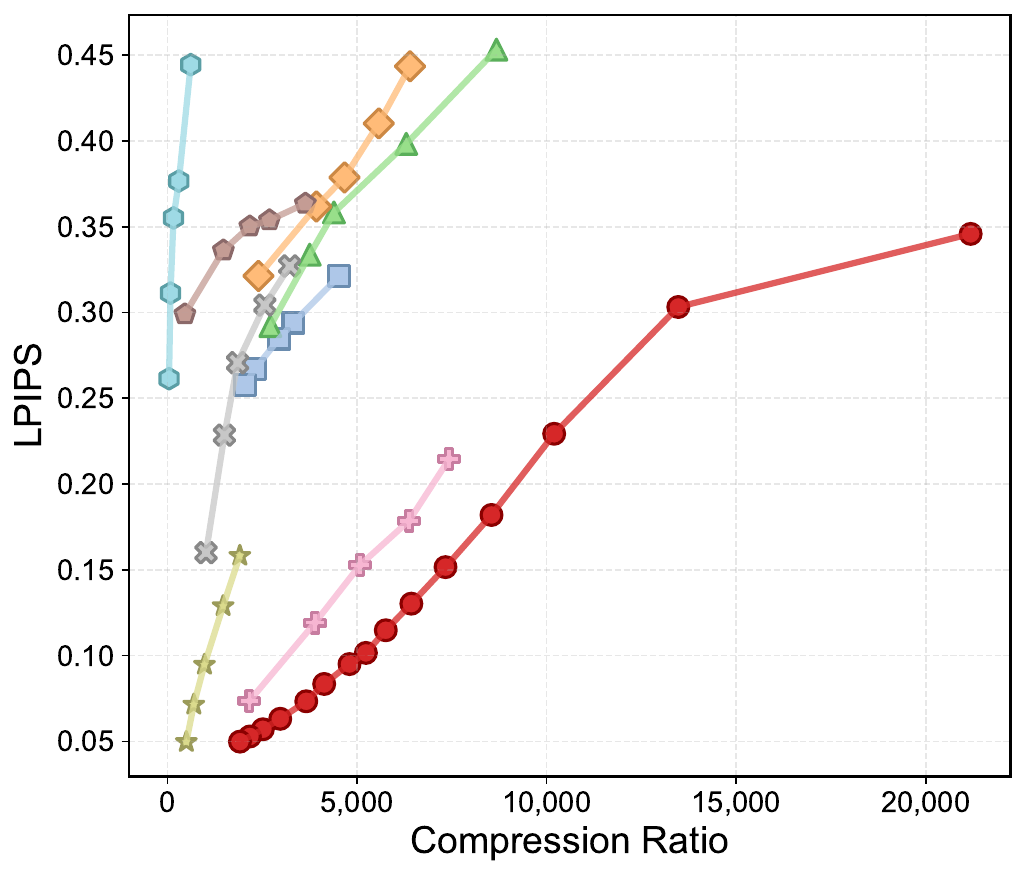}\\
 \mbox{\small \textsf{asteroids}} &
 \mbox{\small \textsf{combustion (MF)}}\\
\end{array}$
\vspace{-.125in}
\caption{Rate-distortion curves of different compression approaches.}
\label{fig:RD-curves}
\vspace{-0.05in}
\end{figure}

Figure~\ref{fig:RD-curves} presents the resulting rate-distortion curves measured in both PSNR and LPIPS.
First, at comparable CRs, EVOLVE consistently achieves the highest PSNR among all evaluated methods on both datasets, demonstrating that the learned representations generalize well to unseen data across a wide range of rate-distortion operating points.
Second, a single EVOLVE model spans a wide, continuous CR range via the learnable gain mechanism.
The densely and smoothly sampled points along the EVOLVE curves confirm that the gain interpolation provides stable, fine-grained rate control without abrupt quality jumps.
Third, while conventional compressors can also operate across multiple CRs by tuning their parameters, they exhibit steeper quality degradation as the CR increases.
For instance, on the \textsf{asteroids} dataset, SZ3 achieves high PSNRs at low CRs but drops sharply beyond 3,000$\times$, whereas EVOLVE maintains competitive reconstruction quality even at CRs exceeding 10,000$\times$.
Similarly, INR-based methods are generally confined to narrower CR ranges, and the LPIPS curves further corroborate these findings, showing that EVOLVE preserves superior perceptual quality across the entire evaluated CR range.

\vspace{-0.05in}
\subsection{\hot{Deployment Considerations}}
\label{sec:deployment}

\hot{Volume compression methods differ in where decompression is placed in the visualization pipeline. For some methods, decompression can be performed on demand at each frame during rendering (e.g., wavelet-based GPU compression~\cite{Treib-TVCG12}, HVQ~\cite{Schneider-VIS03}, dictionary-based sparse representations~\cite{Gobbetti-CGF12,Diaz-CG20}, GPU-accelerated time-varying volume exploration~\cite{Marton-CGF19}, and grid-based INRs~\cite{Muller-TOG22, Weiss-CGF22, Wu-TVCG24, Wurster-TVCG24}). In contrast, other methods require decompressing the whole volume before rendering.

EVOLVE belongs to the latter category. Its decoding latency, typically a few seconds per volume on a consumer GPU, is incurred only once per volume, not per frame. After decoding, all interactive operations, including transfer function editing, isovalue changes, and spatial navigation, proceed at native rendering rates.
For time-varying data exploration, each timestep must likewise be decoded before visualization; future timesteps can be prefetched and decoded asynchronously while the current timestep is being inspected, hiding part of the switching latency.
EVOLVE is therefore best suited when storage and transmission bandwidth savings outweigh the need for immediate random access to compressed timesteps.
For transient rendering, strict timing constraints may limit the affordable decoding complexity and achievable CR; for offline storage, the autoregressive context model can trade decoding latency (refer to Tables~\ref{tab:neural-baselines}) for substantially higher CRs.}
\pin{Deployment also incurs a one-time overhead from the shared model itself (85.6M parameters), which is amortized over many encoded and decoded volumes.}

%% file: conclusion.tex
\section{Conclusions and Future Work}

We have presented EVOLVE, an AE-based framework toward more generalizable learned volume compression for scientific data, shifting the focus from domain-specific optimization to unified representation learning.
Its efficient, scalable design balances compression efficiency and reconstruction fidelity, and supports continuous variable-rate encoding within a single model, offering a practical path toward \hot{generalizable} data-driven volumetric compression.

Several limitations remain.
First, while our approach differs from prior works~\cite{Han-TVCG-SSR, He-TVCG20, Tang-PVis24} by training a generalizable model on a large-scale database, the domains covered remain limited.
\hot{In particular, our database contains only simulation data and no medical or microscopy volumes, whose noise characteristics and near-lossless requirements differ substantially; applying EVOLVE to such domains would require retraining with representative data.}
\hot{Second, training EVOLVE is computationally intensive, requiring a high-end GPU and a three-stage optimization schedule. Once trained on representative data from a domain, however, the model compresses both seen and unseen in-domain volumes without per-volume optimization, and inference runs on consumer-level GPUs.}
\hot{Finally,} the current framework treats each volume independently, and leveraging temporal redundancy in time-varying simulations could further improve CRs.

%% file: appendix.tex
\newpage
\clearpage

\setcounter{section}{0}
\setcounter{figure}{0}
\setcounter{table}{0}


\section{Training Database Composition}
\label{sec:database-detail}

Table~\ref{tab:dataset} reports the full composition of the training database curated as described in Section~\ref{sec:database-curation}.
As exceptions, \textsf{FIT} and \textsf{rotstrat} are actually single static volumes with a resolution of 2,048$\times$2,048$\times$2,048, and we partition each volume into 512 subvolumes.

\begin{table}[!ht]
\centering
\caption{The composition of the training database. Dataset diversity is measured as the 95th percentile of the NNS score using pHash.}
\label{tab:dataset}
\vspace{-0.05in}
\resizebox{\columnwidth}{!}{%
\begin{tabular}{cccccc}
 & volume resolution & \# total & \# selected &original & final \\
dataset & ($x \times y \times z$)  & volumes  & volumes  & NNS$\downarrow$  & NNS$\downarrow$  \\
\hline
\textsf{argon-bubble} & 640$\times$256$\times$256 & 165 & 55 & 0.93 & 0.85 \\
\textsf{combustion} & 480$\times$720$\times$120 & 400 & 400 & 0.69 & 0.69 \\
\textsf{earthquake} & 256$\times$256$\times$96 & 599 & 374 & 0.90 & 0.85 \\
\textsf{explosion} & 128$\times$128$\times$128 & 174 & 65 & 0.98 & 0.84 \\
\textsf{FIT} & 256$\times$256$\times$256 & 512 & 512 & 0.50 & 0.50 \\
\textsf{five-jet} & 128$\times$128$\times$128 & 2,000 & 105 & 0.96 & 0.84 \\
\textsf{gravity} & 128$\times$128$\times$128 & 833 & 833 & 0.66 & 0.66 \\
\textsf{half-cylinder} & 640$\times$240$\times$80 & 600 & 600 & 0.77 & 0.77 \\
\textsf{hurricane} & 500$\times$500$\times$100 & 192 & 133 & 0.90 & 0.85 \\
\textsf{ionization} & 600$\times$248$\times$248 & 600 & 513 & 0.88 & 0.85 \\
\textsf{mantle} & 360$\times$201$\times$180 & 251 & 90 & 0.93 & 0.84 \\
\textsf{MHD} & 256$\times$256$\times$256 & 297 & 297 & 0.68 & 0.68 \\
\textsf{neutron-star} & 192$\times$128$\times$66 & 400 & 400 & 0.54 & 0.54 \\
\textsf{radiative-layer} & 256$\times$128$\times$128 & 600 & 600 & 0.53 & 0.53 \\
\textsf{Rayleigh-Taylor} & 128$\times$128$\times$128 & 100 & 56 & 0.99 & 0.85 \\
\textsf{rotstrat} & 256$\times$256$\times$256 & 512 & 512 & 0.51 & 0.51 \\
\textsf{solar-plume} & 128$\times$128$\times$512 & 28 & 28 & 0.76 & 0.76 \\
\textsf{supercurrent} & 256$\times$128$\times$32 & 908 & 60 & 0.99 & 0.85 \\
\textsf{supernova} & 432$\times$432$\times$432 & 60 & 60 & 0.56 & 0.56 \\
\textsf{Tangaroa} & 300$\times$180$\times$120 & 600 & 593 & 0.85 & 0.85 \\
\textsf{vortex} & 128$\times$128$\times$128 & 90 & 90 & 0.70 & 0.70 \\
\hline
total & --- & 9,921 & 6,376 & --- & --- \\
\end{tabular}%
}
\end{table}

Nine datasets are multivariate or ensemble, each containing multiple physical variables or generated by different simulation parameters.
Specifically, 
\hot{\textsf{combustion}~\cite{Hawkes-PCI07}} includes CHI, HR, VORTS, and YOH;
\hot{\textsf{explosion}~\cite{explosion}} includes RHO, P, and T; 
\hot{\textsf{half-cylinder}~\cite{half-cylinder}} provides VLM and VTM at three Reynolds numbers (160, 320, 640); 
\hot{\textsf{hurricane}~\cite{SciVis-Contest}} includes CLOUD, P, VAPOR, and WSMAG; 
\hot{\textsf{ionization}~\cite{Whalen-ApJ08}} includes GT, H, H$_2$, He, He+, and PD;
\hot{\textsf{MHD}~\cite{MHD}} comprises RHO, MFM, and VLM; and 
\hot{\textsf{Tangaroa}~\cite{Tangaroa}} includes ACC, DIV, VLM, and VTM.
In addition, \hot{\textsf{gravity}~\cite{gravity}} and \hot{\textsf{radiative-layer}~\cite{radiative-layer}} contribute 17 and six distinct parameter combinations, respectively.

\hot{Each variable name denotes a physical quantity of the corresponding simulation. 
For \textsf{combustion}~\cite{Hawkes-PCI07}, CHI, HR, VORTS, and YOH refer to the scalar dissipation rate, heat release, vorticity magnitude, and OH mass fraction. 
For \textsf{hurricane}~\cite{SciVis-Contest}, a simulation of Hurricane Isabel, CLOUD, P, VAPOR, and WSMAG denote the total cloud moisture mixing ratio, atmospheric pressure, water vapor mixing ratio, and wind speed magnitude (computed from the U/V/W wind components), respectively. 
For \textsf{ionization}~\cite{Whalen-ApJ08}, which simulates radiation-driven ionization fronts in primordial gas with a nine-species chemistry network, GT and PD denote the gas temperature and particle density. In contrast, H, H$_2$, He, He+, and H+ denote the abundances of neutral atomic hydrogen, molecular hydrogen, neutral helium, singly ionized helium, and ionized hydrogen, respectively. 
The remaining abbreviations follow common conventions (e.g., P for pressure, T for temperature, RHO for density, VLM/VTM for velocity/vorticity magnitude). 
For detailed information on these datasets, we refer readers to the corresponding entries in the Open SciVis Datasets~\cite{Klacansky-OSV}, IEEE SciVis Contest archives~\cite{SciVis-Contest}, ETH Z{\"u}rich visualization datasets~\cite{ETH-CGL-Data}, and Well~\cite{Ohana-NeurIPS24}.}

For \textsf{ionization}, we remove the early and late timesteps that are largely empty and retain 100 content-rich timesteps for each variable.

\vspace{-0.05in}
\section{Compression and Decompression API}
\label{sec:api}


EVOLVE provides a compression command and a decompression command.
Tables~\ref{tab:api-compress}~and~\ref{tab:api-decompress} summarize the key parameters.

\textbf{Compression.}
To compress a volume, the user specifies the input data directory, the pretrained model checkpoint, and the desired quality level.
EVOLVE uses a continuous gain factor (\texttt{-{}-factor}, float in $[0.0, 10.0]$) for fine-grained bitrate control: larger values yield higher quality at lower CR.
The volume is divided into overlapping blocks with configurable size and stride, enabling memory-efficient processing of arbitrary-resolution volumes.

\textbf{Decompression.}
For decompression, only the compressed bitstream directory, the same model checkpoint, and the matching model configuration are needed.
The volume resolution, block layout, and normalization ranges are automatically recovered from the stored meta info (refer to Appendix~\ref{sec:bitstream-composition}), so no additional data-specific parameters are required.

\begin{table}[!ht]
  \centering
  \caption{Key parameters of the compression command.}
  \label{tab:api-compress}
  \vspace{-0.05in}
  \resizebox{\columnwidth}{!}{%
  \begin{tabular}{l l p{5.0cm}}
  parameter & type & description \\ \hline
  \texttt{-{}-data\_path} & str & input volume file path \\
  \texttt{-{}-checkpoint} & str & path to the pretrained EVOLVE model checkpoint \\
  \texttt{-{}-block\_size} & int$\times$3 & spatial block resolution ($H\!\times\!W\!\times\!D$) \\
  \texttt{-{}-stride} & int$\times$3 & stride for overlapping blocks \\
  \texttt{-{}-factor} & float & gain factor for quality control ($0.0$--$10.0$) \\
  \texttt{-{}-output\_dir} & str & directory for saving compressed bitstreams \\
  \end{tabular}
  }
\end{table}

\begin{table}[!ht]
  \centering
  \caption{Key parameters of the decompression command.}
  \label{tab:api-decompress}
  \vspace{-0.05in}
  \resizebox{\columnwidth}{!}{%
  \begin{tabular}{l l p{5.0cm}}
  parameter & type & description \\ \hline
  \texttt{-{}-bitstream\_path} & str & compressed bitstream file path \\
  \texttt{-{}-checkpoint} & str & path to the pretrained EVOLVE model checkpoint \\
  \texttt{-{}-output\_dir} & str & directory for saving decompressed volumes \\
  \end{tabular}
  }
\end{table}

\vspace{-0.05in}
\section{Compressed File Composition}
\label{sec:bitstream-composition}

Each EVOLVE compressed bitstream consists of three components:
latent $\hat{\bm{y}}$, hyperlatent $\hat{\bm{z}}$, and meta info.
The latent $\hat{\bm{y}}$ is the entropy-coded output of the encoder, representing the primary signal content.
It is partitioned into five channel slices, each further decomposed into anchor and non-anchor positions via a checkerboard pattern for context-based entropy coding.
Because $\hat{\bm{y}}$ has the same spatial resolution as the downsampled input (i.e., $\frac{1}{16}$ along each axis), it carries the vast majority of the information and dominates the file size.
The hyperlatent $\hat{\bm{z}}$ serves as side information: the hyper-encoder compresses $\hat{\bm{y}}$ into a much coarser representation (an additional 8$\times$ spatial downsampling relative to $\hat{\bm{y}}$) to estimate the mean and scale parameters of the Gaussian entropy model.
As a result, $\hat{\bm{z}}$ is 8$\times$8$\times$8 = 512$\times$ smaller than $\hat{\bm{y}}$ in voxel count, and its contribution to the total bitstream is inherently small.
The meta info is a fixed-cost overhead required for correct decoding.
It includes: (1)~a global header storing the volume resolution, number of blocks, and number of channel slices;
(2)~per-block headers recording spatial position, block size, normalization range ($v_\text{min}$, $v_\text{max}$), and the byte lengths of $\hat{\bm{y}}$ and $\hat{\bm{z}}$ payloads; and
(3)~per-slice descriptors storing the byte lengths of anchor and non-anchor bitstreams within each slice.
These fields are stored as fixed-width integers and floats, so meta info scales linearly with the number of patches but is independent of the data content.

Table~\ref{tab:bitstream-composition} reports the composition breakdown on the \textsf{combustion (MF)} and \textsf{isotropic} datasets at two different CRs.
Across all configurations, $\hat{\bm{y}}$ dominates (75.5\%--98.0\% of the total file size).
$\hat{\bm{z}}$ accounts for only 1.3\%--14.5\%, and meta info contributes 0.7\%--10.0\%.
At higher CRs (smaller files), the fixed-cost meta info and $\hat{\bm{z}}$ occupy a proportionally larger share.
At lower CRs (larger files), $\hat{\bm{y}}$ grows disproportionately, reaching 95--98\% of the total, confirming that the overhead of $\hat{\bm{z}}$ and meta info is negligible.

\begin{table}[!ht]
  \centering
  \caption{Compressed bitstream composition of EVOLVE at different CRs.}
  \label{tab:bitstream-composition}
  \vspace{-0.05in}
  \resizebox{\columnwidth}{!}{%
  \begin{tabular}{ccccc}
  dataset & CR & latent $\hat{\bm{y}}$ (\%) & hyperlatent $\hat{\bm{z}}$ (\%) & meta info (\%) \\
  \hline
  \textsf{combustion (MF)} & 7,595 & 75.5 & 14.5 & 10.0 \\
  \textsf{combustion (MF)} & 1,432 & 95.3 & 2.8 & 1.9 \\
  \textsf{isotropic}  & 3,812 & 87.5 & 8.1 & 4.4 \\
  \textsf{isotropic}  & 607   & 98.0 & 1.3 & 0.7 \\
  \end{tabular}
  }
\end{table}

\vspace{-0.05in}
\section{\hot{Inference Memory Footprint}}
\label{sec:memory-footprint}

\hot{EVOLVE compresses a volume by partitioning it into 128$\times$128$\times$128 blocks, each encoded and decoded independently, streaming a single block to the GPU at a time. Table~\ref{tab:memory} reports the peak GPU and host (CPU) memory during compression on a single NVIDIA RTX 4090, measured with \texttt{nvidia-smi} (per-process GPU memory) and the peak resident set (\texttt{VmHWM}), for two test volumes, \textsf{asteroids} (1,000$\times$1,000$\times$1,000, 3.7~GB) and \textsf{gas} (512$\times$512$\times$512, 512~MB), at two different compression rates.
Since only one block resides on the GPU at any time, the GPU memory holds just the network parameters and the activations of the current block, and is therefore constant ($\sim$2.7~GB) regardless of the volume size. The host memory instead holds the full input volume, its reconstruction, and the accumulated bitstream, and thus grows with the volume size. For a given volume, the footprint is independent of the target rate. This constant and modest GPU footprint allows EVOLVE to compress even gigabyte-scale volumes on a single consumer GPU (see Section~\ref{sec:deployment}). For volumes whose reconstruction exceeds the host memory, one can partition them into smaller subvolumes and process them sequentially.}

\begin{table}[t]
\centering
\caption{\hot{Peak GPU and host (CPU) memory during block-based (128$\times$128$\times$128 block) compression on a single NVIDIA RTX 4090, for two test volumes at two different compression rates. Because blocks are streamed to the GPU individually, the GPU footprint remains constant regardless of volume size, whereas the host memory grows with volume size.}}
\label{tab:memory}
\resizebox{\columnwidth}{!}{%
\begin{tabular}{l c c c c c}
\hot{dataset} & \hot{resolution} & \hot{CR} & \hot{PSNR (dB)} & \hot{GPU (GB)} & \hot{CPU (GB)} \\
\hline
\multirow{2}{*}{\hot{\textsf{asteroids}}} & \multirow{2}{*}{\hot{1,000$\times$1,000$\times$1,000}} & \hot{6,523.7} & \hot{49.72} & \hot{2.7} & \hot{21.2} \\
 & & \hot{4,527.6} & \hot{50.87} & \hot{2.7} & \hot{21.2} \\
\hline
\multirow{2}{*}{\hot{\textsf{gas}}} & \multirow{2}{*}{\hot{512$\times$512$\times$512}} & \hot{6,784.1} & \hot{49.14} & \hot{2.7} & \hot{4.0} \\
 & & \hot{4,597.9} & \hot{50.86} & \hot{2.7} & \hot{4.0} \\
\end{tabular}%
}
\end{table}

\vspace{-0.05in}
\section{Latent Space Exploration and Analysis}
\label{sec:lse}

\begin{figure}[!h]
\centering
$\begin{array}{c@{\hspace{0.05in}}c}
 \includegraphics[width=0.47\linewidth]{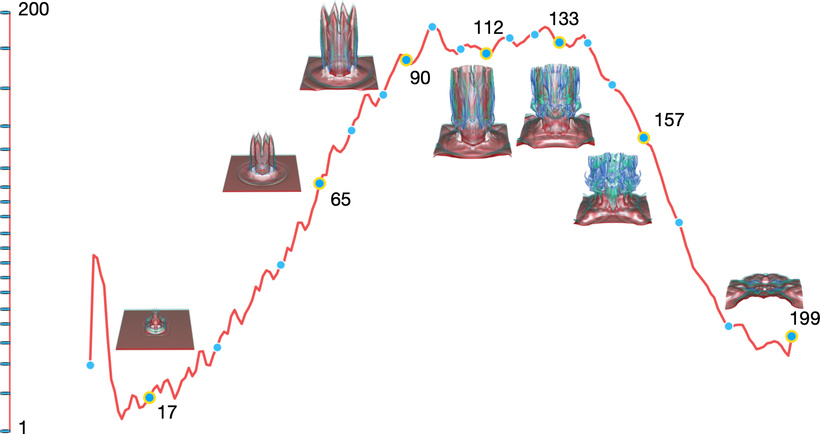}&
 \includegraphics[width=0.47\linewidth]{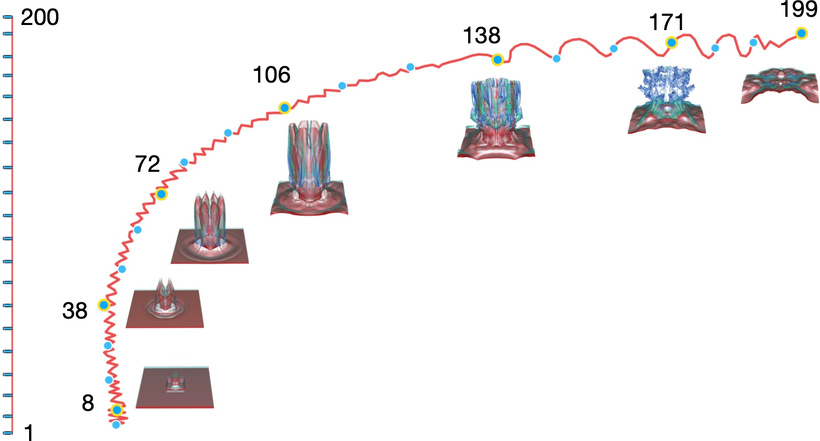}\\
 
 \mbox{\small (a) Wang et al.~\cite{Wang-TVCG08}} &
 \mbox{\small (b) EVOLVE}\\
\end{array}$
\vspace{-.125in}
\caption{Comparison of representative timestep selection results using the \textsf{ionization-T (H+)} dataset. Each method selects 20 timesteps (marked along the vertical timeline) from 200 timesteps with seven rendering snapshots highlighted. In (a), the horizontal axis represents the timesteps.}
\label{fig:Latent-analysis}
\end{figure}

Similar to other autoencoder-based methods~\cite{Han-TVCG20, Porter-VIS19, Shen-TVCG23}, the latent vectors extracted from the pretrained EVOLVE model are interpretable. In Figure~\ref{fig:Latent-analysis}, we compare the representative timestep selection results for the \textsf{ionization-T (H+)} dataset using the importance-driven framework proposed by Wang et al.~\cite{Wang-TVCG08} and EVOLVE.
For EVOLVE, following~\cite{Porter-VIS19}, we project the latent vectors extracted from all volumes into 2D using t-SNE, with each point corresponding to a timestep. We then connect neighboring timesteps to form a trajectory in the latent space and select representative timesteps based on the arclength and angle metrics computed along this trajectory.
Compared with the information-theoretic selection shown in Figure~\ref{fig:Latent-analysis}(a), we can observe that EVOLVE produces a more balanced temporal coverage, as shown in Figure~\ref{fig:Latent-analysis}(b). Early formation stages, transitions, and stabilization phases remain consistently represented. 
This result highlights the potential of EVOLVE for volume analysis applications.

\vspace{-0.05in}
\section{Compression of Time-Varying and Ensemble Data}
\label{sec:tved}

\begin{figure*}[htb]
\centering
$\begin{array}{@{}c@{\hspace{0.1in}}c@{\hspace{0.1in}}c@{\hspace{0.1in}}c@{\hspace{0.1in}}c}
 \vcenter{\hbox{\rotatebox{90}{\small \mbox{$t$=165}}}} &
 \vcenter{\hbox{\includegraphics[width=0.215\linewidth]{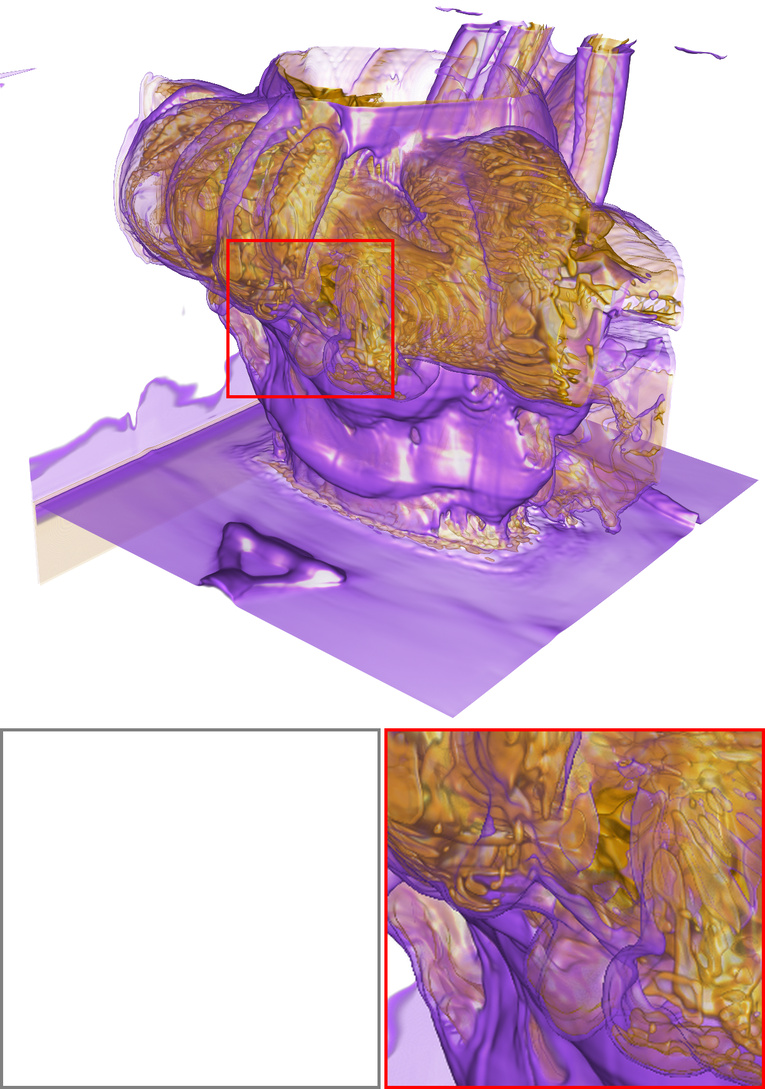}}}&
 \vcenter{\hbox{\includegraphics[width=0.215\linewidth]{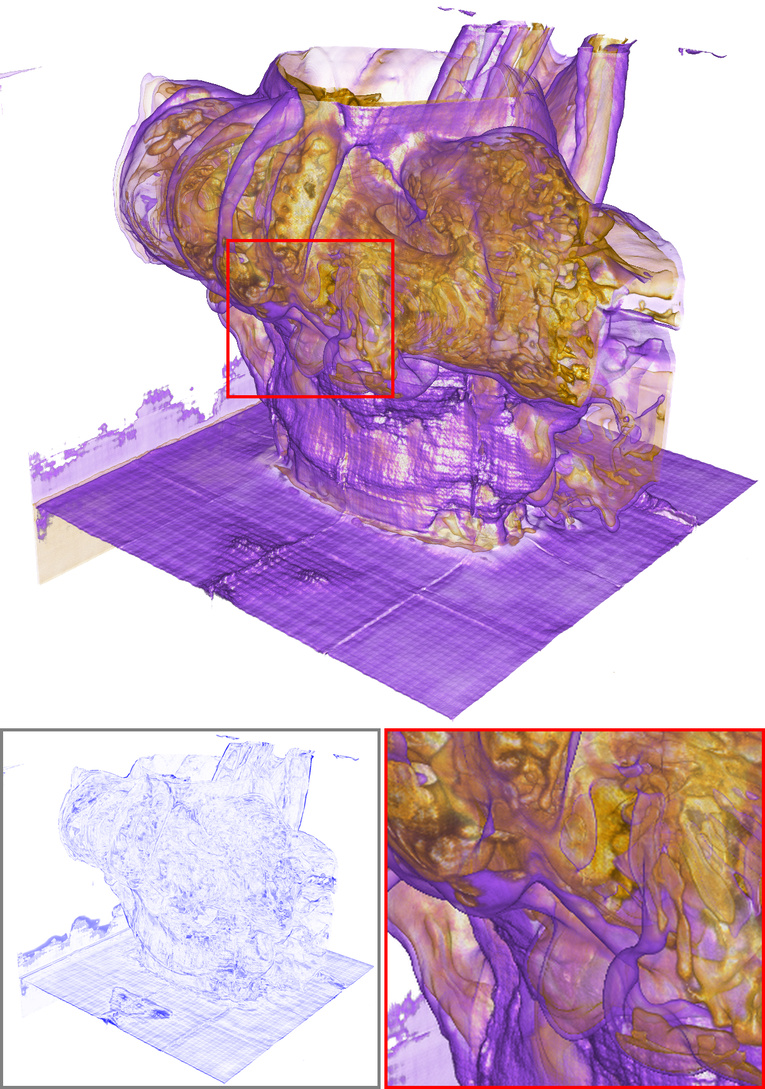}}}&
 \vcenter{\hbox{\includegraphics[width=0.215\linewidth]{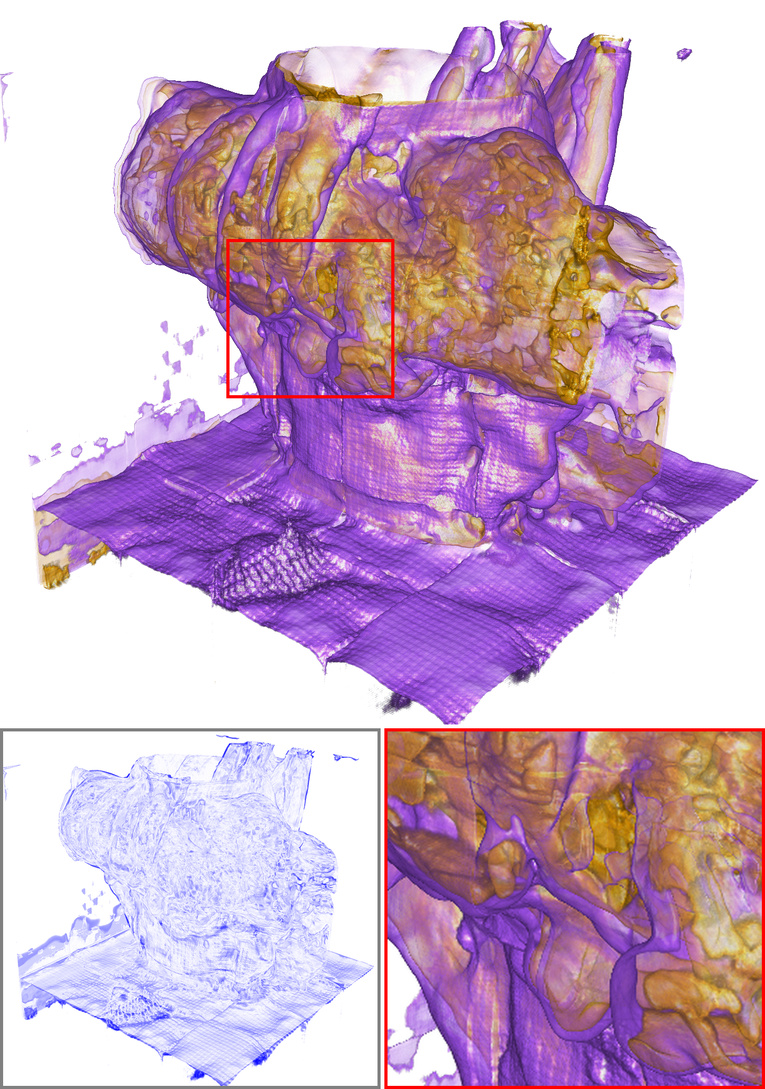}}}&
 \vcenter{\hbox{\includegraphics[width=0.215\linewidth]{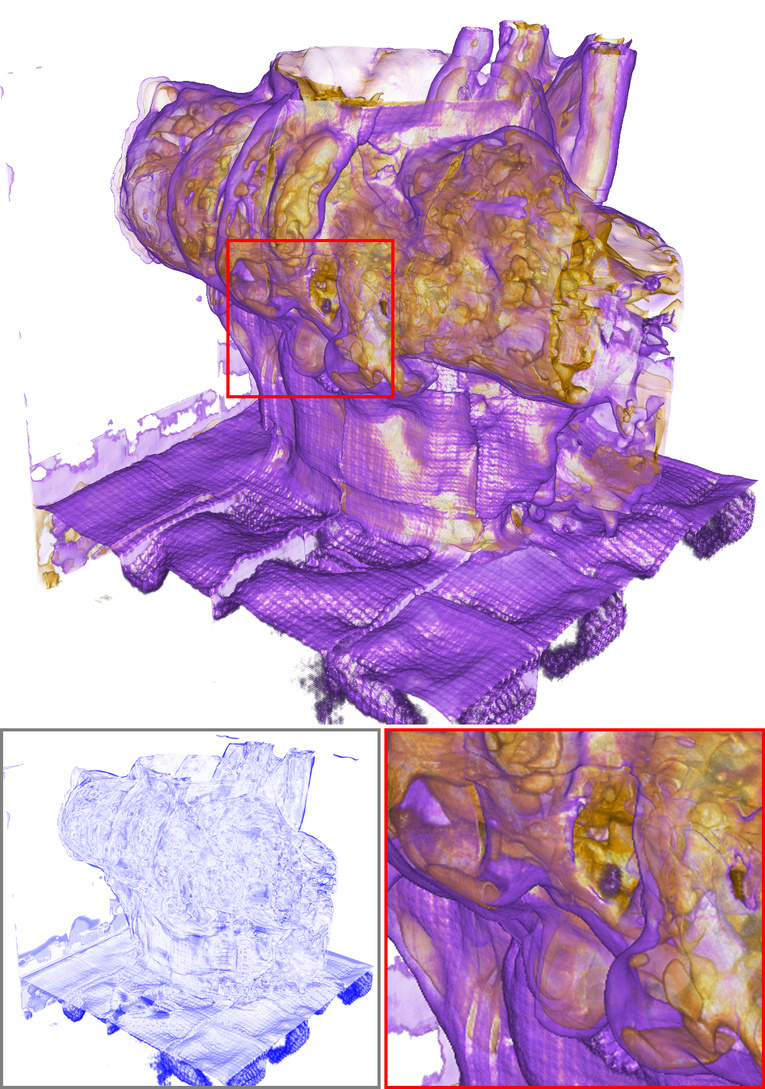}}}\\
 \vcenter{\hbox{\rotatebox{90}{\small \mbox{$t$=219}}}} &
 \vcenter{\hbox{\includegraphics[width=0.215\linewidth]{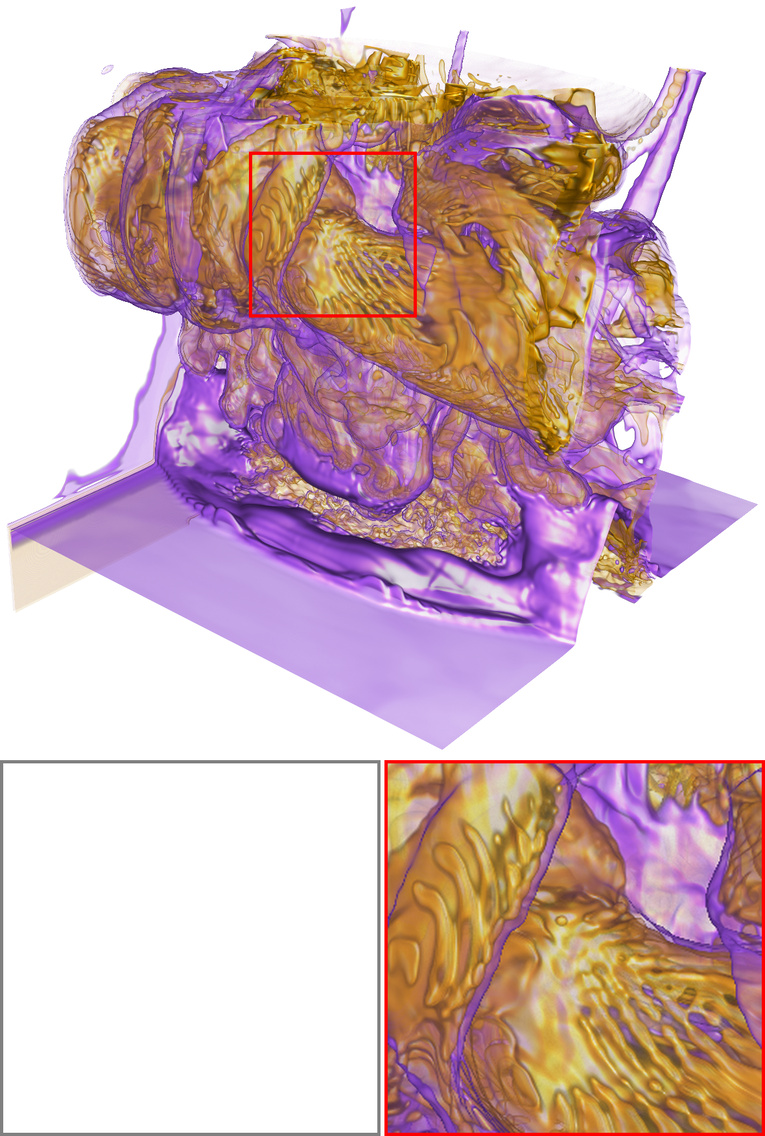}}}&
 \vcenter{\hbox{\includegraphics[width=0.215\linewidth]{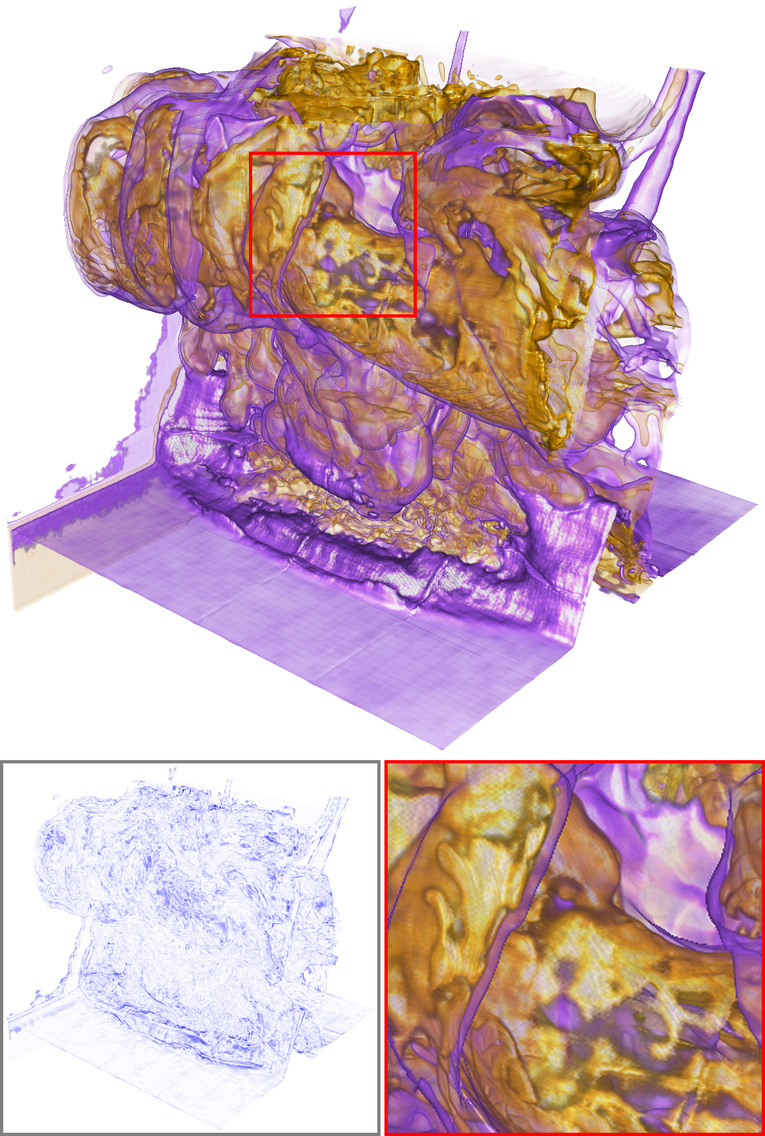}}}&
 \vcenter{\hbox{\includegraphics[width=0.215\linewidth]{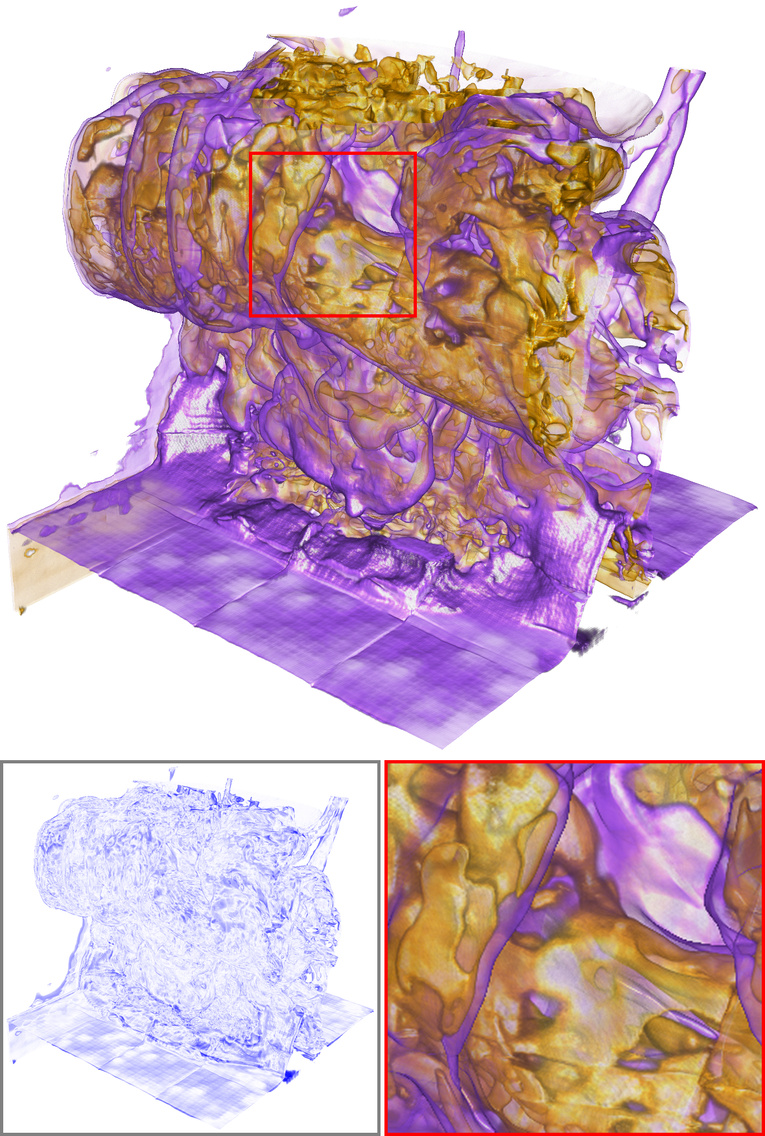}}}&
 \vcenter{\hbox{\includegraphics[width=0.215\linewidth]{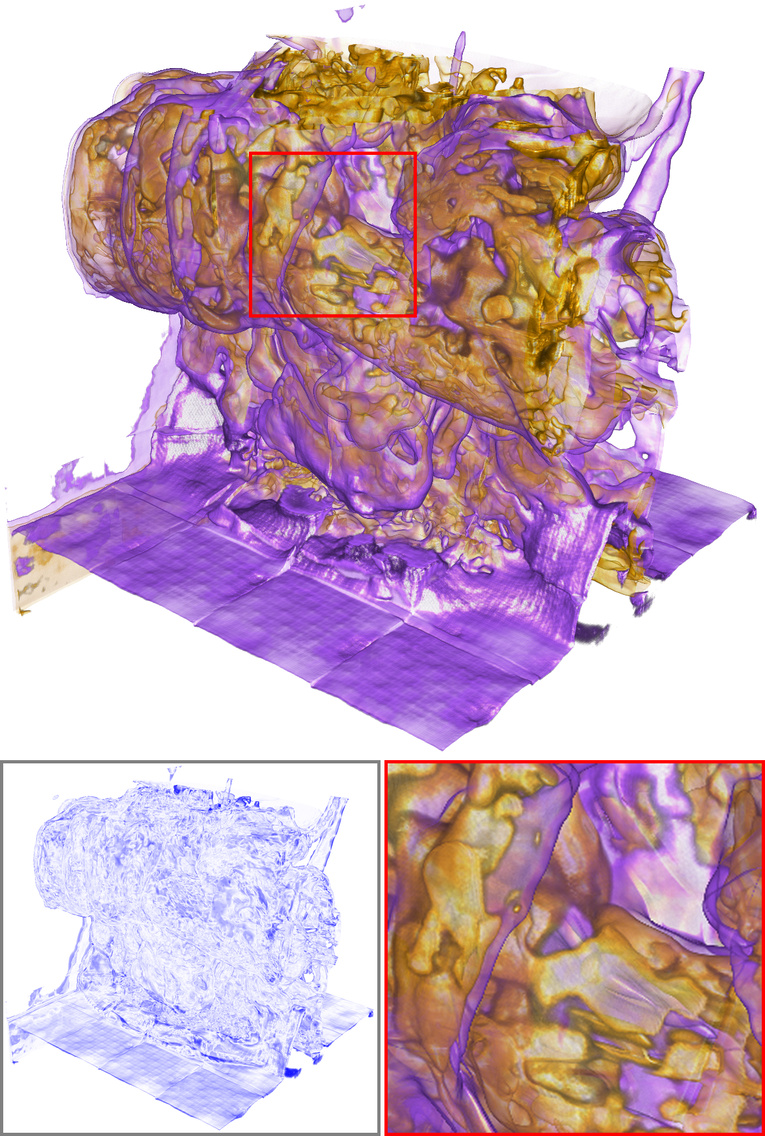}}}\\
 &
 \mbox{\small GT} &
 \mbox{\small high (2,706$\times$)} &
 \mbox{\small medium (4,924$\times$)} &
 \mbox{\small low (12,486$\times$)} \\ 
 \vcenter{\hbox{\rotatebox{90}{\mbox{\small $\Omega_M$=0.126}}}} &
 \vcenter{\hbox{\includegraphics[width=0.215\linewidth]{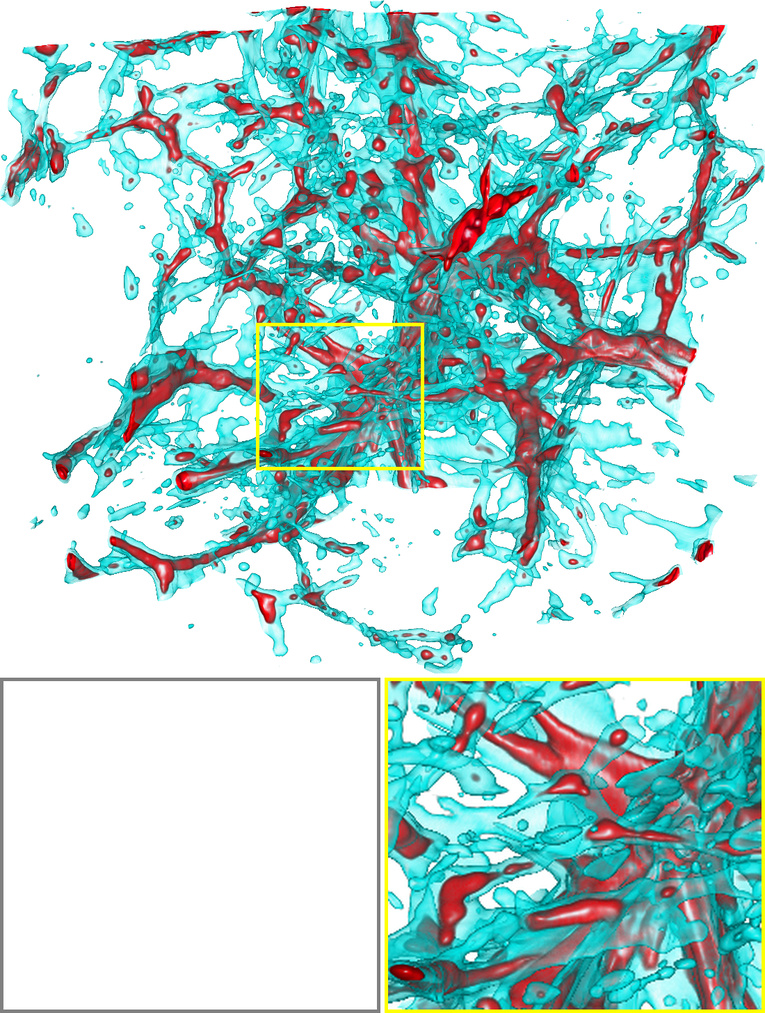}}}&
 \vcenter{\hbox{\includegraphics[width=0.215\linewidth]{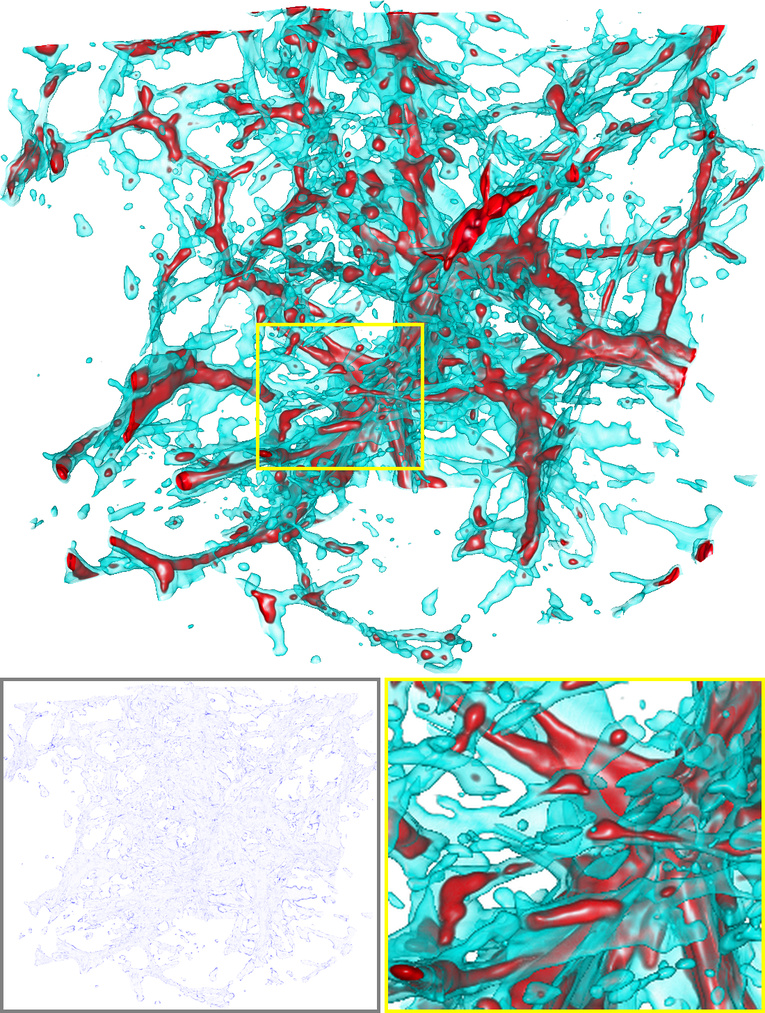}}}&
 \vcenter{\hbox{\includegraphics[width=0.215\linewidth]{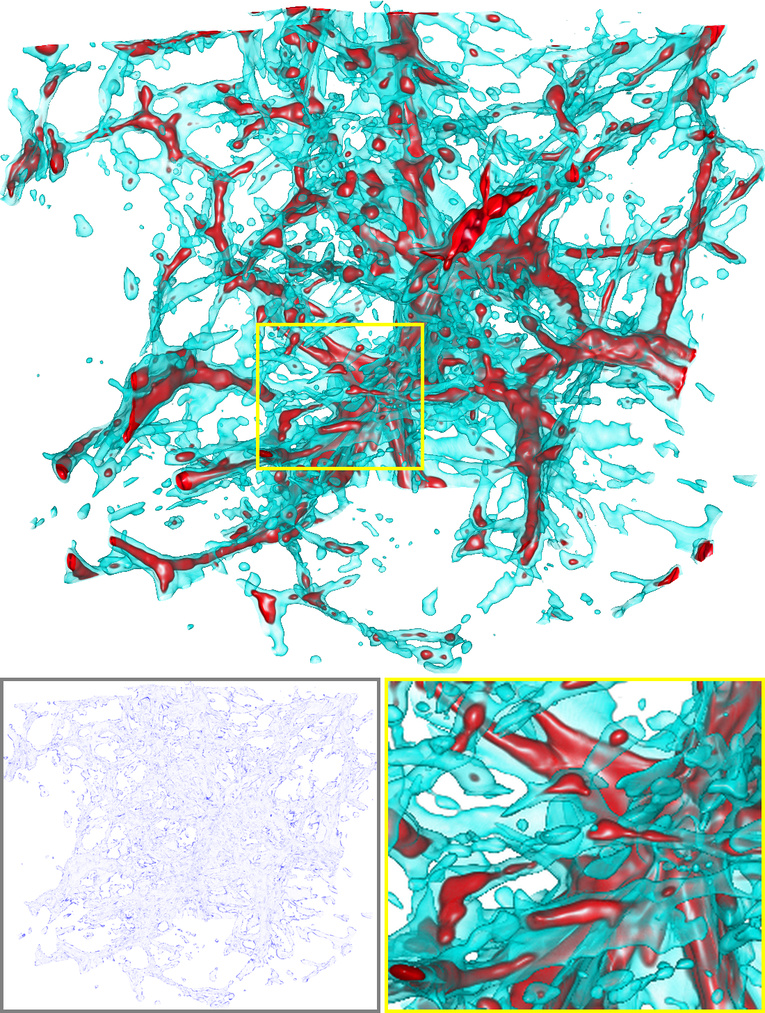}}}&
 \vcenter{\hbox{\includegraphics[width=0.215\linewidth]{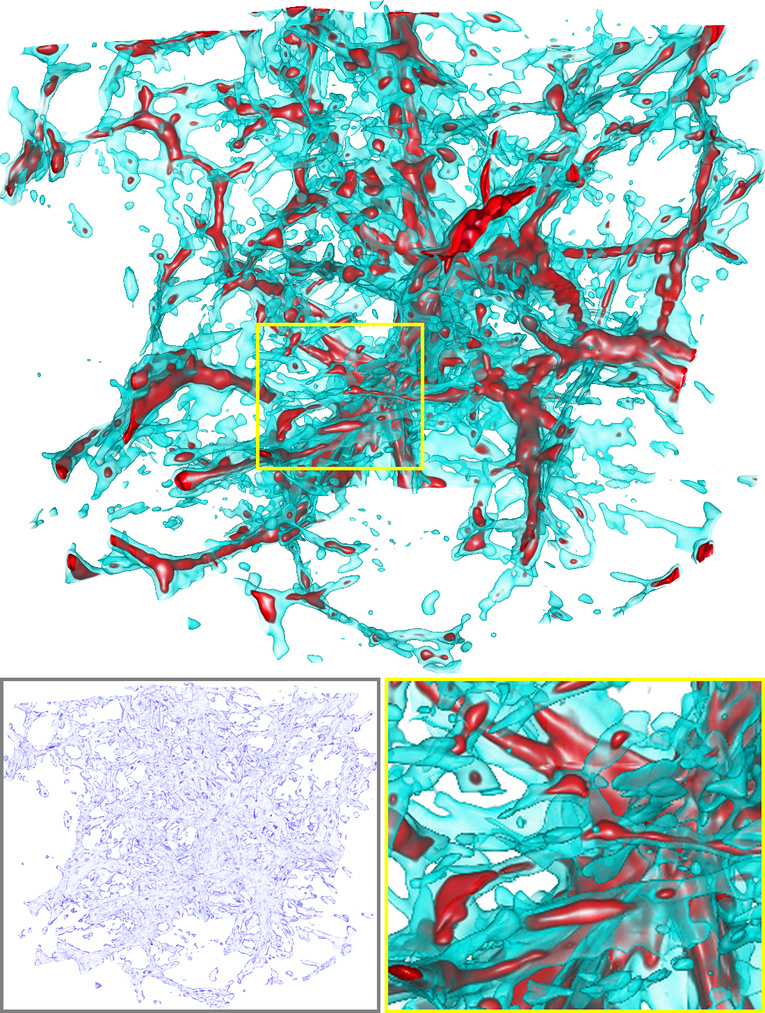}}}\\
 \vcenter{\hbox{\rotatebox{90}{\mbox{\small $\Omega_M$=0.155}}}} &
 \vcenter{\hbox{\includegraphics[width=0.215\linewidth]{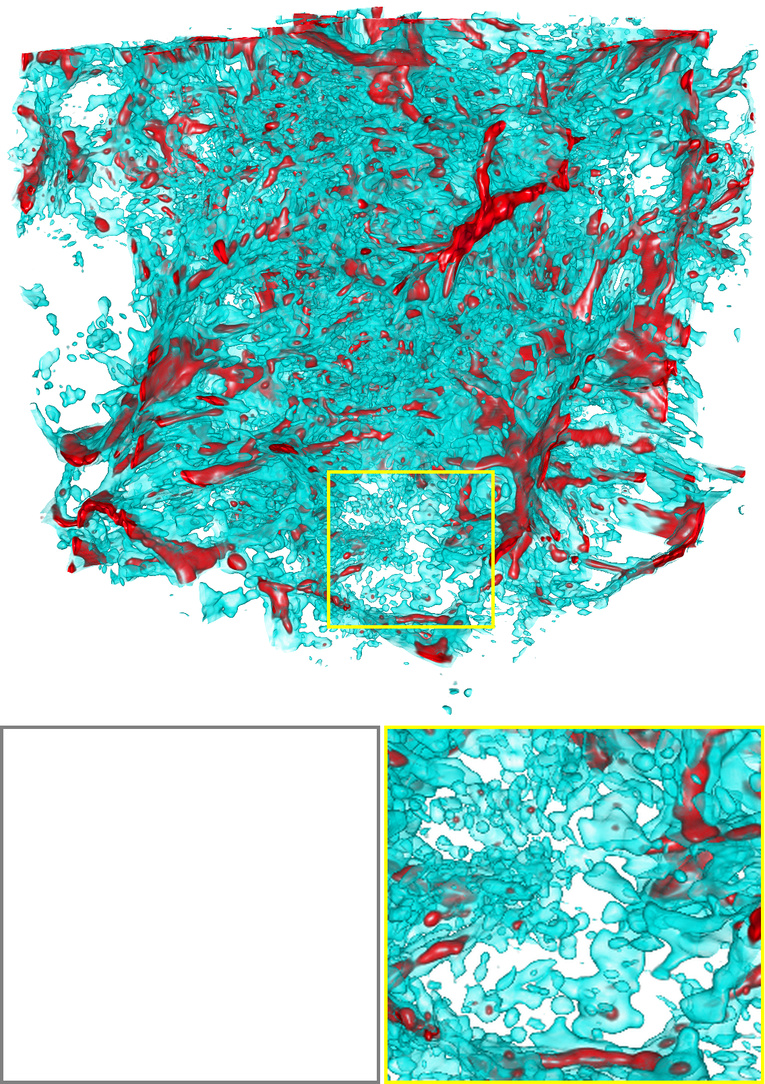}}}&
 \vcenter{\hbox{\includegraphics[width=0.215\linewidth]{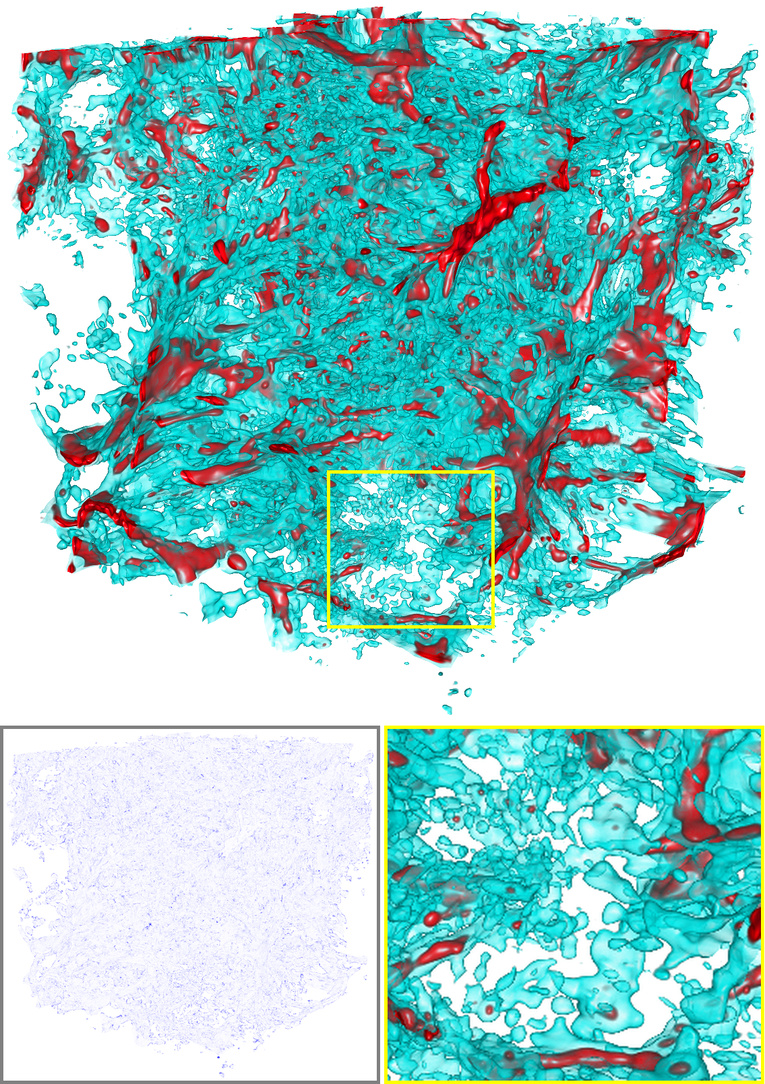}}}&
 \vcenter{\hbox{\includegraphics[width=0.215\linewidth]{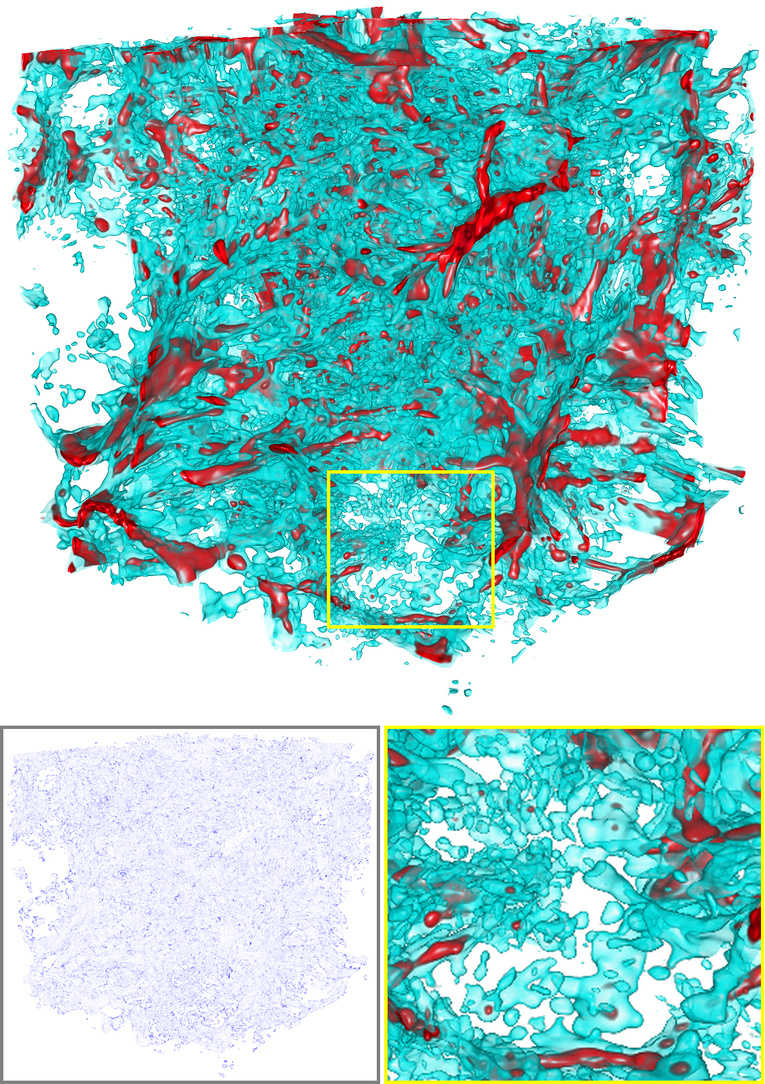}}}&
 \vcenter{\hbox{\includegraphics[width=0.215\linewidth]{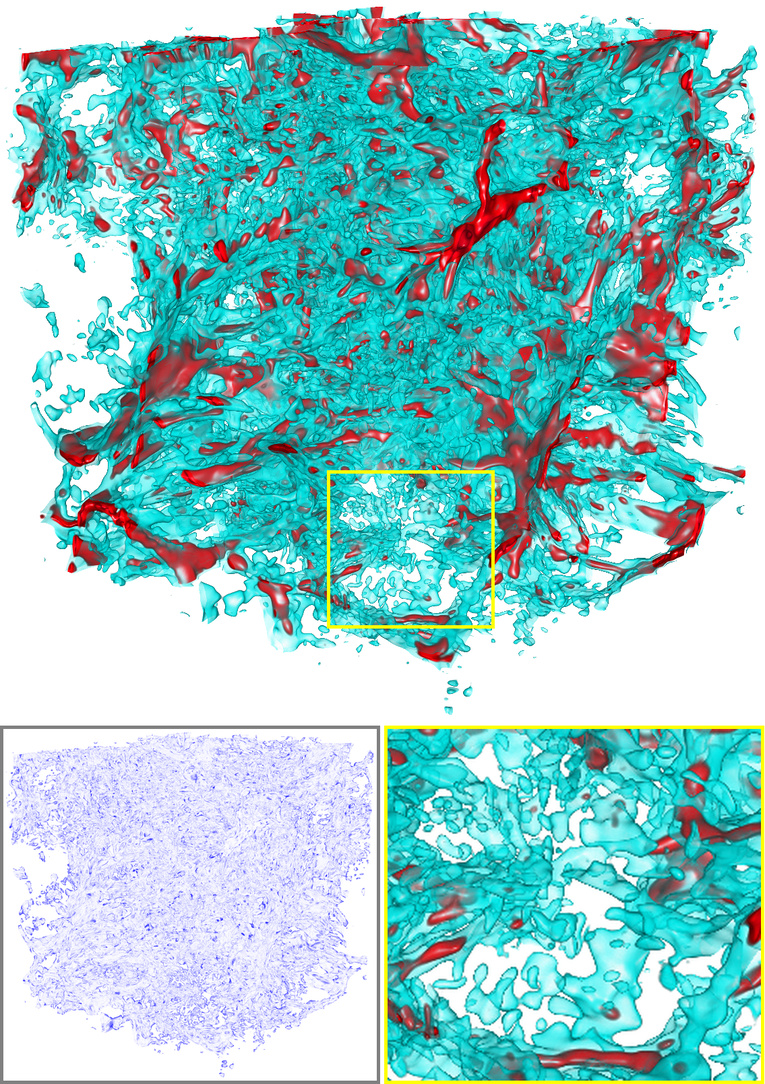}}}\\
 &
 \mbox{\small GT} &
 \mbox{\small high (304$\times$)} &
 \mbox{\small medium (661$\times$)} &
 \mbox{\small low (2,093$\times$)} \\
 \end{array}$
\vspace{-.125in}
\caption{Volume rendering of EVOLVE decompression results on \textsf{asteroids-T} (top) and \textsf{Nyx-E} (bottom), showing selected timesteps and ensemble members at different quality settings.}
\label{fig:time-ensemble}
\end{figure*}

EVOLVE compresses time-varying and ensemble datasets by independently compressing each volume. Table~\ref{tab:time-ensemble} reports the average quantitative results of our method on the \textsf{asteroids-T} and \textsf{Nyx-E} datasets under three quality settings (high, medium, and low), which correspond to decreasing gain factors that trade reconstruction quality for higher CRs. \textsf{asteroids-T} is a time-varying dataset containing 220 timesteps; we select two timesteps ($t$=165 and $t$=219) for visualization. \textsf{Nyx-E} is an ensemble dataset with varying cosmological simulation parameters; we select two members ($\Omega_M$=0.126 and $\Omega_M$=0.155) for visualization. The volume rendering results are shown in Figure~\ref{fig:time-ensemble}.
We observe that our method faithfully reconstructs volumes across all three quality levels. Future work will focus on incorporating additional temporal or ensemble information into the context model to achieve higher CRs.

\begin{table}[!ht]                                                
\centering             
\caption{Average PSNR (dB), LPIPS, ET (sec), DT (sec), and CR per timestep or ensemble using EVOLVE with different quality settings.}
\label{tab:time-ensemble}
\vspace{-0.05in}
\resizebox{0.8\columnwidth}{!}{%
\begin{tabular}{c c c c c c c}
dataset & quality & PSNR$\uparrow$ & LPIPS$\downarrow$ & ET$\downarrow$ & DT$\downarrow$ & CR$\uparrow$ \\
\hline
\multirow{3}{*}{\textsf{asteroids-T}}
& high  & 59.06 & 0.299 & 7.97 & 6.27 & 2,706 \\
& medium   & 57.51 & 0.320 & 7.12 & 8.15 & 4,924 \\
& low   & 50.27 & 0.415 & 6.43 & 6.29 & 12,486 \\
\hline

\multirow{3}{*}{\textsf{Nyx-E}}
& high  & 50.60 & 0.024 & 6.43 & 6.43 & 304 \\
& medium   & 46.56 & 0.043 & 6.09 & 8.39 & 661 \\
& low   & 39.19 & 0.104 & 6.72 & 8.39 & 2,093 \\
\end{tabular}
}
\end{table}

\vspace{-0.05in}
\section{Additional Rendering Results}

\begin{figure*}[htb]
\centering
$\begin{array}{c@{\hspace{0.05in}}c@{\hspace{0.05in}}c@{\hspace{0.05in}}c@{\hspace{0.05in}}c}
\mbox{\small CR / PSNR} &
 \mbox{\small 9,803$\times$ / 47.16 dB} &
 \mbox{\small 2,482$\times$ / 40.85 dB} &
 \mbox{\small 2,401$\times$ / 43.44 dB} &
 \mbox{\small \hot{1,164$\times$ / 40.00 dB}} \\
 \includegraphics[width=0.19\linewidth]{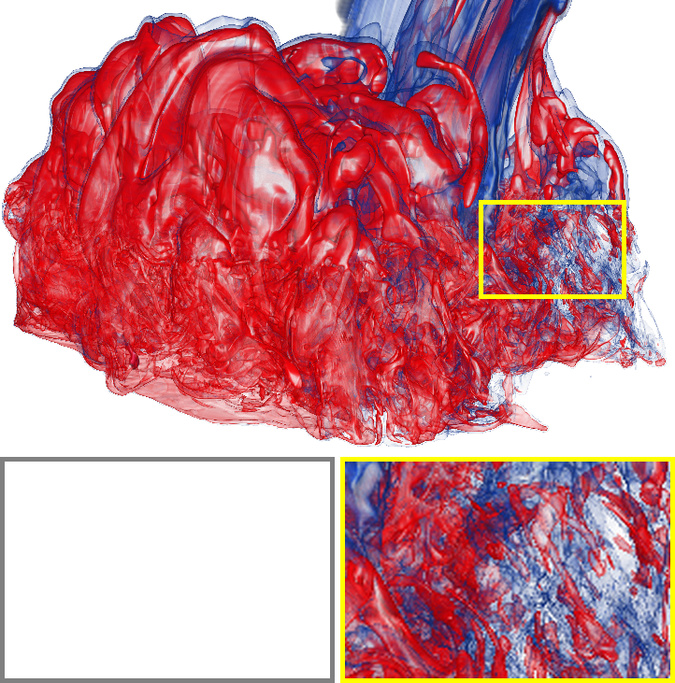}&
 \includegraphics[width=0.19\linewidth]{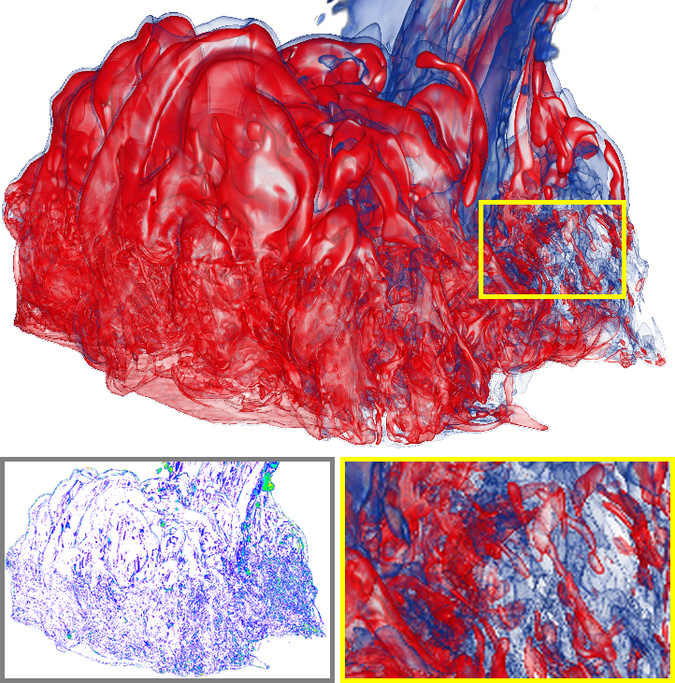}&
 \includegraphics[width=0.19\linewidth]{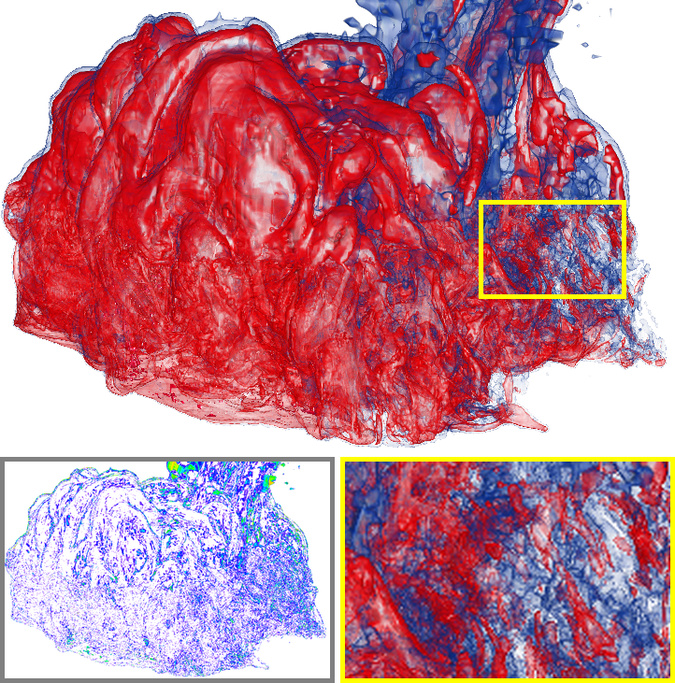}&
 \includegraphics[width=0.19\linewidth]{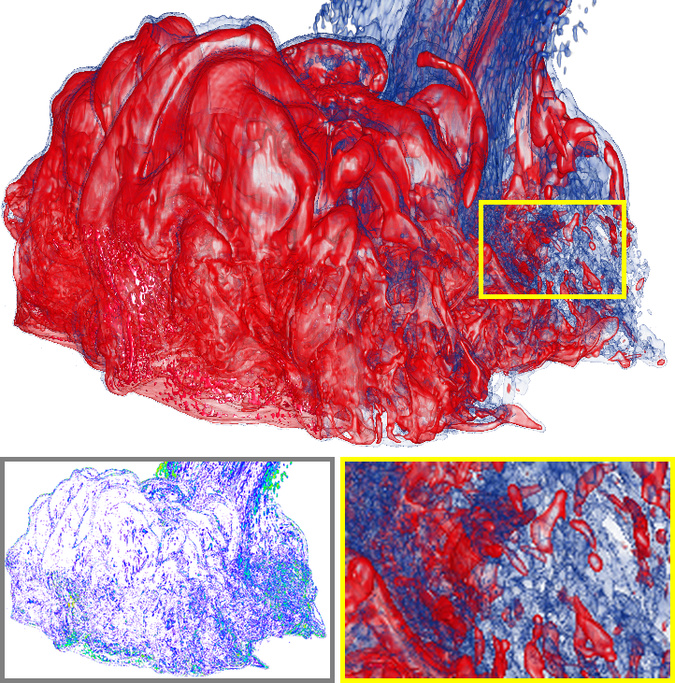}&
 \includegraphics[width=0.19\linewidth]{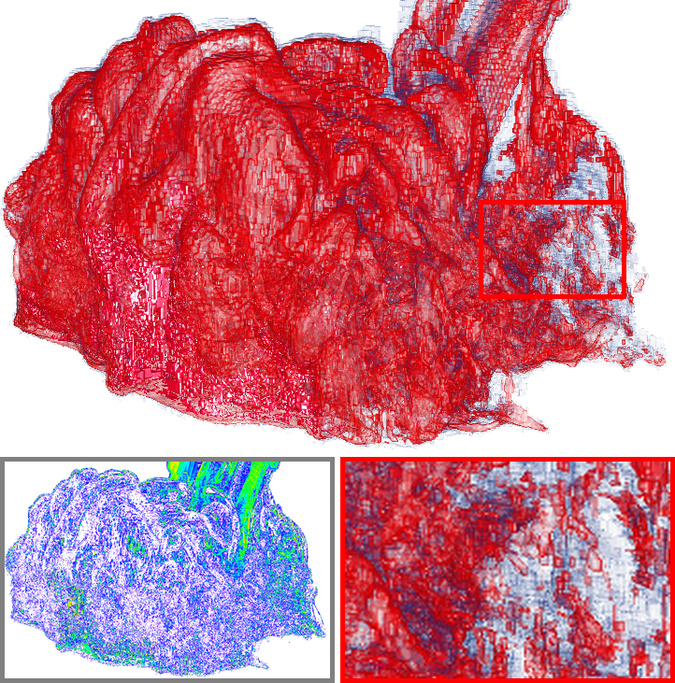}\\
\mbox{\small CR / PSNR} &
 \mbox{\small 2,846$\times$ / 45.18 dB} &
 \mbox{\small 985$\times$ / 40.03 dB} &
 \mbox{\small 870$\times$ / 43.51 dB} &
 \mbox{\small \hot{64$\times$ / 43.79 dB}} \\
 \includegraphics[width=0.19\linewidth]{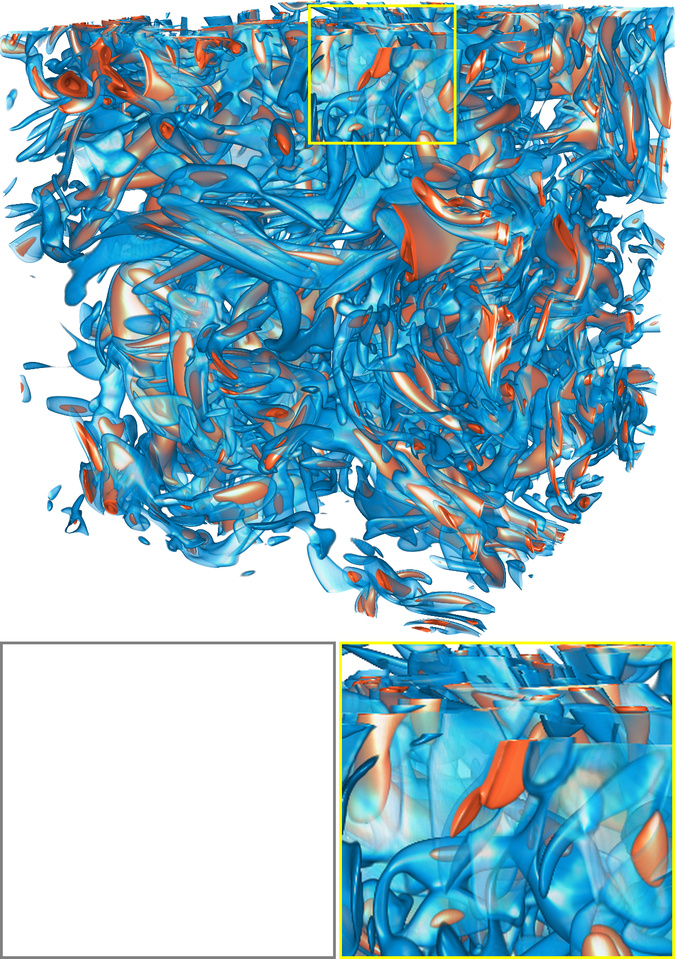}&
 \includegraphics[width=0.19\linewidth]{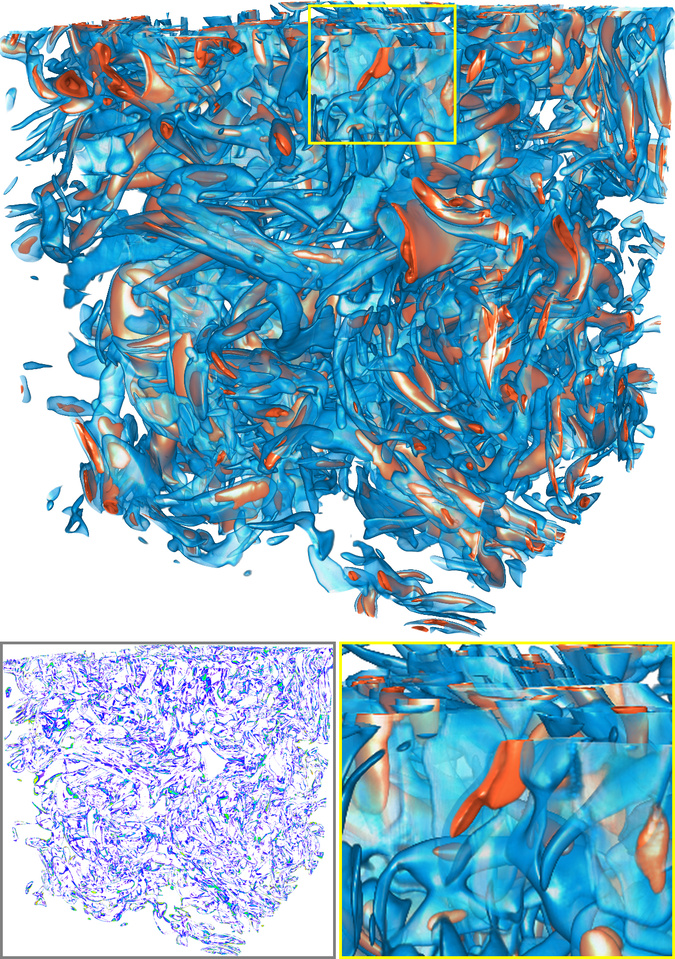}&
 \includegraphics[width=0.19\linewidth]{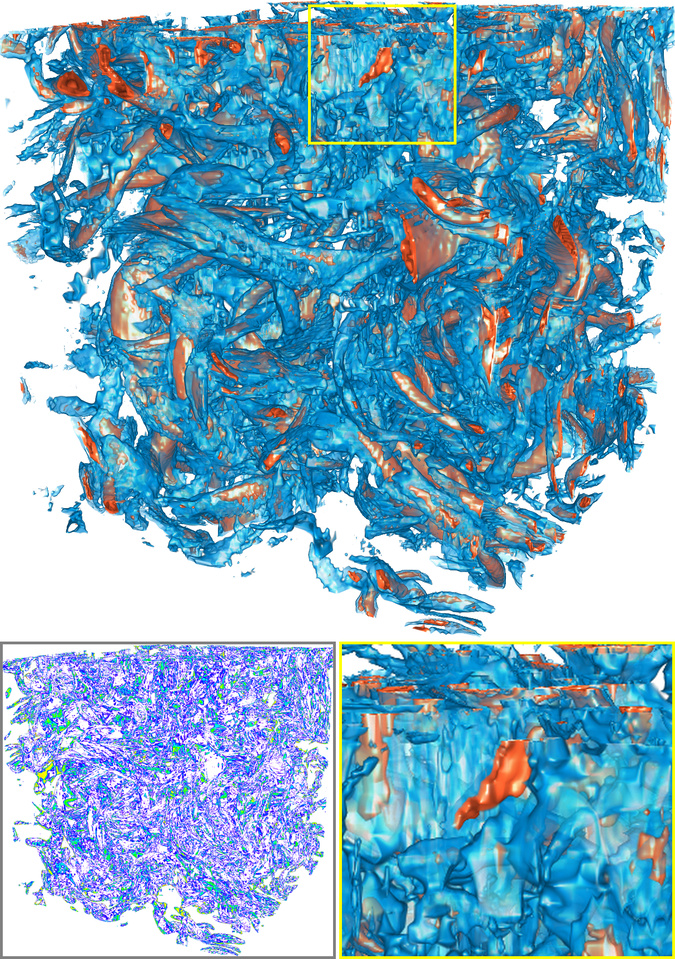}&
 \includegraphics[width=0.19\linewidth]{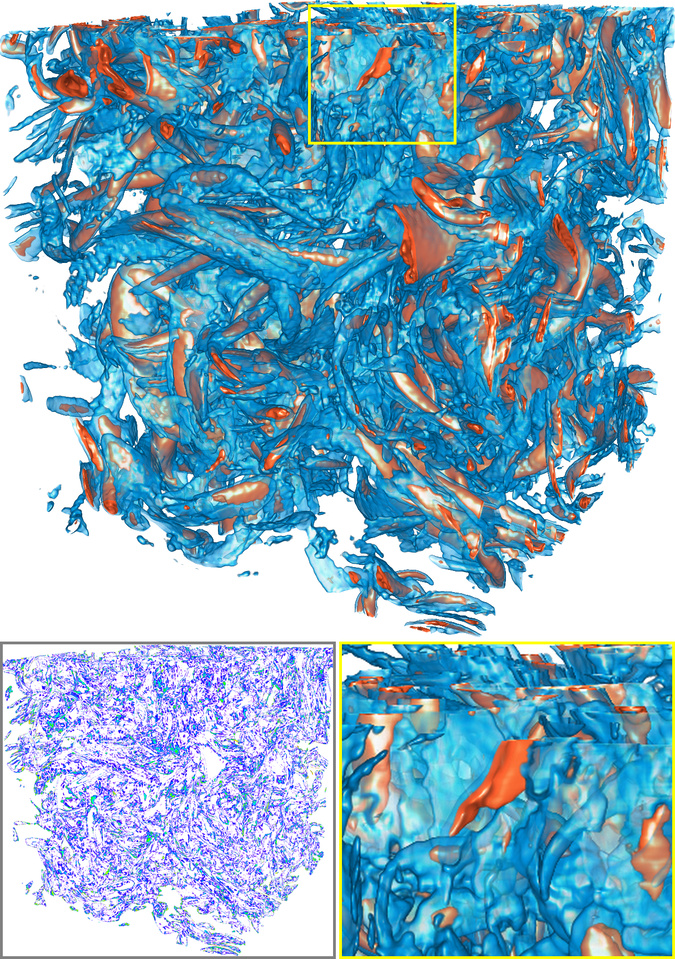}&
 \includegraphics[width=0.19\linewidth]{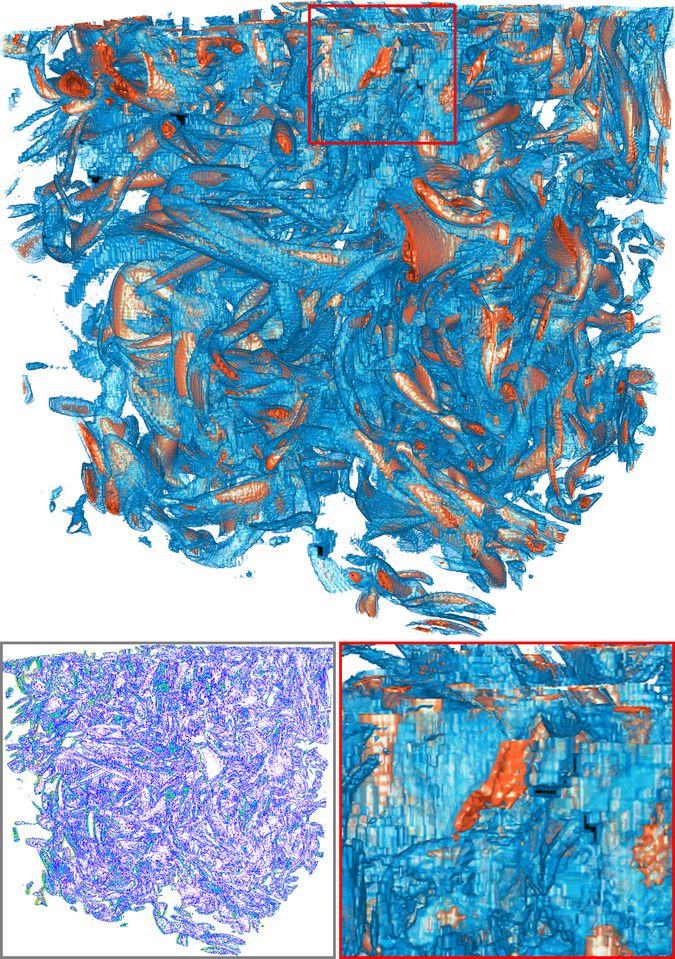}\\
 \mbox{\small GT} &
 \mbox{\small EVOLVE} &
 \mbox{\small SZ3} &
 \mbox{\small TTHRESH} &
 \mbox{\small ZFP}\\
\end{array}$
\vspace{-.125in}
\caption{Comparison of volume rendering results between EVOLVE and conventional lossy compressors. Top and bottom: \textsf{asteroids} and \textsf{isotropic}. 
}
\label{fig:traditional-vol-appx}
\end{figure*}

\begin{figure*}[htb]
\centering
$\begin{array}{c@{\hspace{0.01in}}c@{\hspace{0.01in}}c@{\hspace{0.01in}}c@{\hspace{0.01in}}c@{\hspace{0.01in}}c@{\hspace{0.01in}}c@{\hspace{0.01in}}c@{\hspace{0.01in}}c}
\mbox{\tiny CR / PSNR} &
 \mbox{\tiny 10,517$\times$ / 45.62 dB} &
 \mbox{\tiny 1,714$\times$ / 43.35 dB} &
 \mbox{\tiny 5,954$\times$ / 41.56 dB} &
 \mbox{\tiny 1,979$\times$ / 41.37 dB} &
 \mbox{\tiny 1,575$\times$ / 40.78 dB} &
 \mbox{\tiny 936$\times$ / 40.24 dB} &
 \mbox{\tiny 1,839$\times$ / 42.22 dB} &
 \mbox{\tiny 1,808$\times$ / 41.28 dB} \\
 \includegraphics[width=0.1075\linewidth]{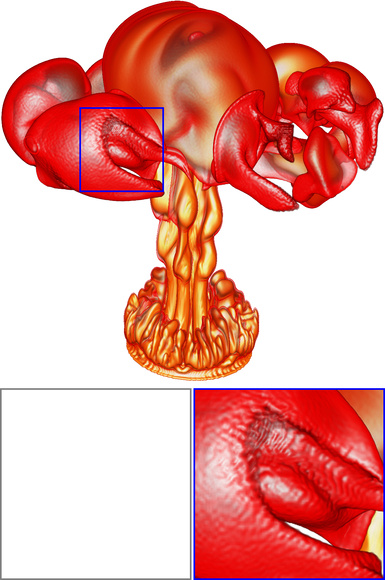}&
 \includegraphics[width=0.1075\linewidth]{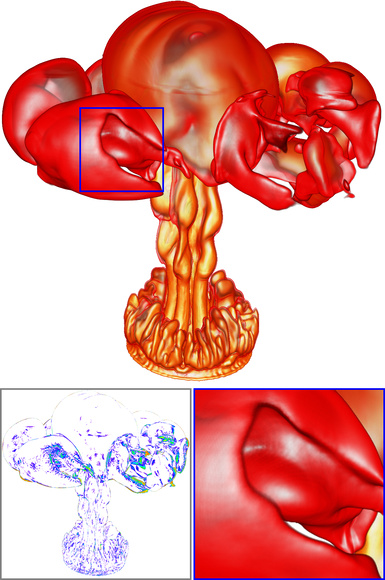}&
 \includegraphics[width=0.1075\linewidth]{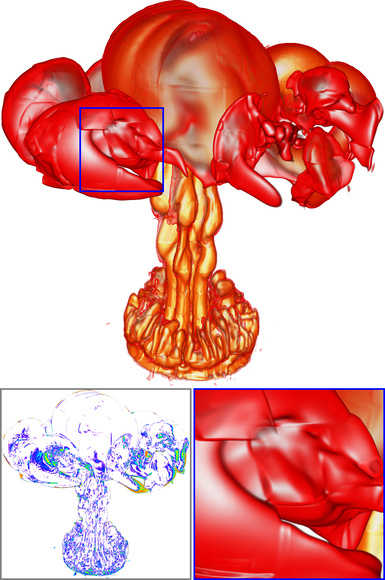}&
 \includegraphics[width=0.1075\linewidth]{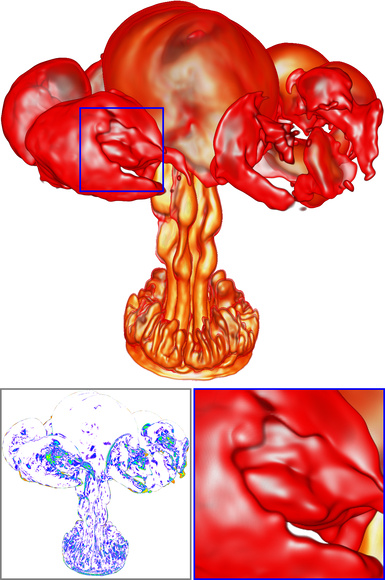}&
 \includegraphics[width=0.1075\linewidth]{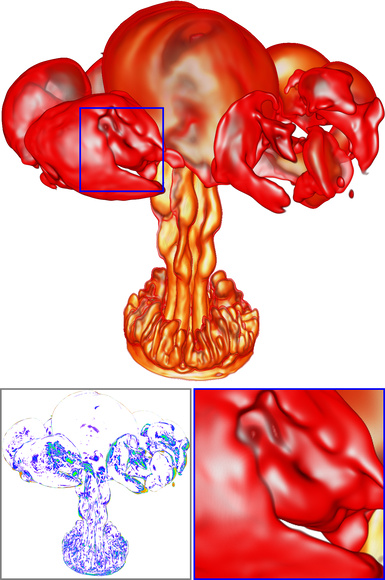}&
 \includegraphics[width=0.1075\linewidth]{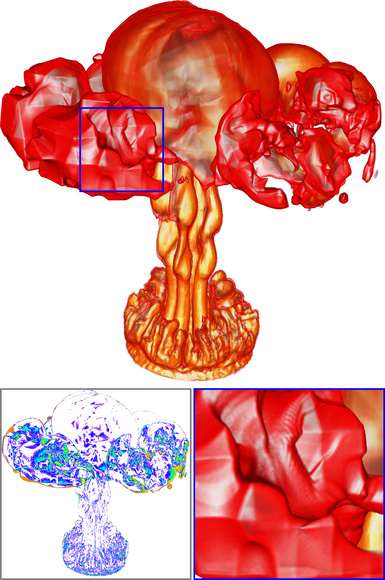}&
 \includegraphics[width=0.1075\linewidth]{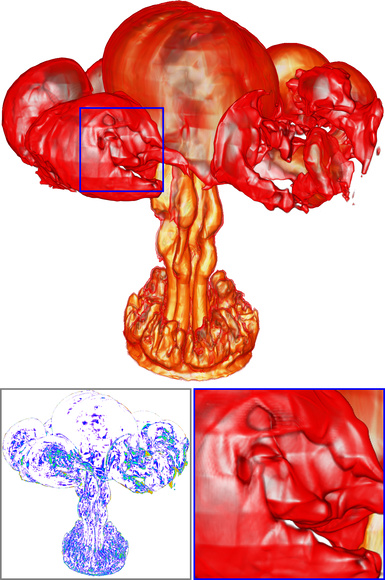}&
 \includegraphics[width=0.1075\linewidth]{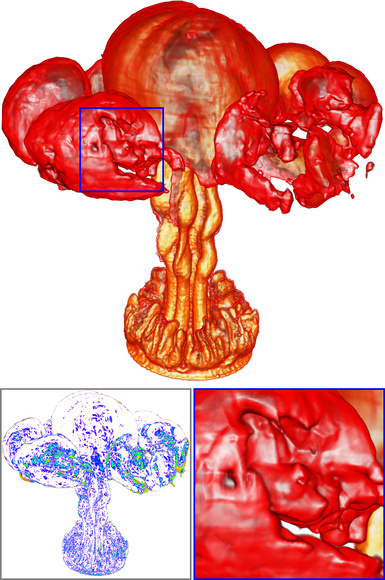}&
 \includegraphics[width=0.1075\linewidth]{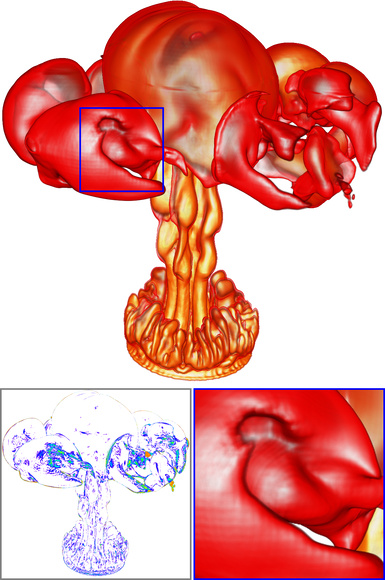}\\
\mbox{\tiny CR / PSNR} &
 \mbox{\tiny 2,596$\times$ / 44.91 dB} &
 \mbox{\tiny 446$\times$ / 41.50 dB} &
 \mbox{\tiny 1,044$\times$ / 42.29 dB} &
 \mbox{\tiny 633$\times$ / 40.44 dB} &
 \mbox{\tiny 371$\times$ / 40.90 dB} &
 \mbox{\tiny 166$\times$ / 42.23 dB} &
 \mbox{\tiny 168$\times$ / 41.35 dB} &
 \mbox{\tiny 1,114$\times$ / 42.17 dB} \\
 \includegraphics[width=0.1075\linewidth]{figures/vol_half-cylinder_gt.jpg}&
 \includegraphics[width=0.1075\linewidth]{figures/vol_half-cylinder_evolve.jpg}&
 \includegraphics[width=0.1075\linewidth]{figures/vol_half-cylinder_ecnr.jpg}&
 \includegraphics[width=0.1075\linewidth]{figures/vol_half-cylinder_neurcomp.jpg}&
 \includegraphics[width=0.1075\linewidth]{figures/vol_half-cylinder_siren.jpg}&
 \includegraphics[width=0.1075\linewidth]{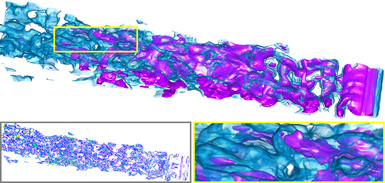}&
 \includegraphics[width=0.1075\linewidth]{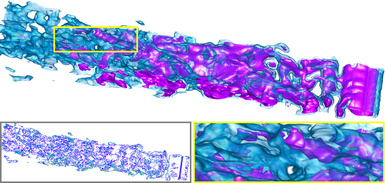}&
 \includegraphics[width=0.1075\linewidth]{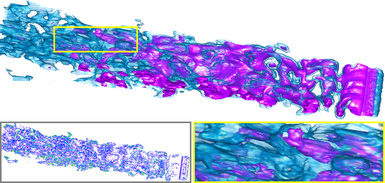}&
 \includegraphics[width=0.1075\linewidth]{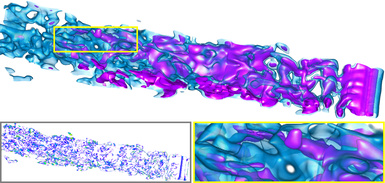}\\
\mbox{\tiny CR / PSNR} &
 \mbox{\tiny 10,774$\times$ / 45.20 dB} &
 \mbox{\tiny 3,047$\times$ / 43.29 dB} &
 \mbox{\tiny 6,649$\times$ / 43.48 dB} &
 \mbox{\tiny 5,256$\times$ / 42.09 dB} &
 \mbox{\tiny 1,575$\times$ / 44.23 dB} &
 \mbox{\tiny 1,813$\times$ / 42.16 dB} &
 \mbox{\tiny 1,839$\times$ / 43.91 dB} &
 \mbox{\tiny 3,748$\times$ / 38.84 dB} \\
 \includegraphics[width=0.1075\linewidth]{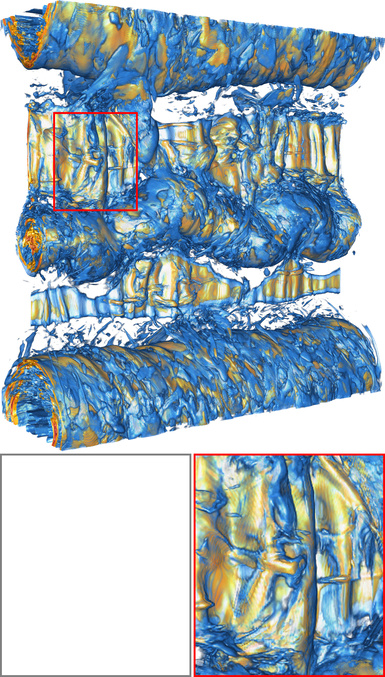}&
 \includegraphics[width=0.1075\linewidth]{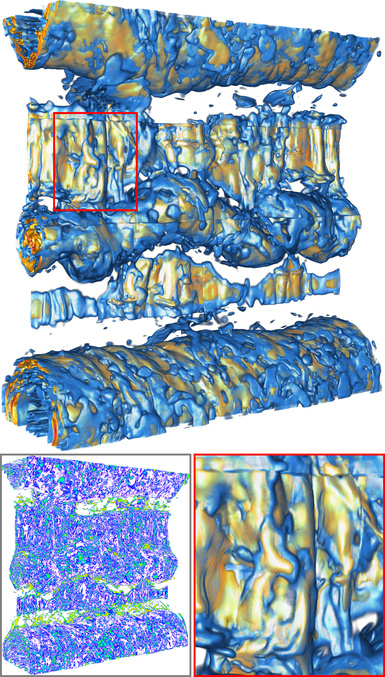}&
 \includegraphics[width=0.1075\linewidth]{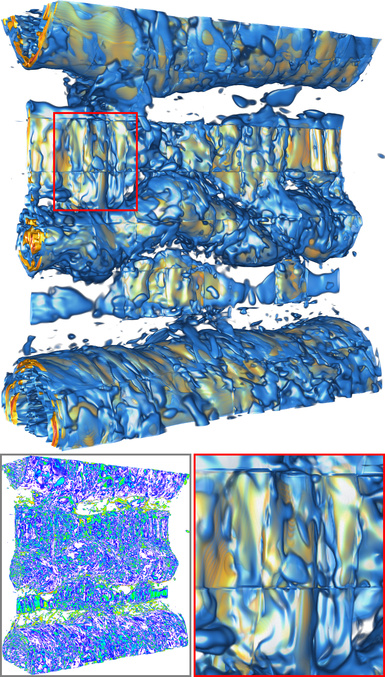}&
 \includegraphics[width=0.1075\linewidth]{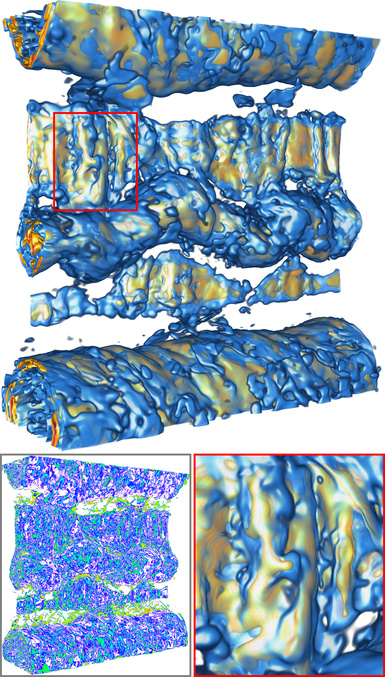}&
 \includegraphics[width=0.1075\linewidth]{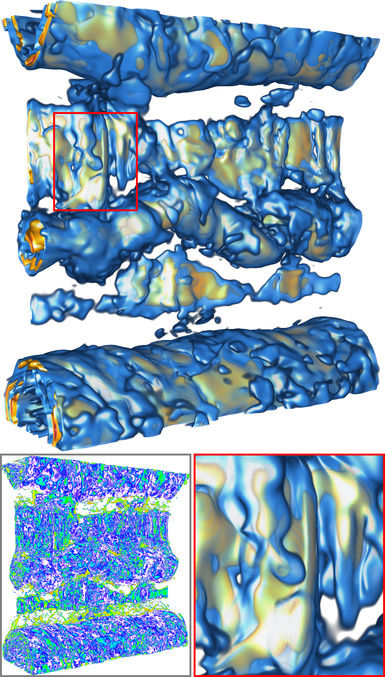}&
 \includegraphics[width=0.1075\linewidth]{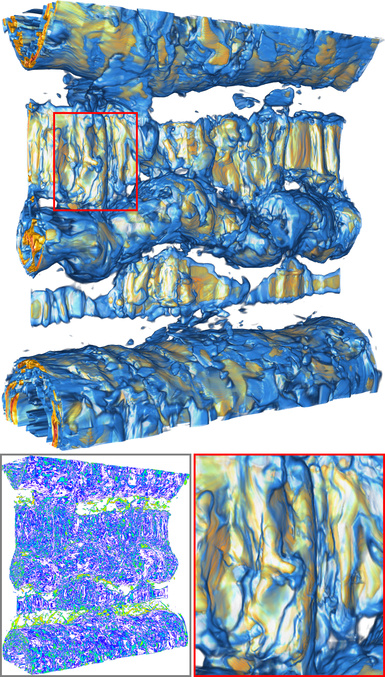}&
 \includegraphics[width=0.1075\linewidth]{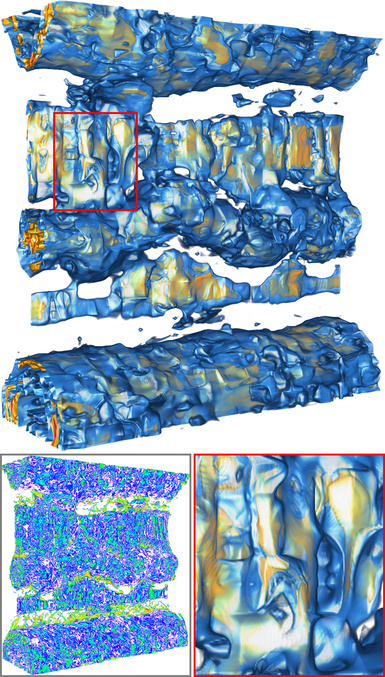}&
 \includegraphics[width=0.1075\linewidth]{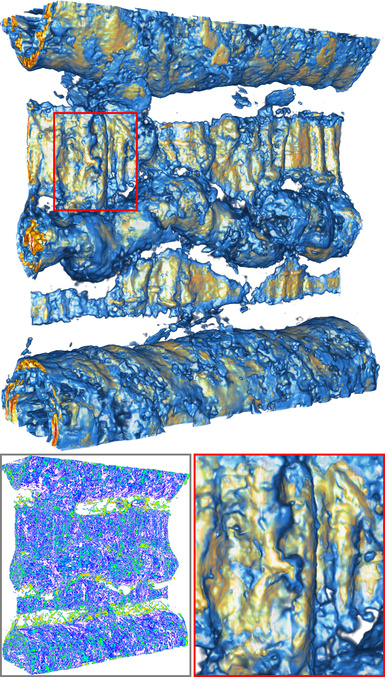}&
 \includegraphics[width=0.1075\linewidth]{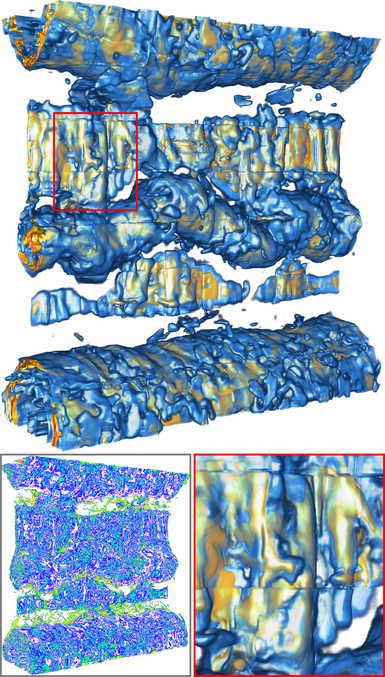}\\
 \mbox{\small GT} &
 \mbox{\small EVOLVE} &
 \mbox{\small ECNR} &
 \mbox{\small NeurComp} &
 \mbox{\small SIREN} &
 \mbox{\small AMGSRN++} &
 \mbox{\small fV-SRN} &
 \mbox{\small Instant-NGP} &
 \mbox{\small IDLat}\\
\end{array}$
\vspace{-.125in}
\caption{Comparison of volume rendering results between EVOLVE and \hot{deep-learning-based} compressors. From top to bottom: \textsf{gas}, \textsf{half-cylinder (VLM, 6,400)}, and \textsf{magnetic}. 
}
\label{fig:neural-vol-appx}
\end{figure*}

\begin{figure*}[htb]
\centering
$\begin{array}{c@{\hspace{0.05in}}c@{\hspace{0.05in}}c@{\hspace{0.05in}}c@{\hspace{0.05in}}c}
\mbox{\small CR / PSNR} &
 \mbox{\small 6,047$\times$ / 49.31 dB} &
 \mbox{\small 2,948$\times$ / 41.10 dB} &
 \mbox{\small 2,714$\times$ / 42.98 dB} &
 \mbox{\small \hot{129$\times$ / 41.92 dB}} \\
 \includegraphics[width=0.19\linewidth]{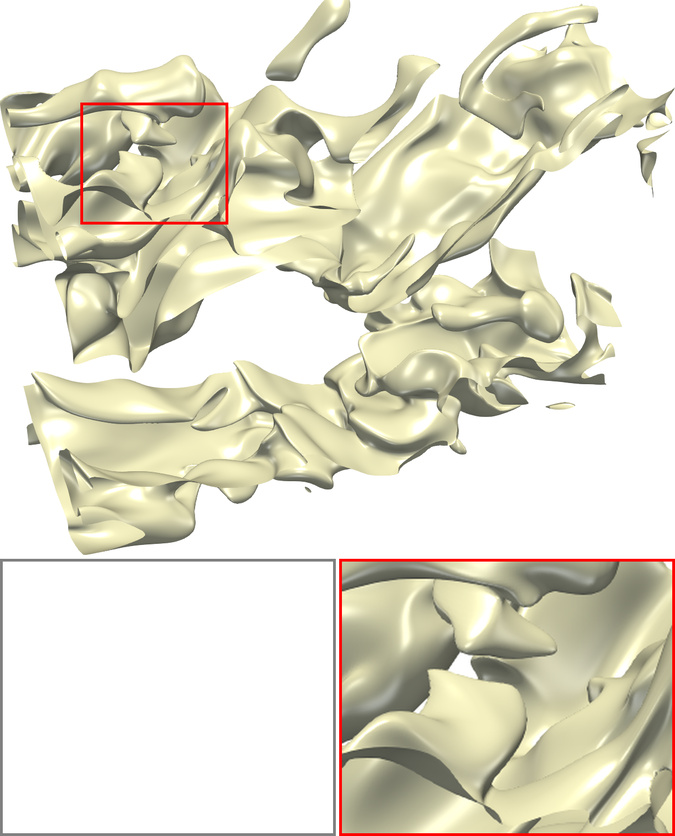}&
 \includegraphics[width=0.19\linewidth]{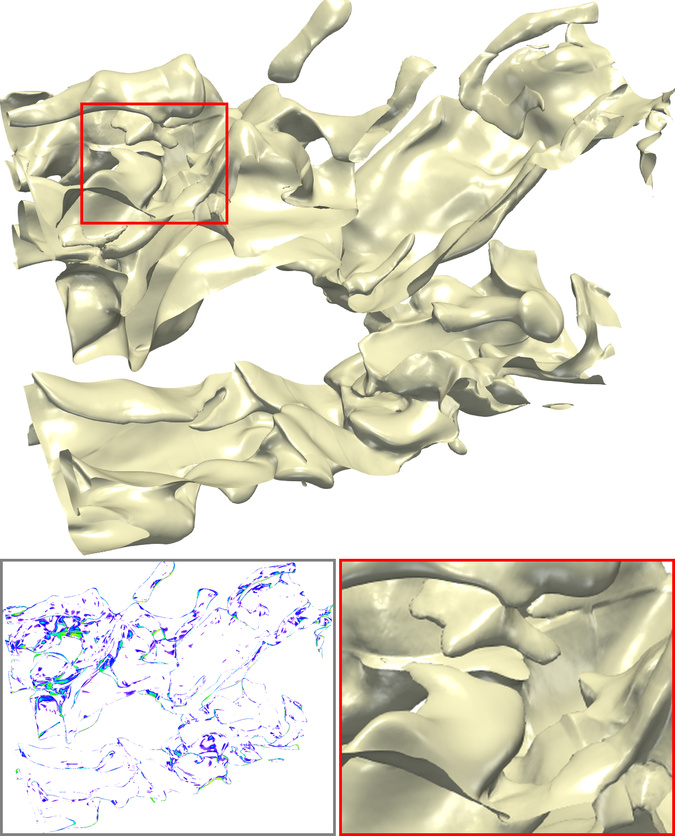}&
 \includegraphics[width=0.19\linewidth]{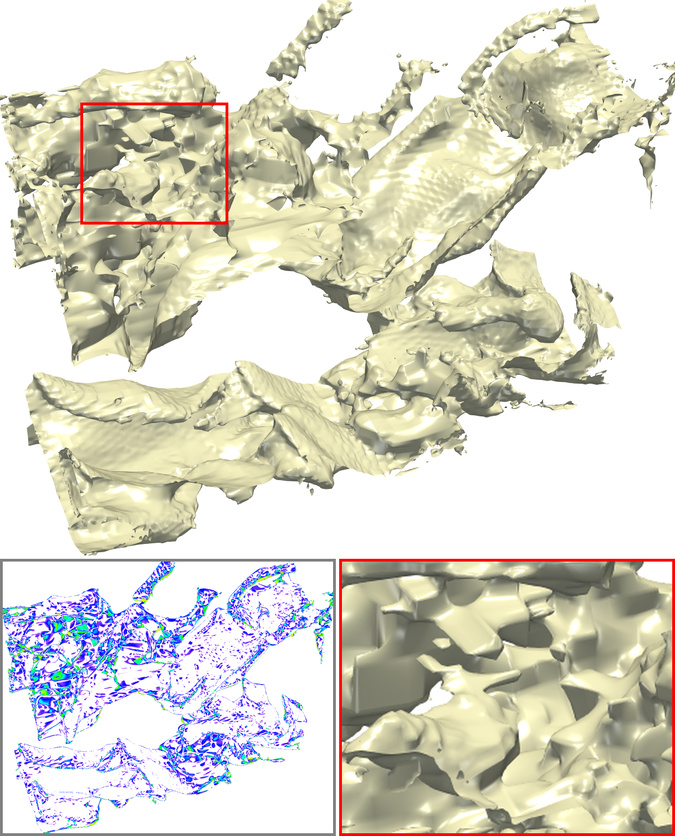}&
 \includegraphics[width=0.19\linewidth]{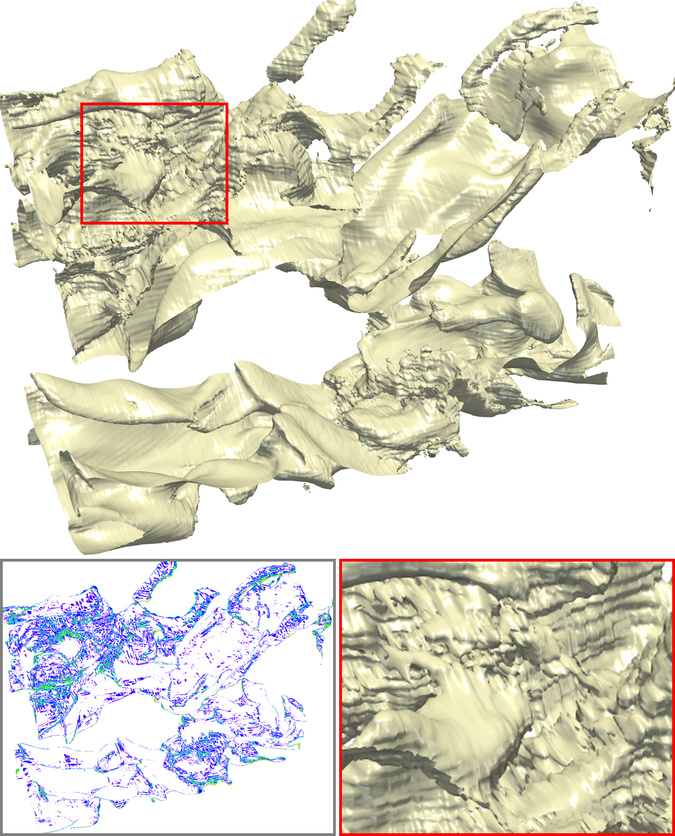}&
 \includegraphics[width=0.19\linewidth]{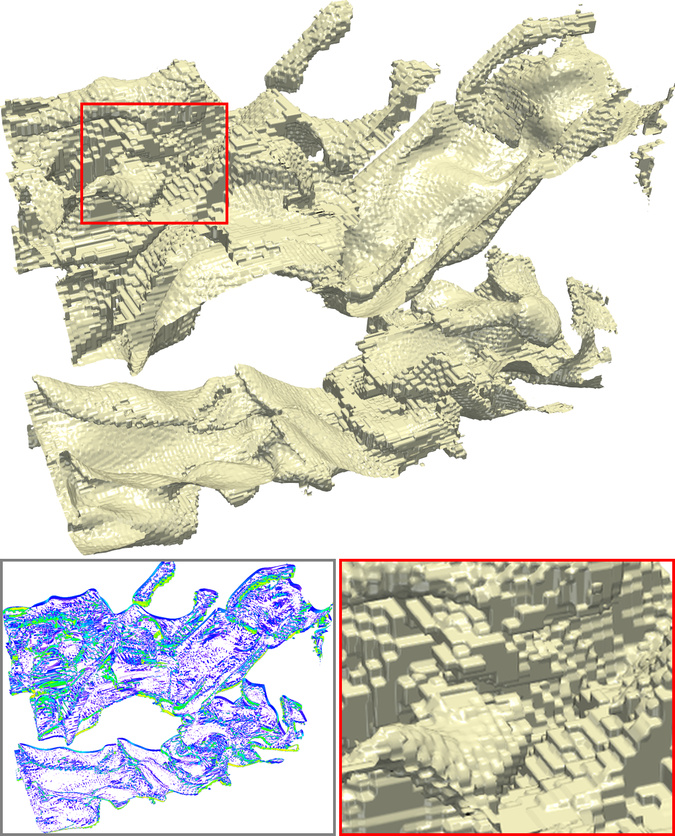}\\
\mbox{\small CR / PSNR} &
 \mbox{\small 7,843$\times$ / 47.58 dB} &
 \mbox{\small 1,051$\times$ / 41.43 dB} &
 \mbox{\small 6,334$\times$ / 40.11 dB} &
 \mbox{\small \hot{107$\times$ / 41.44 dB}} \\
 \includegraphics[width=0.19\linewidth]{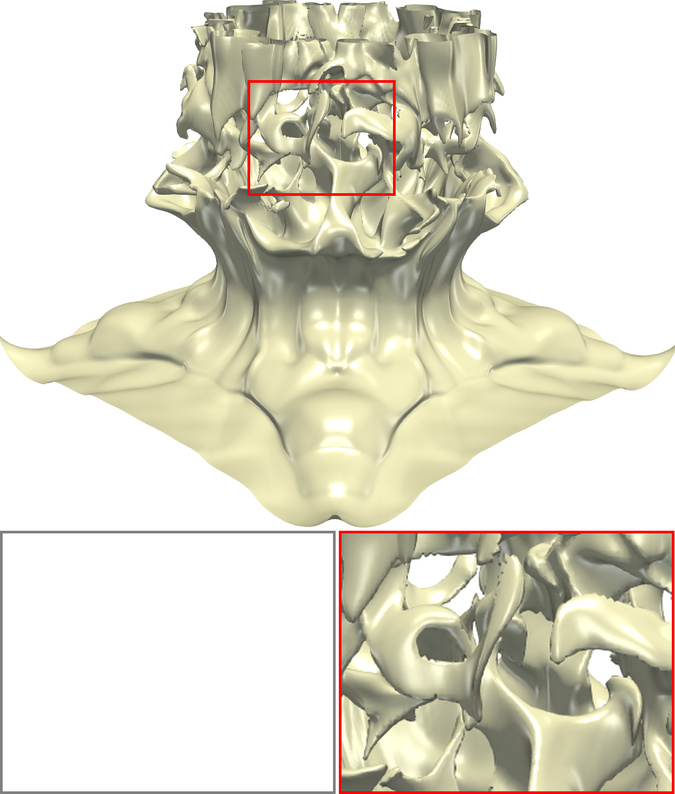}&
 \includegraphics[width=0.19\linewidth]{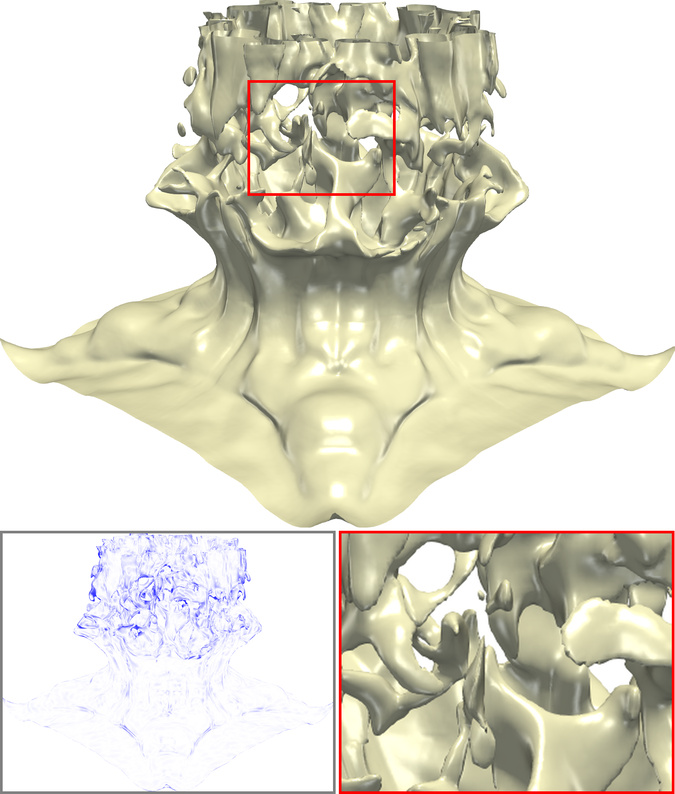}&
 \includegraphics[width=0.19\linewidth]{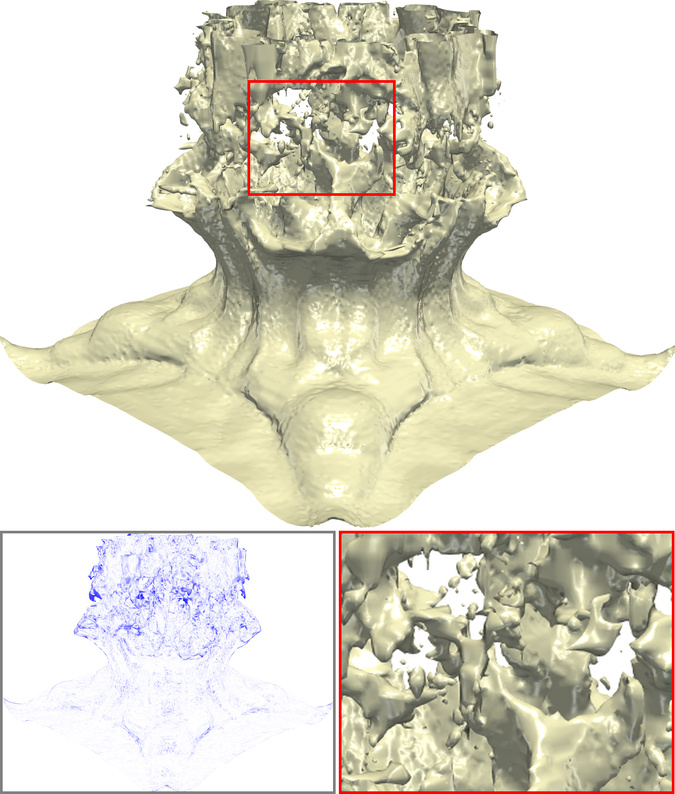}&
 \includegraphics[width=0.19\linewidth]{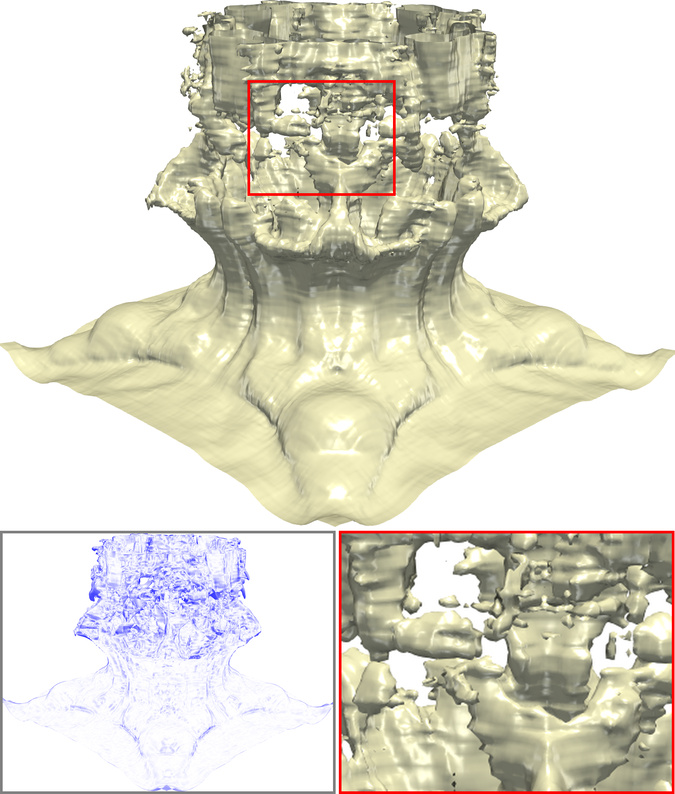}&
 \includegraphics[width=0.19\linewidth]{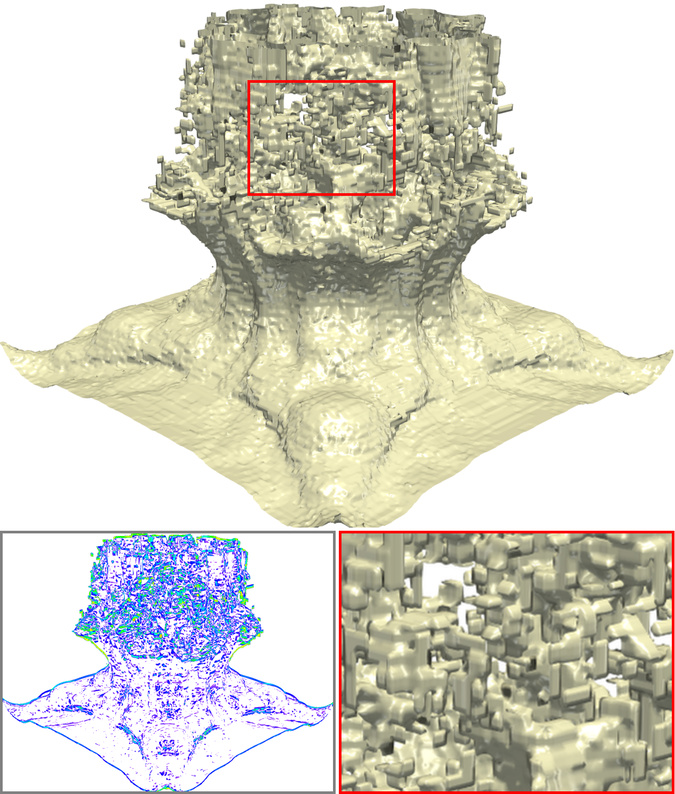}\\
 \mbox{\small GT} &
 \mbox{\small EVOLVE} &
 \mbox{\small SZ3} &
 \mbox{\small TTHRESH} &
 \mbox{\small ZFP}\\
\end{array}$
\vspace{-.125in}
\caption{Comparison of isosurface rendering results between EVOLVE and conventional lossy compressors. Top and bottom: \textsf{combustion (MF)} and \textsf{ionization (H+)}. The chosen isovalues are 0.5 and 0.1, respectively.}
\label{fig:traditional-iso-appx}
\end{figure*}

\begin{figure*}[htb]
\centering
$\begin{array}{c@{\hspace{0.01in}}c@{\hspace{0.01in}}c@{\hspace{0.01in}}c@{\hspace{0.01in}}c@{\hspace{0.01in}}c@{\hspace{0.01in}}c@{\hspace{0.01in}}c@{\hspace{0.01in}}c}
\mbox{\tiny CR / PSNR} &
 \mbox{\tiny 10,517$\times$ / 45.62 dB} &
 \mbox{\tiny 1,714$\times$ / 43.35 dB} &
 \mbox{\tiny 5,954$\times$ / 41.56 dB} &
 \mbox{\tiny 1,979$\times$ / 41.37 dB} &
 \mbox{\tiny 1,575$\times$ / 40.78 dB} &
 \mbox{\tiny 936$\times$ / 40.24 dB} &
 \mbox{\tiny 1,839$\times$ / 42.22 dB} &
 \mbox{\tiny 1,808$\times$ / 41.28 dB} \\
 \includegraphics[width=0.1075\linewidth]{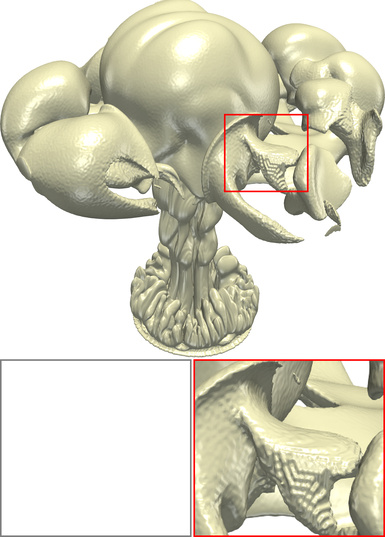}&
 \includegraphics[width=0.1075\linewidth]{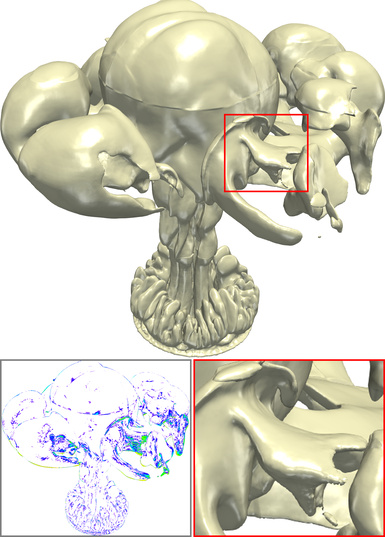}&
 \includegraphics[width=0.1075\linewidth]{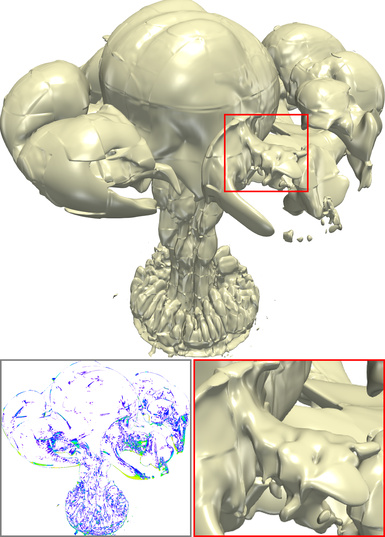}&
 \includegraphics[width=0.1075\linewidth]{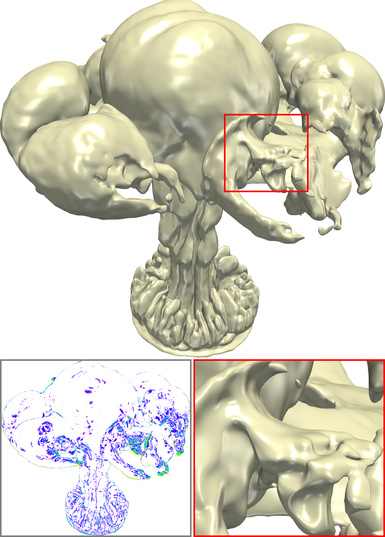}&
 \includegraphics[width=0.1075\linewidth]{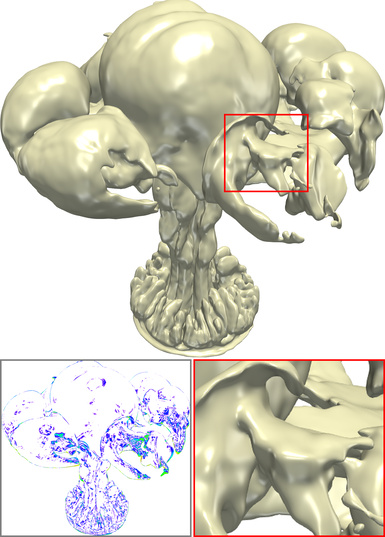}&
 \includegraphics[width=0.1075\linewidth]{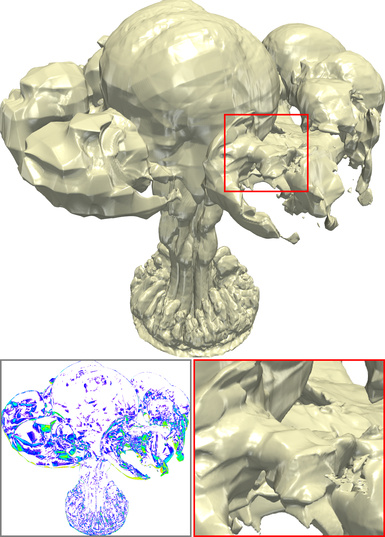}&
 \includegraphics[width=0.1075\linewidth]{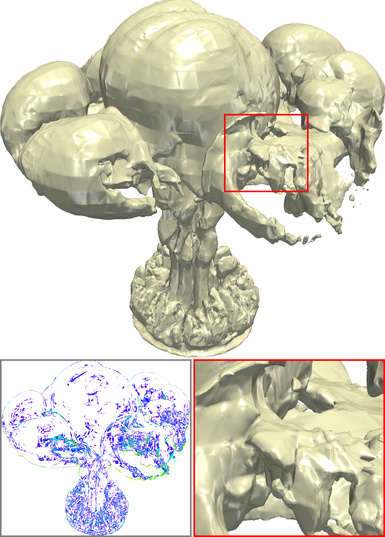}&
 \includegraphics[width=0.1075\linewidth]{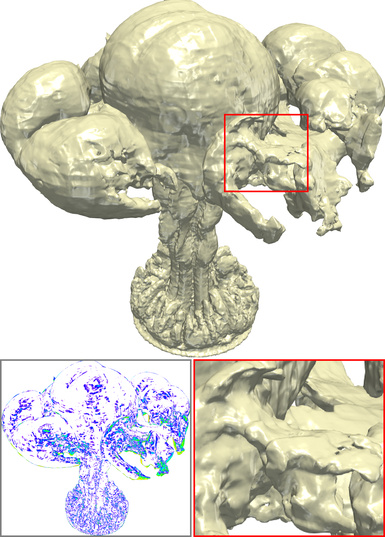}&
 \includegraphics[width=0.1075\linewidth]{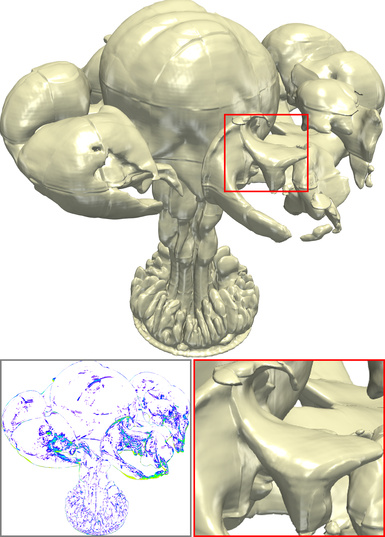}\\
\mbox{\tiny CR / PSNR} &
 \mbox{\tiny 2,596$\times$ / 44.91 dB} &
 \mbox{\tiny 446$\times$ / 41.50 dB} &
 \mbox{\tiny 1,044$\times$ / 42.29 dB} &
 \mbox{\tiny 633$\times$ / 40.44 dB} &
 \mbox{\tiny 371$\times$ / 40.90 dB} &
 \mbox{\tiny 166$\times$ / 42.23 dB} &
 \mbox{\tiny 168$\times$ / 41.35 dB} &
 \mbox{\tiny 1,114$\times$ / 42.17 dB} \\
 \includegraphics[width=0.1075\linewidth]{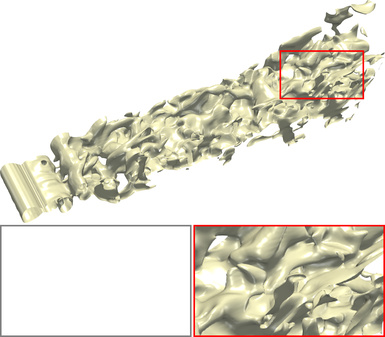}&
 \includegraphics[width=0.1075\linewidth]{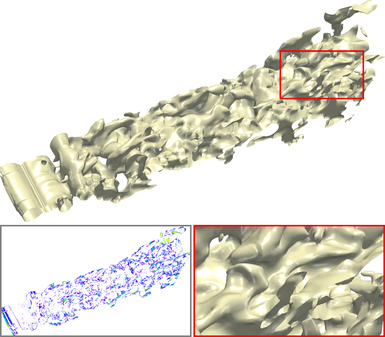}&
 \includegraphics[width=0.1075\linewidth]{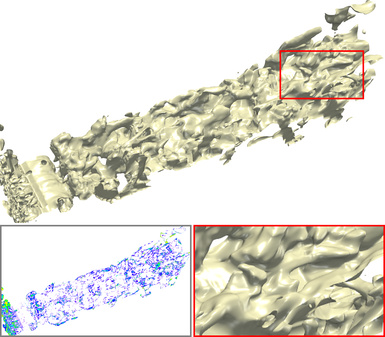}&
 \includegraphics[width=0.1075\linewidth]{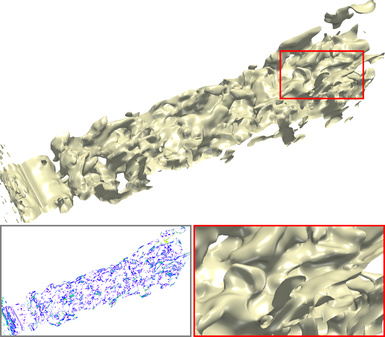}&
 \includegraphics[width=0.1075\linewidth]{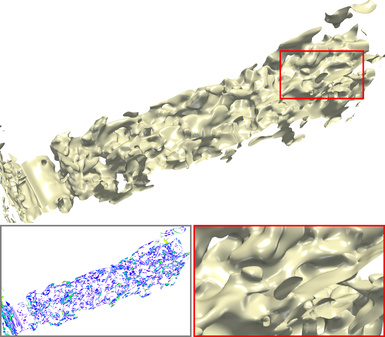}&
 \includegraphics[width=0.1075\linewidth]{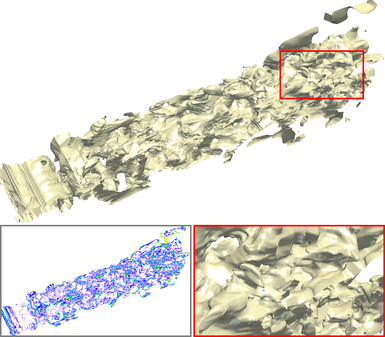}&
 \includegraphics[width=0.1075\linewidth]{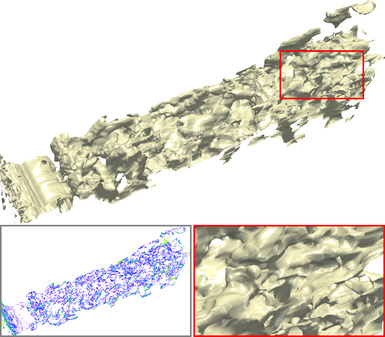}&
 \includegraphics[width=0.1075\linewidth]{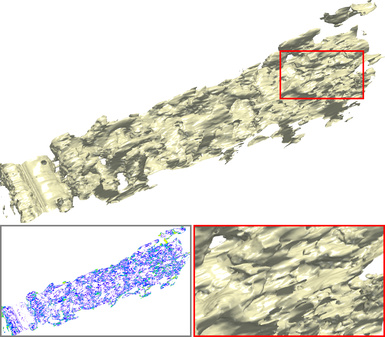}&
 \includegraphics[width=0.1075\linewidth]{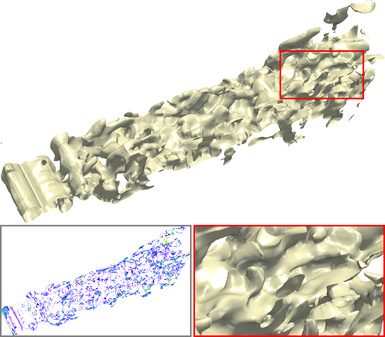}\\
\mbox{\tiny CR / PSNR} &
 \mbox{\tiny 10,774$\times$ / 45.20 dB} &
 \mbox{\tiny 3,047$\times$ / 43.29 dB} &
 \mbox{\tiny 6,649$\times$ / 43.48 dB} &
 \mbox{\tiny 5,256$\times$ / 42.09 dB} &
 \mbox{\tiny 1,575$\times$ / 44.23 dB} &
 \mbox{\tiny 1,813$\times$ / 42.16 dB} &
 \mbox{\tiny 1,839$\times$ / 43.91 dB} &
 \mbox{\tiny 3,748$\times$ / 38.84 dB} \\
 \includegraphics[width=0.1075\linewidth]{figures/iso_magnetic_gt.jpg}&
 \includegraphics[width=0.1075\linewidth]{figures/iso_magnetic_evolve.jpg}&
 \includegraphics[width=0.1075\linewidth]{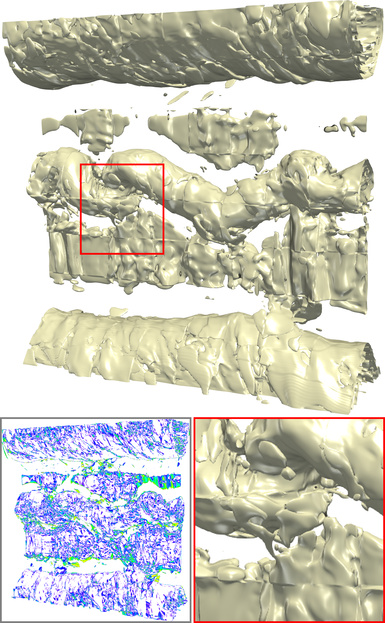}&
 \includegraphics[width=0.1075\linewidth]{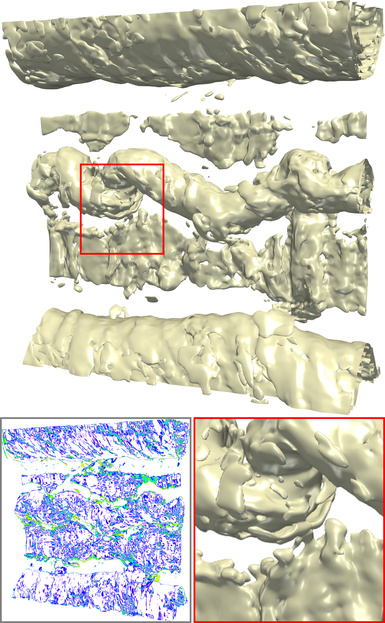}&
 \includegraphics[width=0.1075\linewidth]{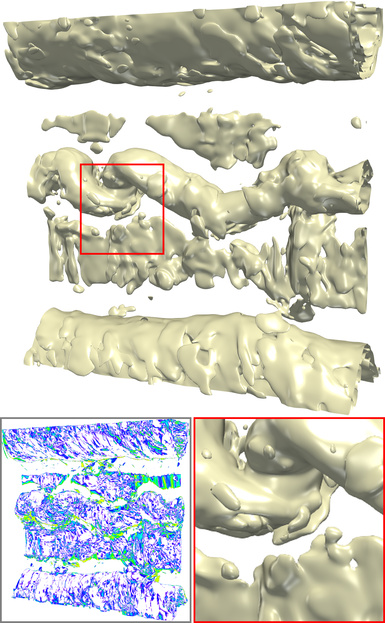}&
 \includegraphics[width=0.1075\linewidth]{figures/iso_magnetic_amgsrn++.jpg}&
 \includegraphics[width=0.1075\linewidth]{figures/iso_magnetic_fv-srn.jpg}&
 \includegraphics[width=0.1075\linewidth]{figures/iso_magnetic_instant-ngp.jpg}&
 \includegraphics[width=0.1075\linewidth]{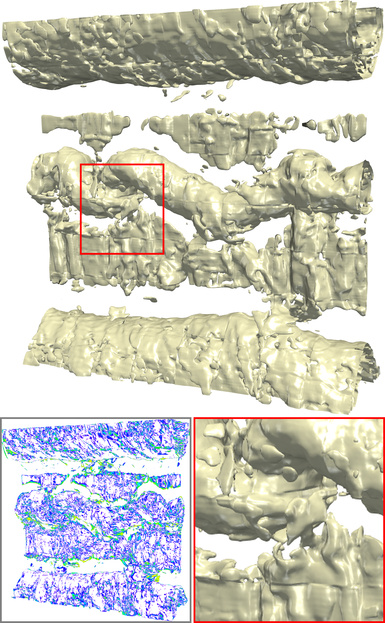}\\
 \mbox{\small GT} &
 \mbox{\small EVOLVE} &
 \mbox{\small ECNR} &
 \mbox{\small NeurComp} &
 \mbox{\small SIREN} &
 \mbox{\small AMGSRN++} &
 \mbox{\small fV-SRN} &
 \mbox{\small Instant-NGP} &
 \mbox{\small IDLat}\\
\end{array}$
\vspace{-.125in}
\caption{Comparison of isosurface rendering results between EVOLVE and \hot{deep-learning-based} compressors. From top to bottom: \textsf{gas}, \textsf{half-cylinder (VLM, 6,400)}, and \textsf{magnetic}. The chosen isovalues are 0.25, 0.6, and 0.05, respectively.}
\label{fig:neural-iso-appx}
\end{figure*}

Due to page limits, the main paper only presents a subset of rendering comparisons. In Figures~\ref{fig:traditional-vol-appx}, \ref{fig:neural-vol-appx}, \ref{fig:traditional-iso-appx}, and \ref{fig:neural-iso-appx}, we provide the complete set of volume rendering and isosurface rendering results across all test datasets, comparing EVOLVE with both conventional lossy compressors and \hot{deep-learning-based} compressors at their respective CRs and reconstruction quality.


\vspace{-0.05in}
\section{\hot{Evaluation of Scanned Volumetric Data}}
\label{sec:scan-eval}

\begin{figure*}[p]
\centering
$\begin{array}{c@{\hspace{0.1in}}c@{\hspace{0.1in}}c@{\hspace{0.1in}}c@{\hspace{0.1in}}c}
\mbox{\small CR / PSNR} &
 \mbox{\small 728.1$\times$ / 49.64 dB} &
 \mbox{\small 701.9$\times$ / 46.22 dB} &
 \mbox{\small 80.4$\times$ / 43.46 dB} &
 \mbox{\small 127.3$\times$ / 43.72 dB} \\
 \includegraphics[width=0.16\linewidth]{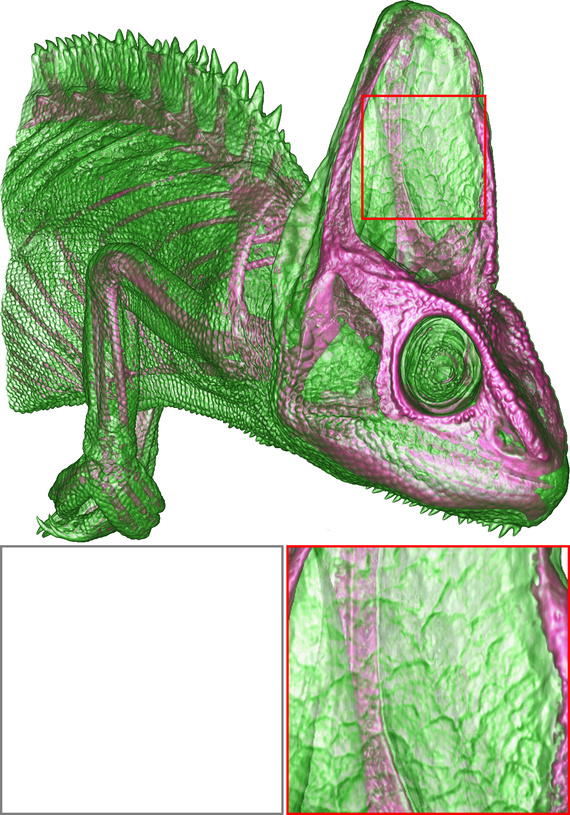}&
 \includegraphics[width=0.16\linewidth]{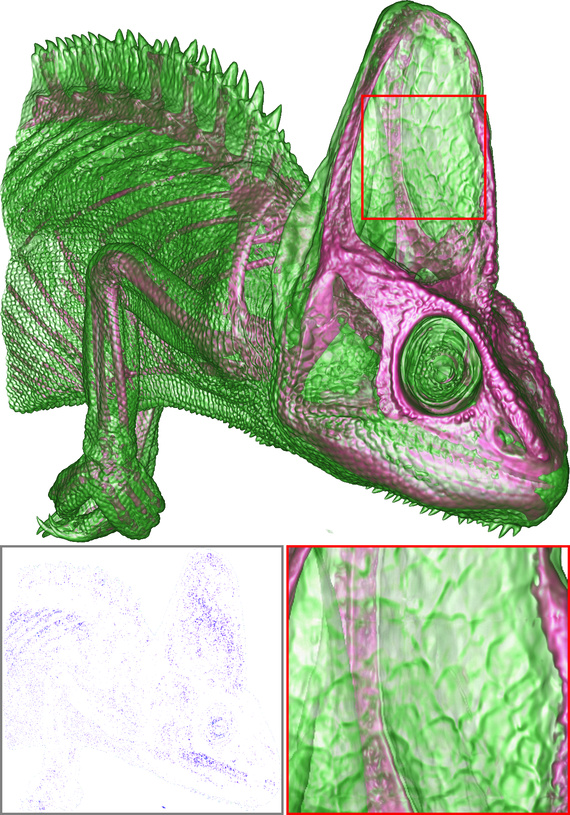}&
 \includegraphics[width=0.16\linewidth]{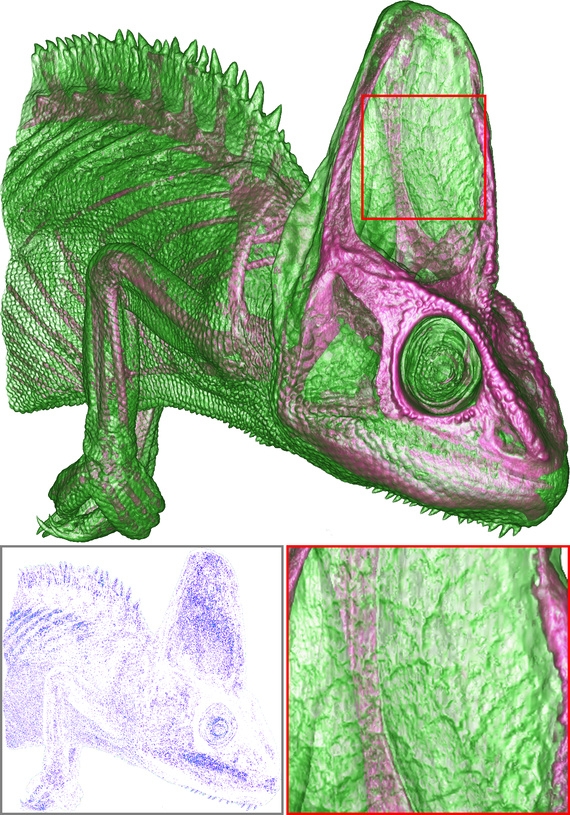}&
 \includegraphics[width=0.16\linewidth]{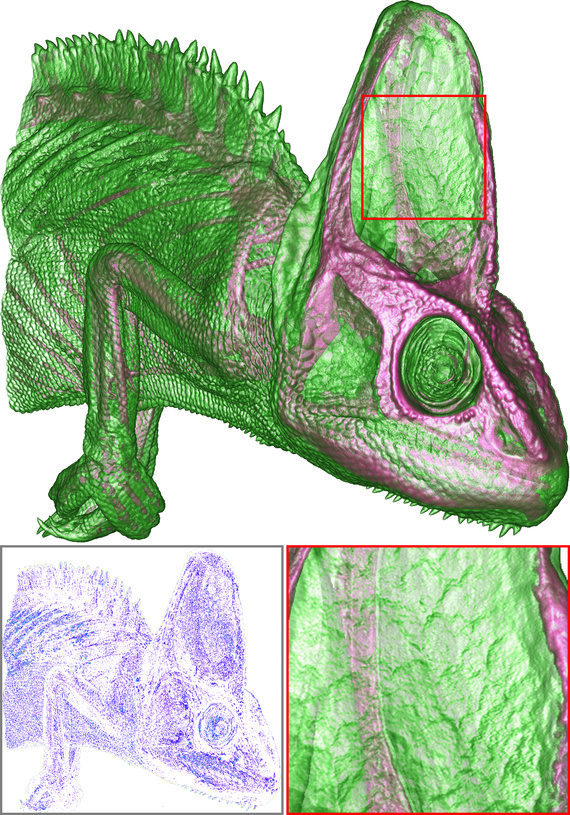}&
 \includegraphics[width=0.16\linewidth]{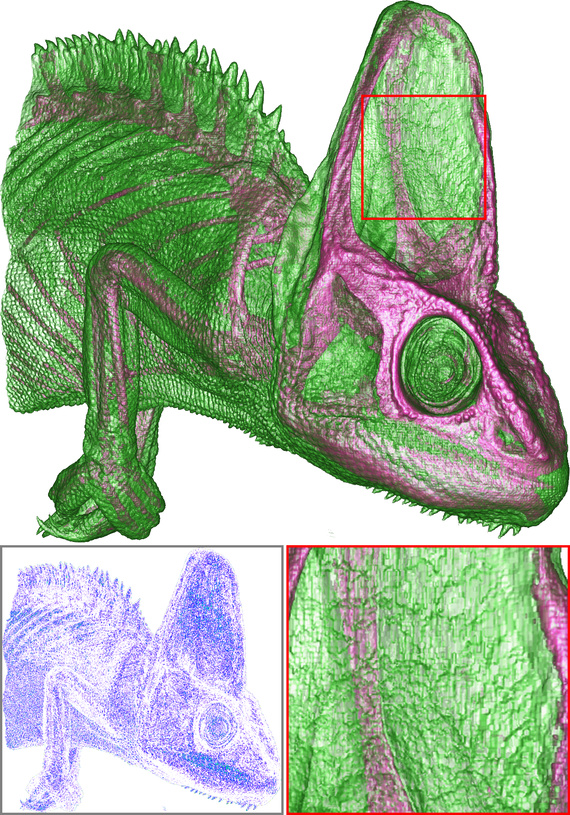}\\

\mbox{\small CR / PSNR} &
 \mbox{\small 326.6$\times$ / 44.08 dB} &
 \mbox{\small 99.7$\times$ / 43.50 dB} &
 \mbox{\small 103.2$\times$ / 43.45 dB} &
 \mbox{\small 79.3$\times$ / 43.13 dB} \\
 \includegraphics[width=0.16\linewidth]{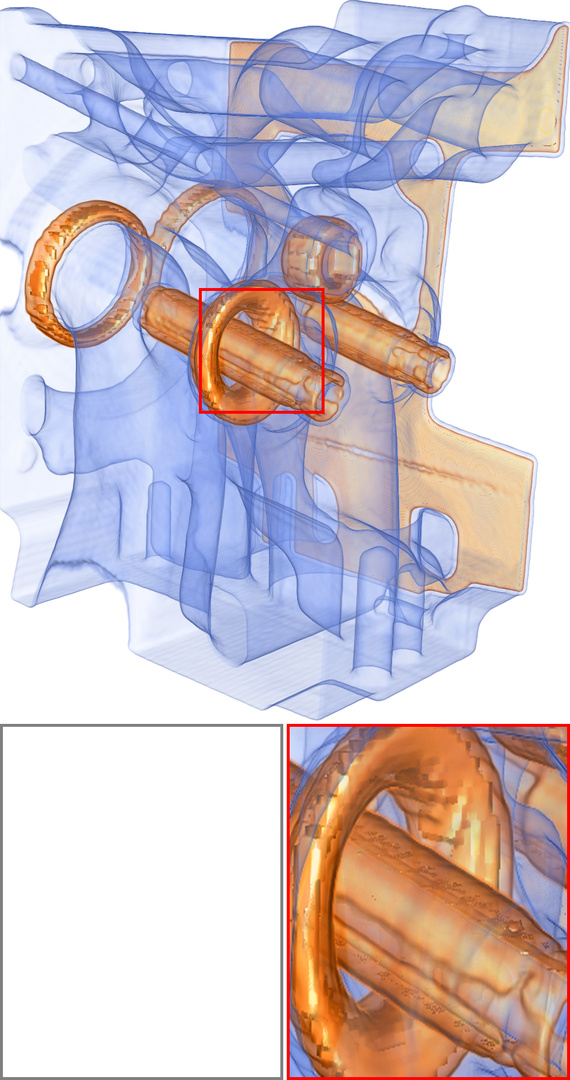}&
 \includegraphics[width=0.16\linewidth]{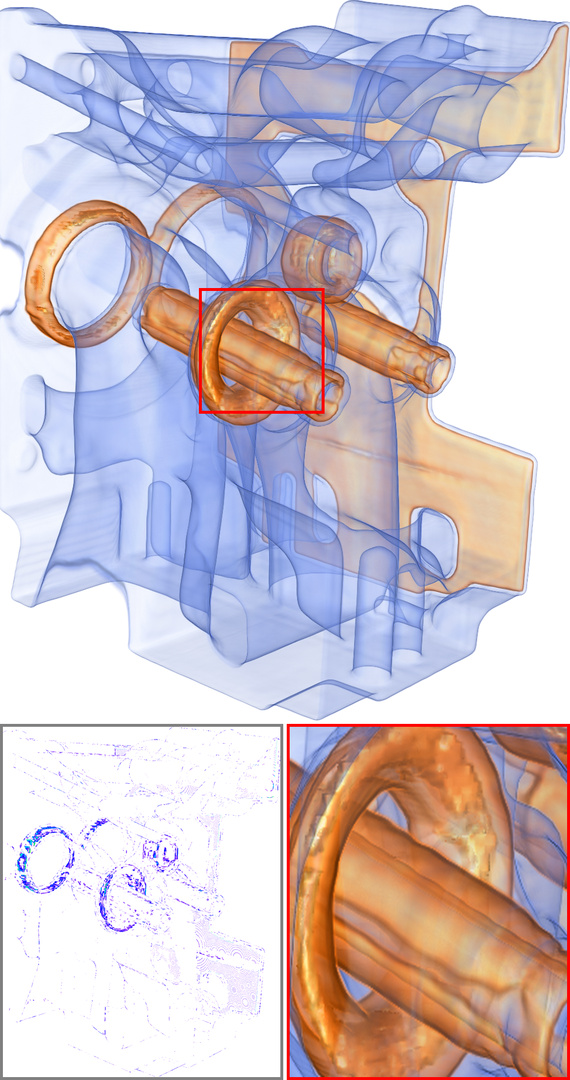}&
 \includegraphics[width=0.16\linewidth]{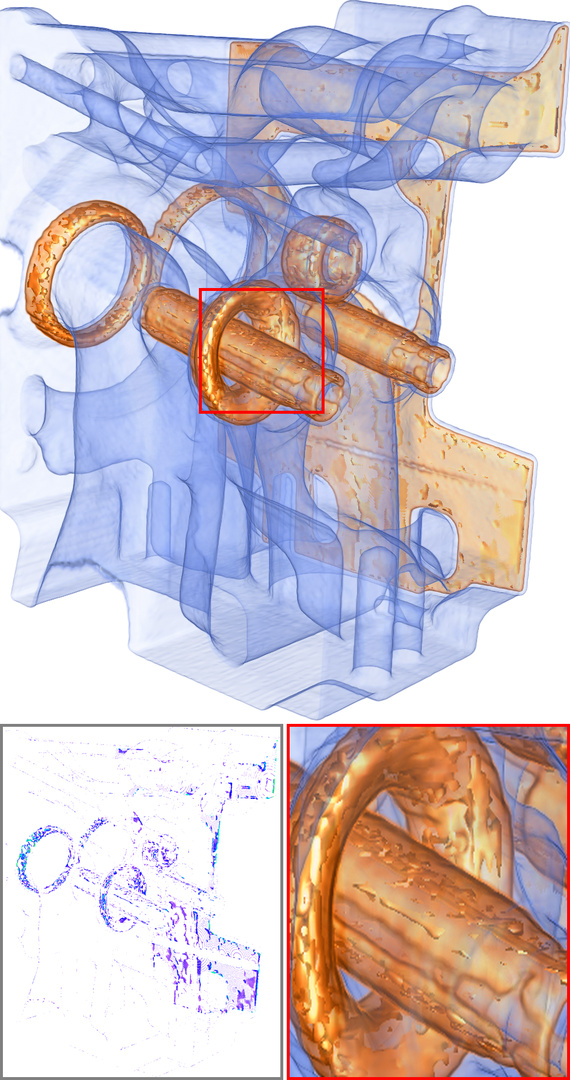}&
 \includegraphics[width=0.16\linewidth]{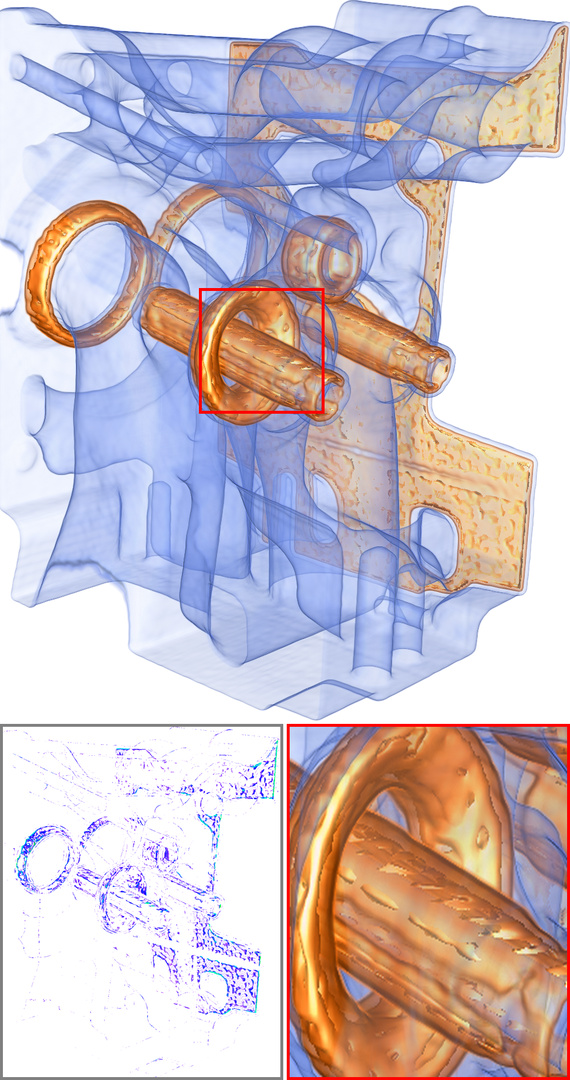}&
 \includegraphics[width=0.16\linewidth]{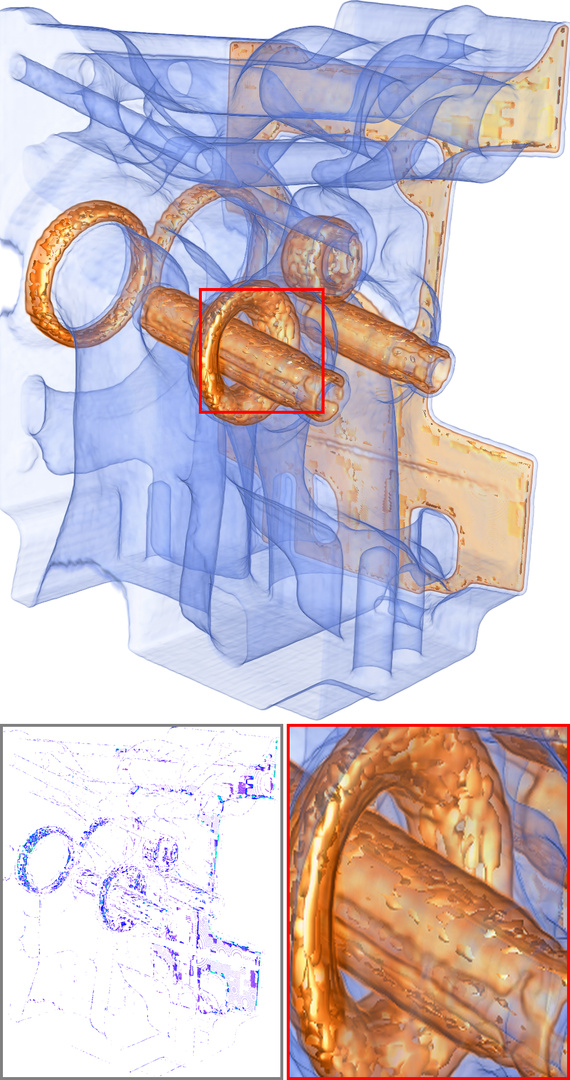}\\

\mbox{\small CR / PSNR} &
 \mbox{\small 257.4$\times$ / 33.71 dB} &
 \mbox{\small 35.4$\times$ / 45.59 dB} &
 \mbox{\small 19.8$\times$ / 45.68 dB} &
 \mbox{\small 29.0$\times$ / 44.11 dB} \\
 \includegraphics[width=0.16\linewidth]{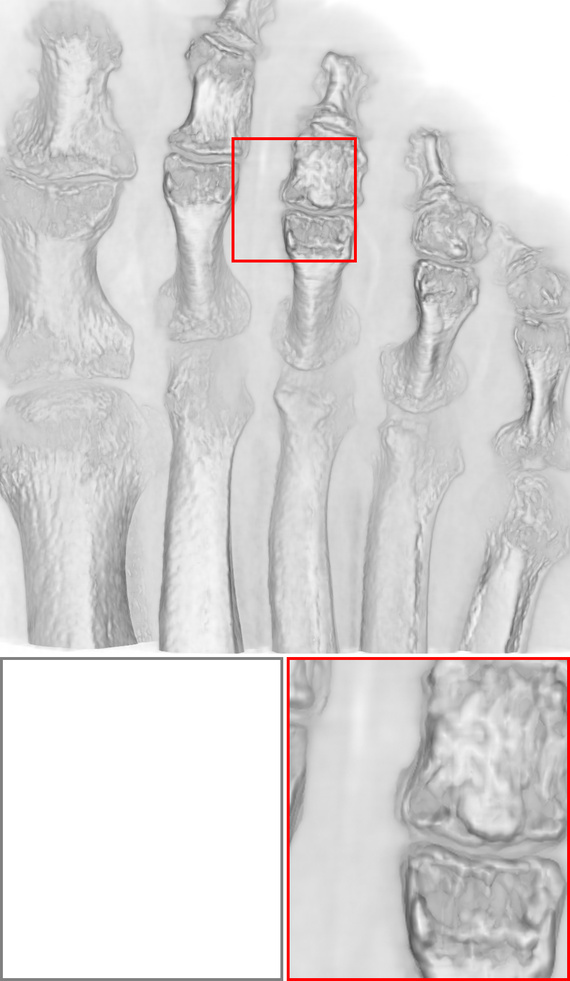}&
 \includegraphics[width=0.16\linewidth]{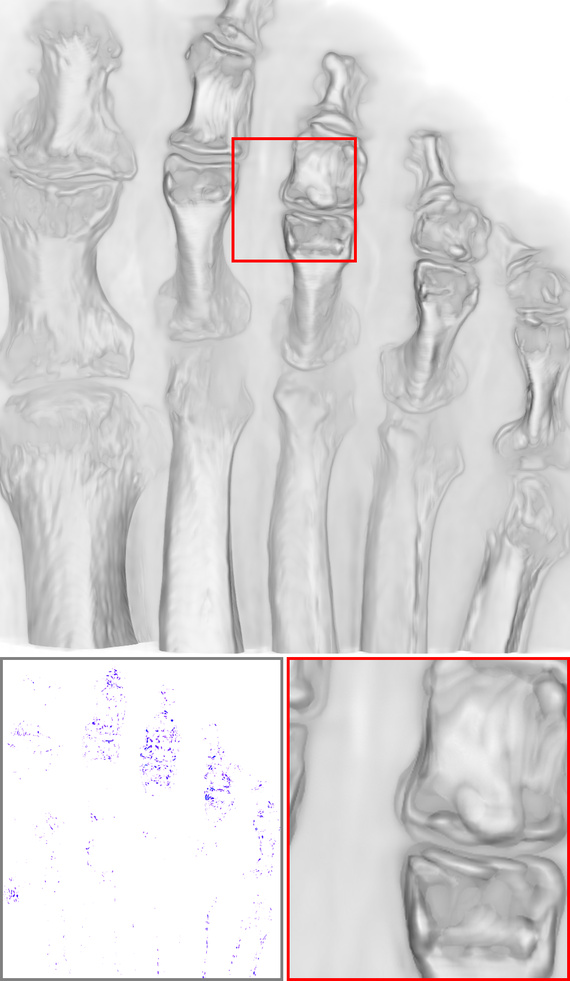}&
 \includegraphics[width=0.16\linewidth]{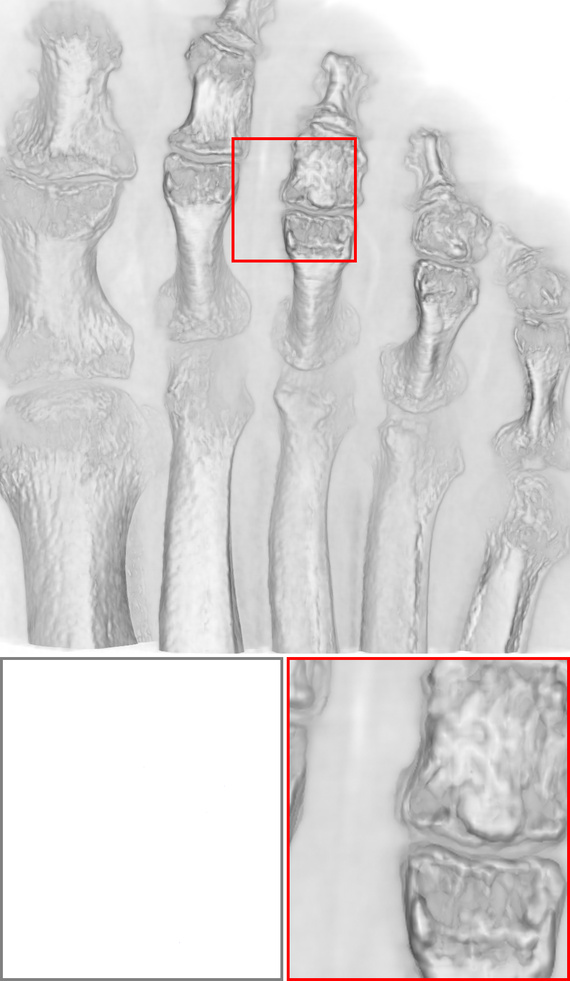}&
 \includegraphics[width=0.16\linewidth]{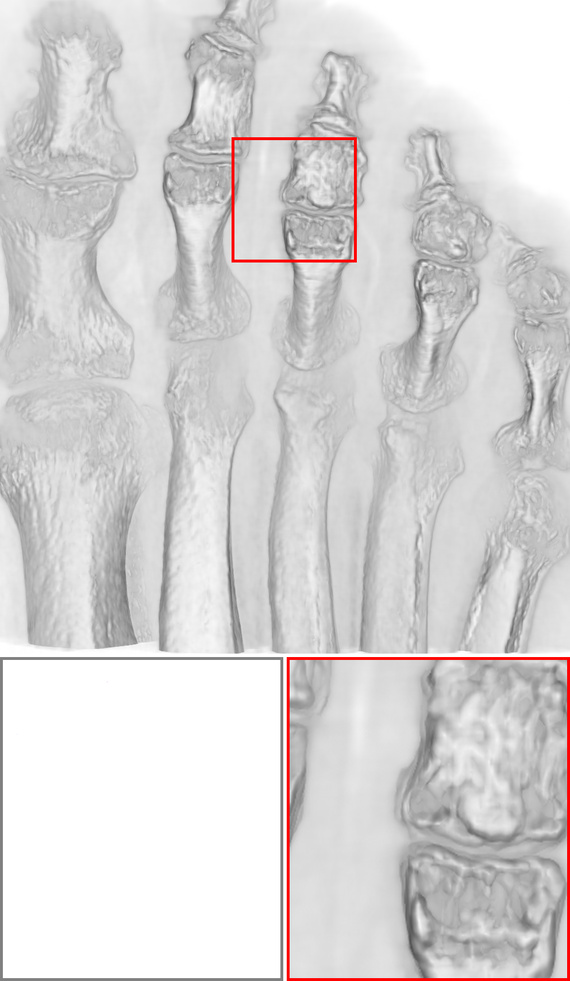}&
 \includegraphics[width=0.16\linewidth]{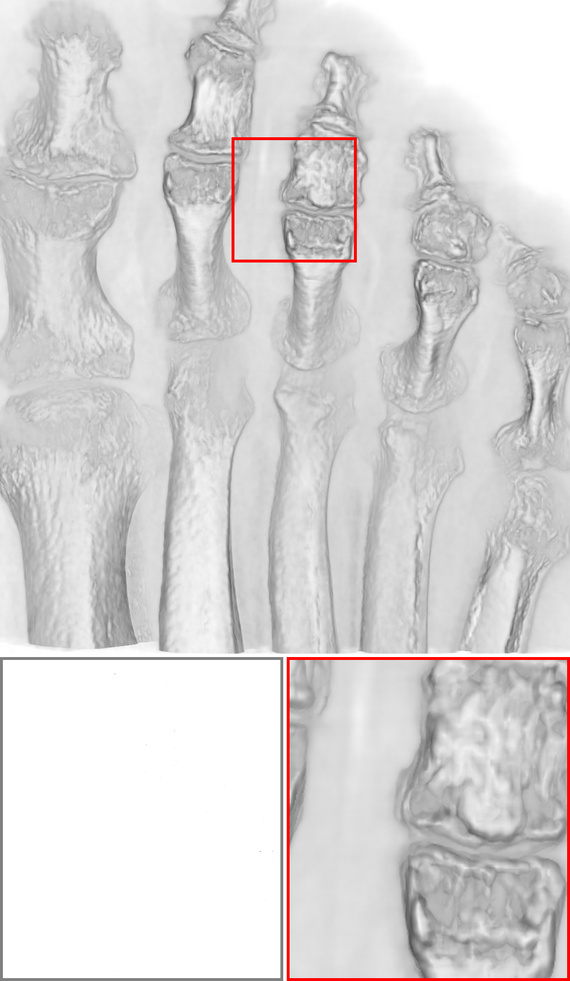}\\
 
\mbox{\small CR / PSNR} &
 \mbox{\small 3,829.7$\times$ / 50.54 dB} &
 \mbox{\small 1,061.1$\times$ / 44.67 dB} &
 \mbox{\small 49.5$\times$ / 44.94 dB} &
 \mbox{\small 919.6$\times$ / 42.92 dB} \\
 \includegraphics[width=0.16\linewidth]{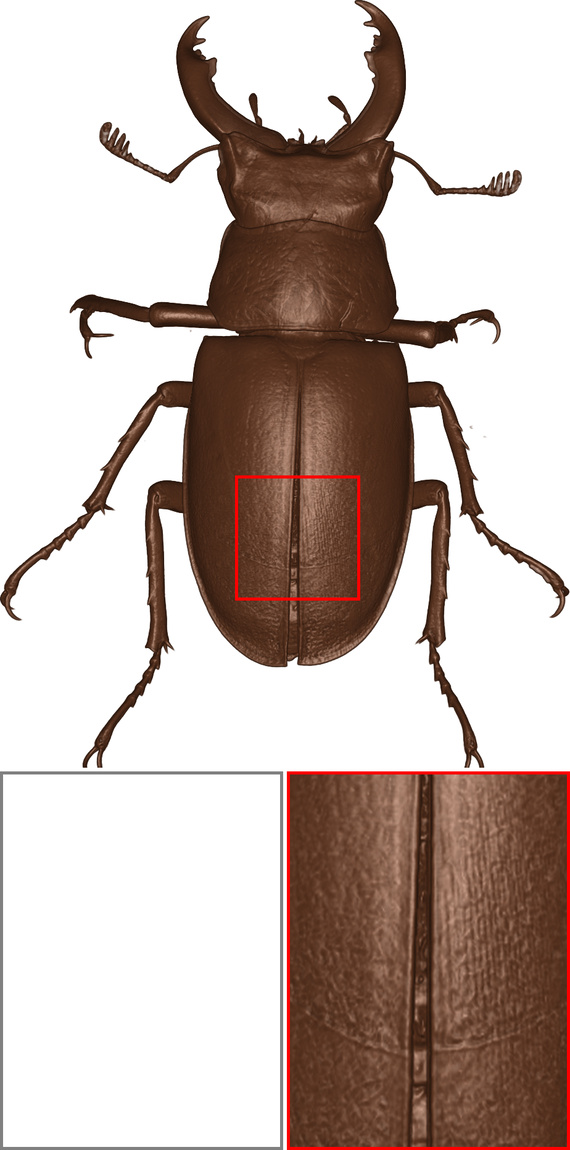}&
 \includegraphics[width=0.16\linewidth]{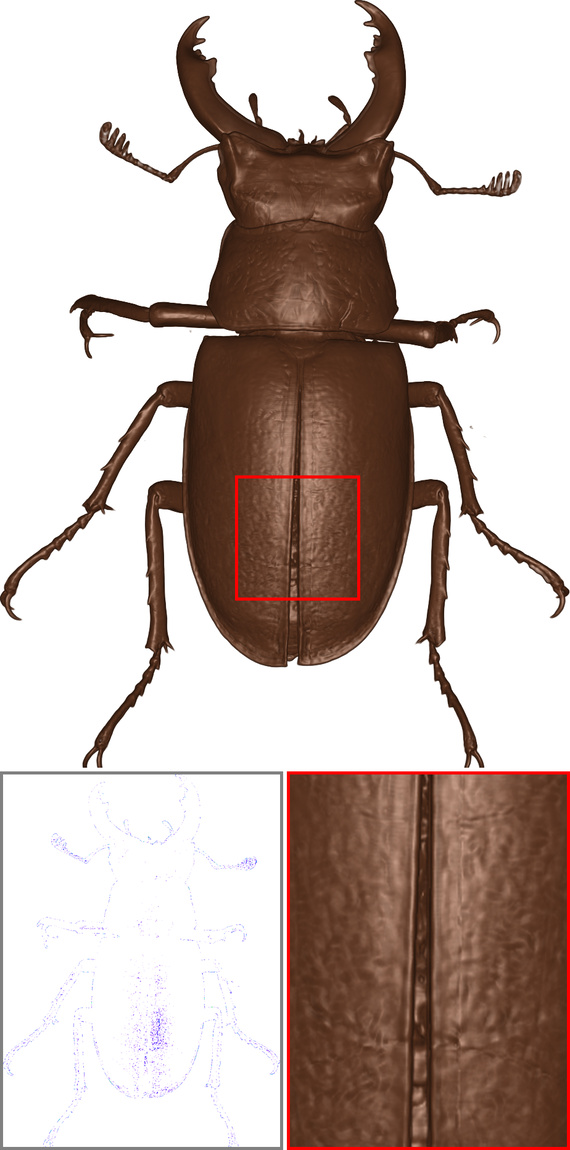}&
 \includegraphics[width=0.16\linewidth]{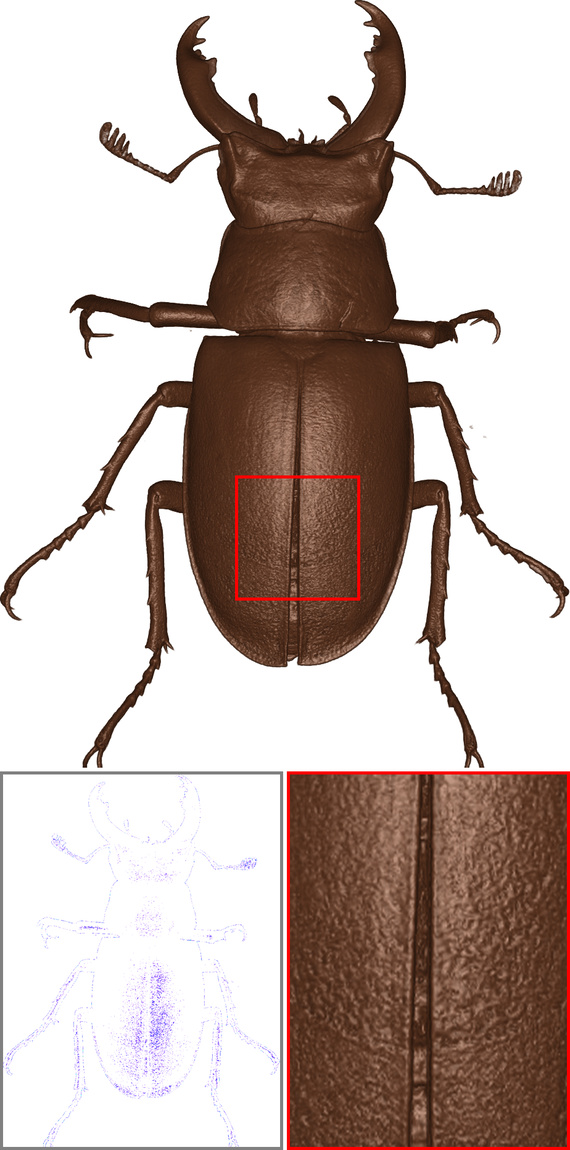}&
 \includegraphics[width=0.16\linewidth]{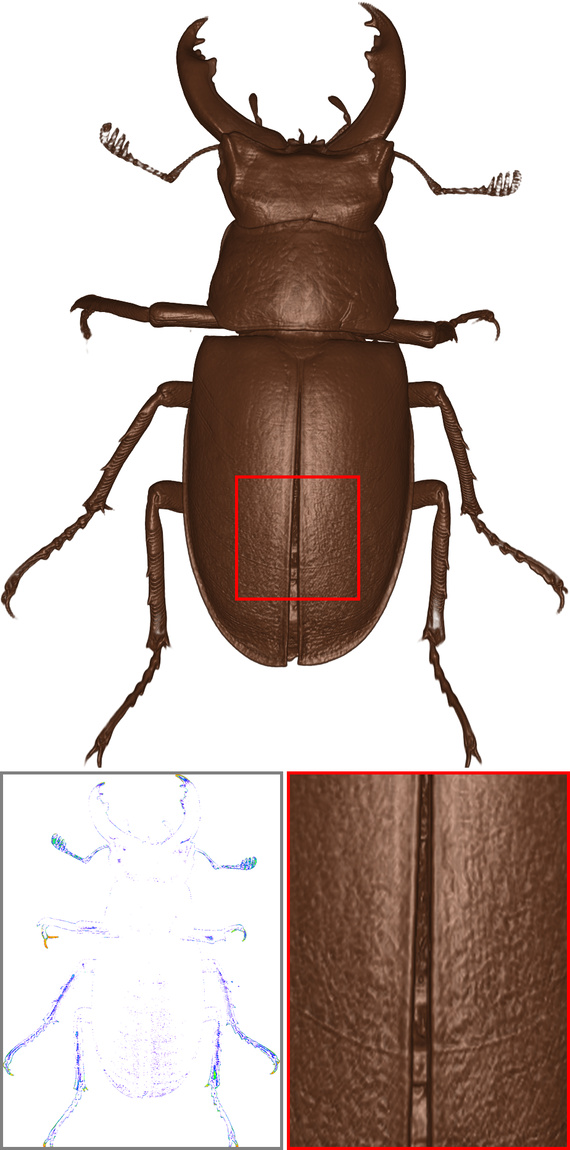}&
 \includegraphics[width=0.16\linewidth]{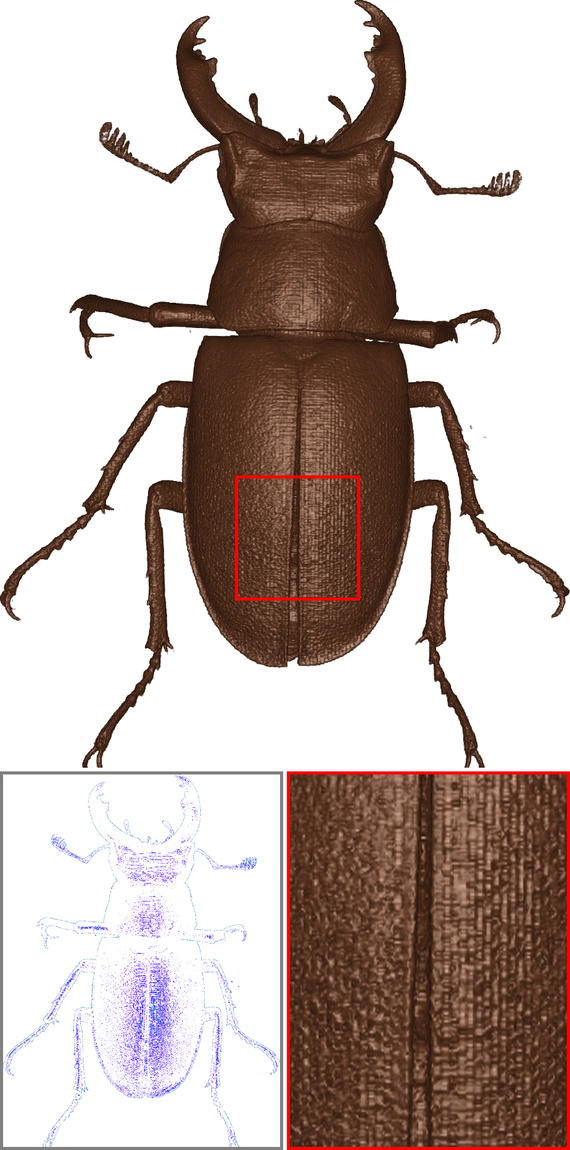}\\
 \mbox{\small GT} &\mbox{\small EVOLVE} &\mbox{\small SZ3} &\mbox{\small TTHRESH} &\mbox{\small ZFP}\\
\end{array}$
\vspace{-.125in}
\caption{\hot{Volume rendering comparison on scanned data (top to bottom: 
\textsf{chameleon}, 
\textsf{engine}, 
\textsf{foot}, 
and \textsf{stag-beetle}). 
For each method, the bottom-left inset shows the pixel-wise difference from GT, and the bottom-right inset shows the zoom-in of the red box. For \textsf{foot}, EVOLVE (33.71 dB) visibly smooths the trabecular bone texture, reflecting its quality saturation on out-of-distribution CT data, while for the other datasets EVOLVE attains the highest CR at matched or higher quality without any retraining.}}
\label{fig:scan-vol}
\end{figure*}

\hot{To assess how EVOLVE behaves outside its training distribution, we evaluate it on four scanned volumes from the Open SciVis Datasets~\cite{Klacansky-OSV}: 
\textsf{chameleon} (1,024$\times$1,024$\times$1,080), 
\textsf{engine} (256$\times$256$\times$128), 
\textsf{foot} (256$^3$), and
\textsf{stag-beetle} (832$\times$832$\times$494), all normalized to $[0,1]$ in FP32.
We compare EVOLVE (the released model, without any retraining) against ZFP, TTHRESH, and SZ3.
For each dataset, EVOLVE operates at the high-quality level.
Table~\ref{tab:scan-results} reports the quantitative results, and Figure~\ref{fig:scan-vol} presents the corresponding rendering results.
Because these scanned volumes fall outside EVOLVE's training distribution, its overall benefit here is limited: although EVOLVE still edges out the conventional compressors on \textsf{chameleon}, \textsf{engine}, and \textsf{stag-beetle}, the margin on \textsf{engine} is small. 
For \textsf{foot}, its quality saturates at 33.7 dB and cannot be improved further by increasing the rate budget.
This demonstrates the necessity of training on broader, more diverse data to further improve EVOLVE's performance and generalization across a wider range of volumes.}

\begin{table}[!ht]
\centering
\caption{\hot{PSNR (dB), LPIPS, ET (sec), DT (sec), and CR on scanned volumetric data. For \textsf{foot} and \textsf{engine}, EVOLVE reports its highest achievable quality.}}
\label{tab:scan-results}
\vspace{-0.05in}
\resizebox{0.8\columnwidth}{!}{%
\begin{tabular}{c c c c c c c}
\hot{dataset} & \hot{method} & \hot{PSNR$\uparrow$} & \hot{LPIPS$\downarrow$} & \hot{ET$\downarrow$} & \hot{DT$\downarrow$} & \hot{CR$\uparrow$} \\
\hline
\multirow{4}{*}{\hot{\textsf{chameleon}}}
& \hot{ZFP}  & \hot{43.72} & \hot{0.1685} & \hot{\textbf{6.22}} & \hot{\textbf{3.68}} & \hot{127.3} \\
& \hot{TTHRESH}  & \hot{43.46} & \hot{0.1196} & \hot{225.75} & \hot{83.18} & \hot{80.4} \\
& \hot{SZ3}      & \hot{46.22} & \hot{0.0831} & \hot{16.38} & \hot{8.71} & \hot{701.9} \\
& \hot{EVOLVE}   & \hot{\textbf{49.64}} & \hot{\textbf{0.0390}} & \hot{67.61} & \hot{55.04} & \hot{\textbf{728.1}} \\
\hline
\multirow{4}{*}{\hot{\textsf{engine}}}
& \hot{ZFP}  & \hot{43.13} & \hot{0.0663} & \hot{\textbf{0.07}} & \hot{\textbf{0.03}} & \hot{79.3} \\
& \hot{TTHRESH}  & \hot{43.45} & \hot{0.0663} & \hot{2.98} & \hot{0.30} & \hot{103.2} \\
& \hot{SZ3}      & \hot{43.50} & \hot{0.0623} & \hot{0.14} & \hot{0.09} & \hot{99.7} \\
& \hot{EVOLVE}   & \hot{\textbf{44.08}} & \hot{\textbf{0.0600}} & \hot{0.51} & \hot{0.66} & \hot{\textbf{326.6}} \\
\hline
\multirow{4}{*}{\hot{\textsf{foot}}}
& \hot{ZFP}  & \hot{44.11} & \hot{0.0063} & \hot{\textbf{0.12}} & \hot{\textbf{0.10}} & \hot{29.0}  \\
& \hot{TTHRESH}  & \hot{\textbf{45.68}} & \hot{\textbf{0.0031}} & \hot{5.61} & \hot{1.14} & \hot{19.8} \\
& \hot{SZ3}      & \hot{45.59} & \hot{0.0036} & \hot{0.28} & \hot{0.21} & \hot{35.4} \\
& \hot{EVOLVE}   & \hot{33.71} & \hot{0.0929} & \hot{1.07} & \hot{0.94} & \hot{\textbf{257.4}} \\
\hline
\multirow{4}{*}{\hot{\textsf{stag-beetle}}}
& \hot{ZFP}  & \hot{42.92} & \hot{0.0801} & \hot{\textbf{0.78}} & \hot{\textbf{0.85}} & \hot{919.6} \\
& \hot{TTHRESH}  & \hot{44.94} & \hot{0.0357} & \hot{64.71} & \hot{22.27} & \hot{49.5}  \\
& \hot{SZ3}      & \hot{44.67} & \hot{0.0292}  & \hot{3.89} & \hot{2.05} & \hot{1,061.1} \\
& \hot{EVOLVE}   & \hot{\textbf{50.54}} & \hot{\textbf{0.0247}} & \hot{30.11} & \hot{26.49} & \hot{\textbf{3,829.7}} \\
\end{tabular}
}
\end{table}

